**Unraveling the Planar-Globular Transition in Gold Nanoclusters through Evolutionary Search**

*Alper Kinaci,[1] Badri Narayanan,[1] Fatih G. Sen,[1] Michael J. Davis,[2] Stephen K. Gray,[1] Subramanian K. R. S. Sankaranarayanan,[1] Maria K. Y. Chan[1*]*

[1]Center for Nanoscale Materials, Argonne National Laboratory, Lemont, IL 60439 USA

[2]Chemical Sciences Division, Argonne National Laboratory, Lemont, IL 60439 USA

**Corresponding author*

**ABSTRACT**

Au nanoclusters are of technological relevance for catalysis, photonics, sensors, and of fundamental scientific interest owing to planar to globular structural transformation at an anomalously high number of atoms i.e. in the range 12-14. The nature and causes of this transition remain a mystery. In order to unravel this conundrum, high throughput density functional theory (DFT) calculations, coupled with a global structural optimization scheme based on a modified genetic algorithm (GA) are conducted. More than 20,000 $Au_{12}$, $Au_{13}$, and $Au_{14}$ nanoclusters are evaluated. With any DFT functional, globular and planar structures coexist across the size range of interest. The planar-globular transition is gradual at room temperature rather than a sharp transition as previously believed. The effects of anionicity, *s-d* band hybridization and long range interactions on the dimensional transition are quantified by using the structures adjacent to minima. Anionicity marginally changes the relative stability of the clusters. The degree of *s*-*d* hybridization is varied via changing the Hubbard U value which corroborate that *s*-*d* hybridization alone does not stabilize planar structures. van der Waals interactions, on the other hand, stabilize globular structures. These results elucidate the balance between the different reasons of the dimensional transition in gold nanoclusters.

Gold catalysts have been the subject of intense research after the demonstration of catalytic activity of gold nanoparticles for CO oxidation a few decades ago.[1] As interest grew in gold nanoparticles, other catalytic properties[2] along with their applications in optoelectronics,[3] molecular assembly,[4] biorecognition and chemical sensors[5] emerged. From an engineering point of view, the control and manipulation of the properties of nanoclusters in these applications are critical. Two important control parameters in nanoclusters are size and shape.[6, 7] Considerable effort has been devoted to gold clusters with fewer than 100 atoms to understand the geometric properties of various sizes, and as a result several interesting observations have been made: In contrast to alkali and other noble metal clusters which exhibit planar (2D) to globular (3D) transformations in the size range of 5-7 atoms,[8, 9] planarity is conserved until the 12-14 atom range for Au nanoparticles.[10, 11, 12, 13] This anomalously high atom number for planar nanostructure is often claimed to be associated with the tug of war between strong 5*d*-6*s* orbital hybridization due to relativistic effects[14, 15] and "aurophilicity",[16] which is broadly defined as an attractive interaction of dispersive character. The structures of the 3D clusters also vary depending on their size. Above the 2D to 3D transformation point, hollow cages, tubes and space-filling (such as tetrahedron, icosahedron, Wulff shaped) structures may form without distinct transition ranges for $Au_n$ clusters with $n < 100$.[13, 17, 18, 19, 20]

In determining $Au_n$ structures, computational approaches[9, 15, 18, 21, 22] are heavily employed in addition to spectroscopic measurements.[10, 17] Among theoretical approaches, density functional theory (DFT) is widely employed since it offers a powerful method for the first-principles prediction of electronic and structural properties of materials at a reasonable computational cost. Using a local minimization scheme such as the conjugate gradient method, the local minimum close to a certain configuration can usually be correctly identified using DFT-derived forces. However, finding the global minimum structure for a given composition requires global minimization techniques such as simulated annealing,[23] genetic algorithm (GA),[24] basin[21] and minima hopping,[25] and particle swarm[26] methods, coupled with a reliable local optimizer. These stochastic global optimization methods have increasingly been utilized in materials discovery and development.[27, 28, 29, 30, 31, 32]

GA is an evolutionary algorithm, which is based on the idea of the survival of the fittest. Inspired from nature, the algorithm starts with a set of parent configurations, which is defined as the population. For structural optimization of materials, the parents may have different lattice types, atomic basis and even composition. Some of the individuals in this population are more fit – which in the context of structural optimization means lower in total energy – based on their structures compared to others. Fit structures survive and are allowed to mate. Mating is accomplished by crossing patterns (genes) between fit individuals. During crossover, random mutations in the genes are also allowed to a certain degree to avoid a stagnant gene pool and a better sampling of the phase space. The offspring individuals form the next generation of parents and this process continues until some pre-defined criteria are met. In creating the genome of a structure, two methods are generally applied. In the first one, the properties of the structure, such as lattice parameters, lattice angles, location of atoms, type of atoms *etc*. are converted to bit-strings[33] which are subjected to crossover. In the second method, two "fit" parent structures are spatially decomposed into smaller fragments. Then, the fragments from different parents are combined to form a child structure, mimicking the process of dividing and recombining chromosome pairs in biology.[34]

Gold nanoparticles have been studied using global optimization algorithms. Basin hopping, which is reminiscent of a Monte Carlo method, combined with an empirical potential (for $Au_{n<110}$)[21] and DFT calculations (for $Au_{n<20}$)[18] have been used to study unsupported clusters. Supported $Au_8$ on MgO was investigated with a genetic algorithm coupled with DFT.[35] These studies, however, did not discuss or provide an ensemble of low energy structures that were close enough in energy to the global minimum to be accessible during typical synthesis conditions. Also, the impacts of ionicity and long-range interactions on the relative stabilities of near minima have not been reported.

In this study, we implement a genetic algorithm employing 4-parent crossover and adaptive genetic manipulations and couple it with DFT to search for global minima in isolated $Au_{12}$, $Au_{13}$, and $Au_{14}$ clusters. Both 2D and 3D evolutions are pursued, *i.e.* with and without restricting the atoms to planar geometries, for each cluster size. Our goals are to obtain the ground state structures in 2D and 3D and generate a collection of near-ground state isomers that could be observed due to thermal excitations or non-equilibrium synthesis techniques. We also investigate the relationships between planarity and total energies from the large set of structures generated from the global optimization for $Au_{12\text{-}14}$. We test our GA-DFT approach on anionic clusters to identify the impact of an excess electron in the near minima energy landscape. Within the context of neutral clusters, we discuss the effect of *s-d* orbital hybridization, spin-orbit coupling and long-range interactions on the 2D-3D transition.

## RESULTS AND DISCUSSION

### Neutral Clusters

The combined GA-DFT algorithm is utilized several times to obtain global minima in 12, 13 and 14 atom planar and globular Au particles. In Figures 1, 2 and 3, we present the evolution for $Au_{12}$, $Au_{13}$ and $Au_{14}$ clusters starting from 3D, 2D and 3D structures as these GA evaluations produce the identified global minima for the corresponding cluster sizes. The total energies of all 20 members of the population are given at each generation, along with lowest energy systems at selected generations, to show the structural evolution and some common structural motifs throughout the optimization process. Similar plots for 2D $Au_{12}$, 3D $Au_{13}$ and 2D $Au_{14}$ are given in Figures S1-S3 in the Supplementary Information (SI). In all these figures, we report the per atom formation energies of the clusters from non-interacting gold atoms (Note that the DFT calculations on the clusters have been performed without spin polarization during GA minimization). The energy of a single gold atom is calculated using the same settings as the clusters except that we also considered spin polarization for the atom. Since the energy differences between spin polarized and non-spin polarized calculations are very small for clusters, we expect the formation energies are reliable. Starting from a randomly constructed (subject to bond length constraints, see Methods) $Au_n$, the coupled optimization scheme results in energy drops of ~2 to 4.6 eV in about 50 generations. Considering that every generation consists of 20 members, structures at or close to the ground state have been identified in approximately 1000 DFT calculations. Earlier studies that utilize basin hopping and DFT for the prediction of global minima in C-H-O clusters with similar numbers of atoms also required about 1000 DFT calculations.[36] An important advantage of GA, compared to basin hopping, lies in the crossover operations where large distances can be traversed in the phase space and sampling from wide apart phase space volumes can be obtained. In basin hopping, the trial movements for the atoms are not drastic[37] and resemble the mutations in GA rather than crossover. Such an advantage may explain the efficiency of GA in finding many near-equilibrium structures which were not obtained by the previous basin-hopping search.

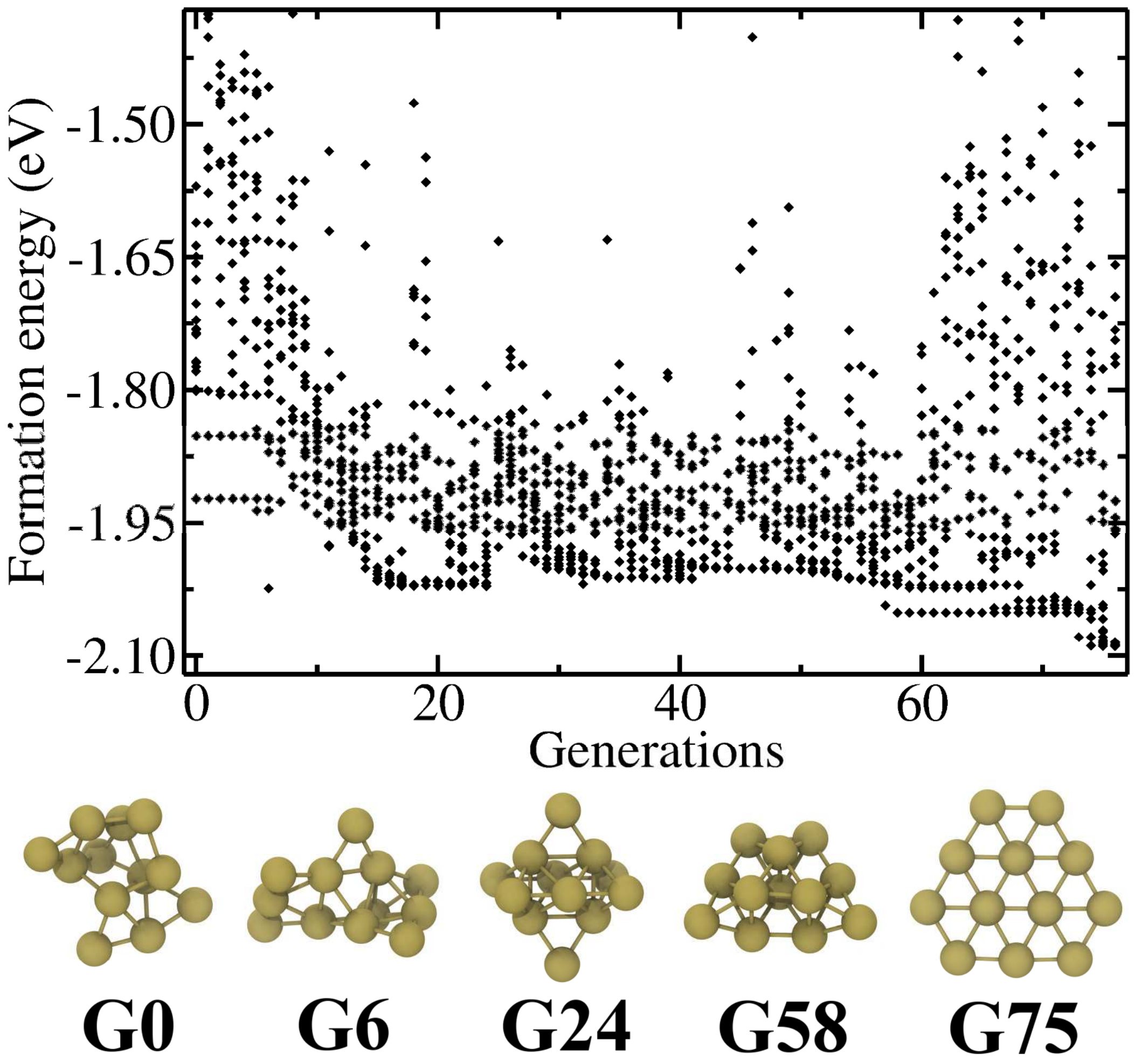

**Figure 1:** Per atom formation energies and structures of 3D $Au_{12}$ clusters as a function of generation number through GA optimization. The lowest energy clusters at selected generations (given by G#) are also presented in the bottom panel.

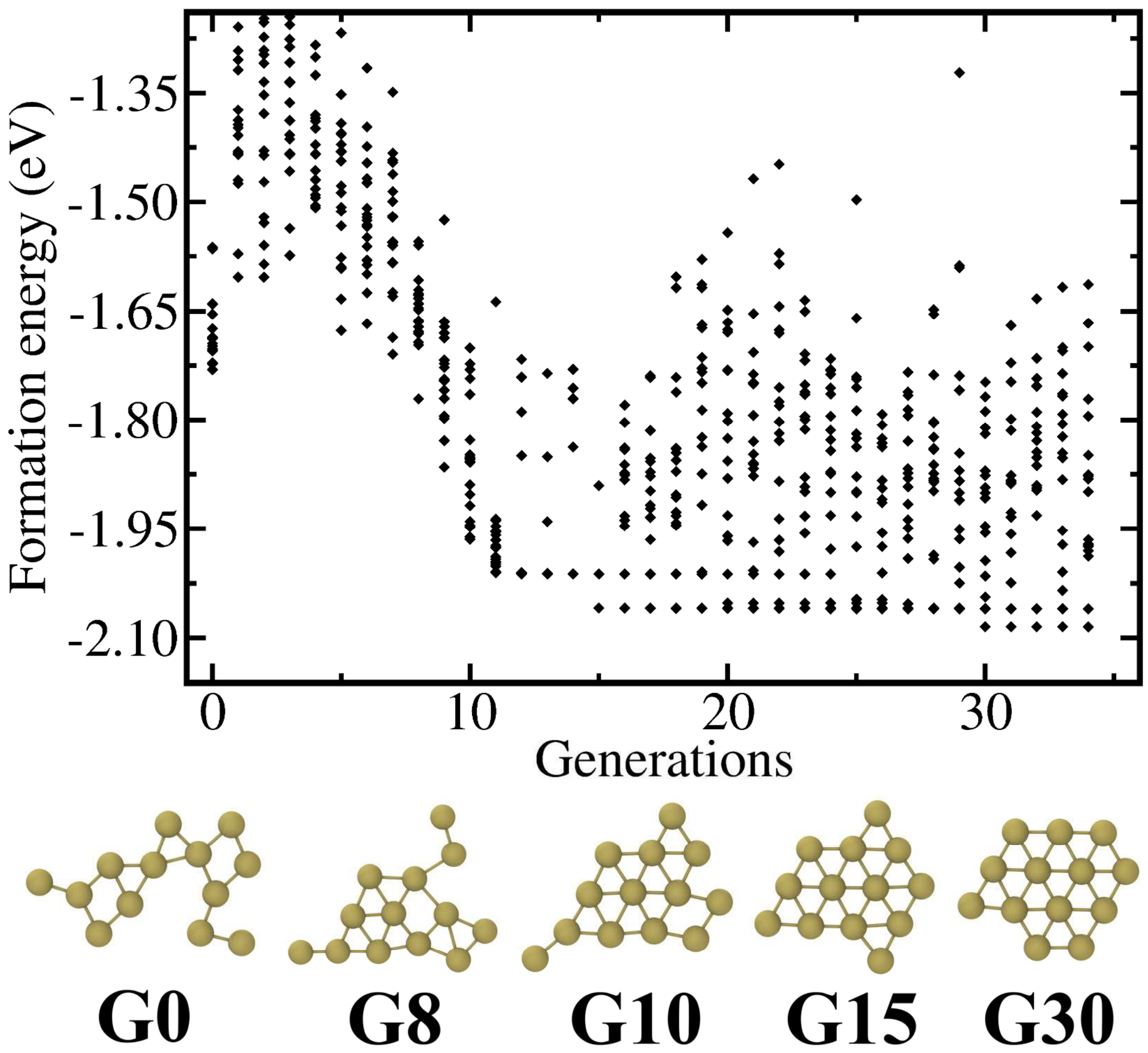

**Figure 2:** Per atom formation energies and structures of 2D $Au_{13}$ clusters as a function of generation number through GA optimization. The lowest energy clusters at selected generations (given by G#) are also presented in the bottom panel.

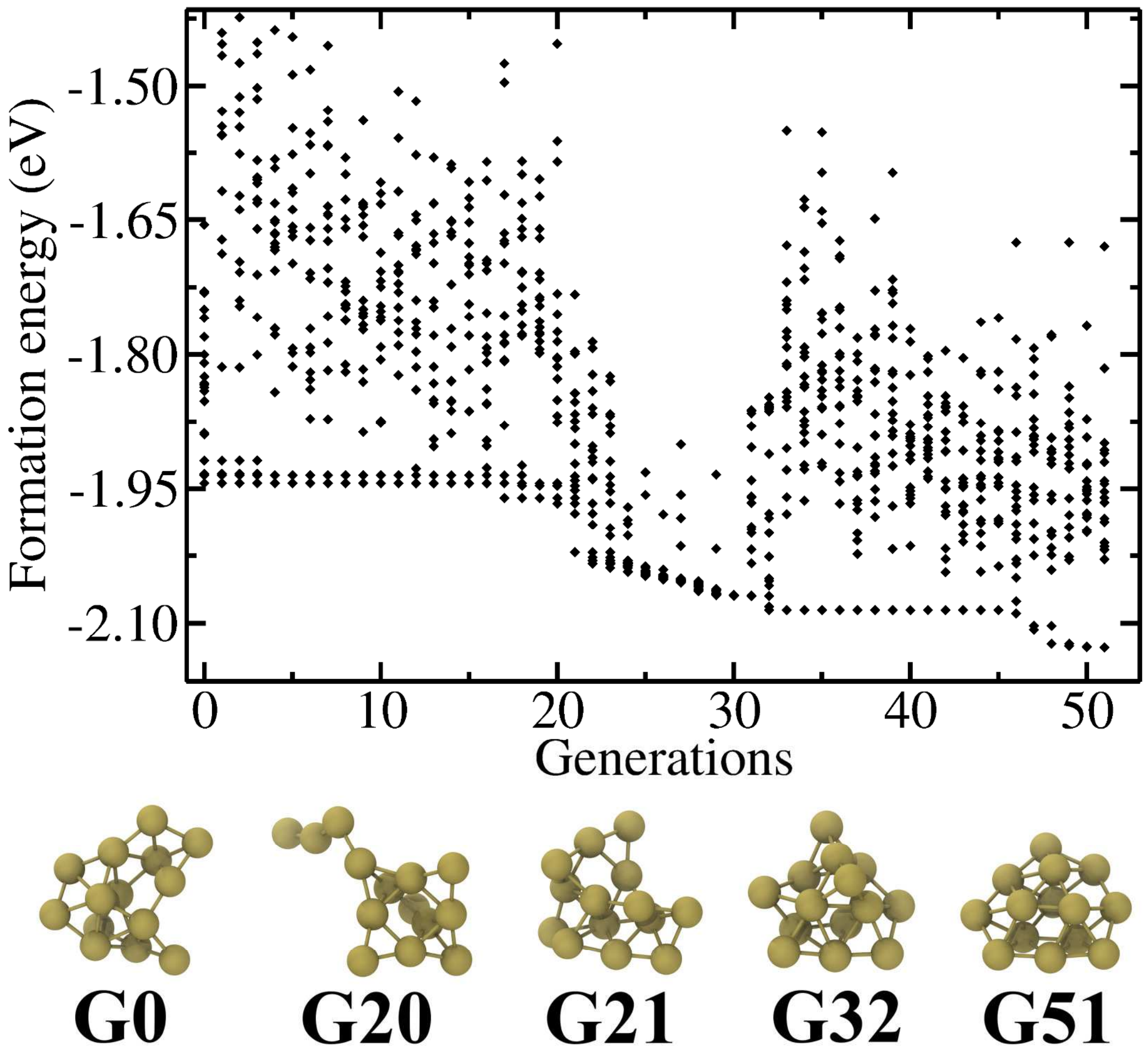

**Figure 3:** Per atom formation energies and structures of 3D $Au_{14}$ clusters as a function of generation number through GA optimization. The lowest energy clusters at selected generations (given by G#) are also presented in the bottom panel.

For 12 and 13 atom clusters, 3D evolution resulted in the planarization of the initially generated globular geometries as shown in Figures 1 and S2 respectively. The lowest energy structures (G75 in Figure 1 and G30 in Figure 2) are the ones that minimize the edge length, simultaneously minimizing the number of dangling bonds and maximizing the average coordination number in the plane. The evolution restricted to 2D produces similar planar geometries for these clusters. The GA predictions for $Au_{12}$ and $Au_{13}$ minima are same as the ones given by Lee *et al.*[22] On the other hand, $Au_{13}(2)$ in Table 2 is proposed as the lowest energy structure for $Au_{13}$ in another study which used basin hopping.[12] For the 14 atom system, the lowest energy (G51 in Figure 3) is obtained for a pouch-like structure which was previously suggested as candidate for global minimum of anionic clusters.[18] Apart from lowest energy structures, we have identified structures that have energies per atom within $2k_BT$ of the minima at room temperature $T$ in each cluster system. Note that we use an increased cutoff energy and consider spin-polarization to re-evaluate these clusters as mentioned in Methods section. These near-minima structures are shown in Tables 1, 2 and 3. The energy differences ($\Delta E$) of these structures from the candidate minima are also included per atom basis below each cluster. The Cartesian coordinates of the clusters in these tables are available respectively as images in xyz format in the Supplementary Dataset (SD). From the tables, it is seen that the 2D structures are generally edge variants of each other. The near-minima include both 2D and 3D clusters, indicating that the transition from 2D to 3D structures is not abrupt for gold clusters in the 12-14 atom range. Accordingly, it is possible to experimentally observe an ensemble of structures mixing 2D and 3D geometries at room temperature.

**Table 1:** $Au_{12}$ clusters that are in $2nk_BT$ ($n$=12, $T$=300K) proximity of the predicted minimum energy structure. The $\Delta E$ is described as the per atom energy difference of the corresponding structure from the minimum which is given as $Au_{12}(1)$. See SD file for atomic coordinates.

| | | | |
|---|---|---|---|
| $Au_{12}(1)$ | $Au_{12}(2)$ | $Au_{12}(3)$ | $Au_{12}(4)$ |
| 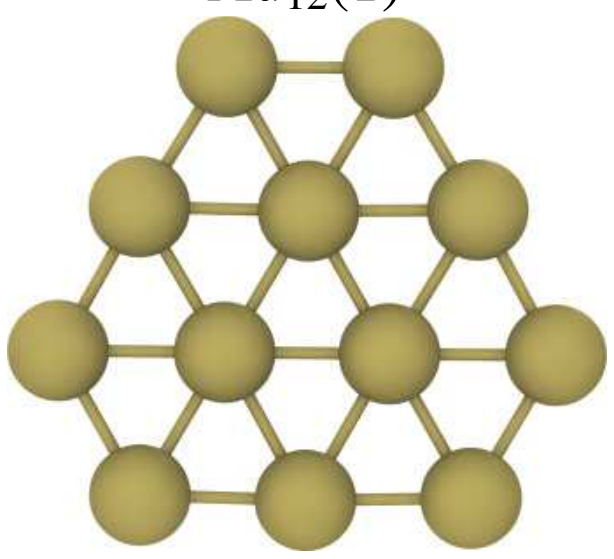 | 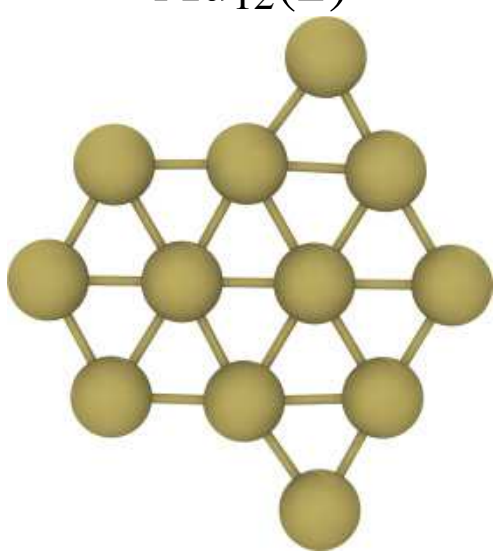 | 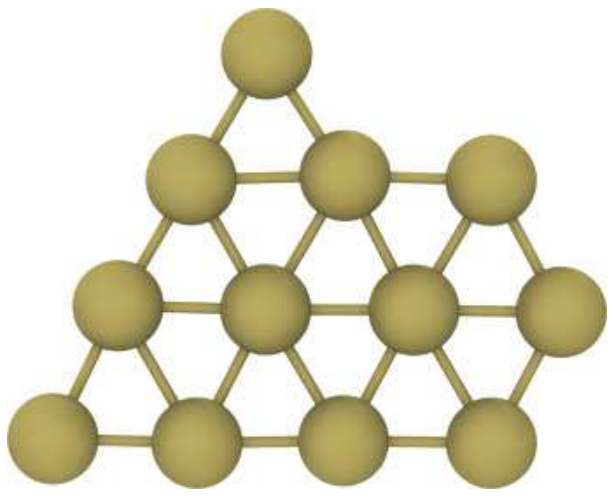 | 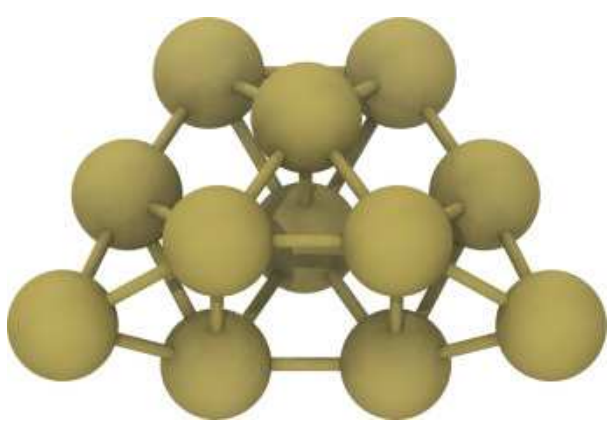 |
| $\Delta E$=0.000 eV | $\Delta E$=0.025 eV | $\Delta E$=0.033 eV | $\Delta E$=0.039 eV |
| $Au_{12}(5)$ | $Au_{12}(6)$ | $Au_{12}(7)$ | $Au_{12}(8)$ |

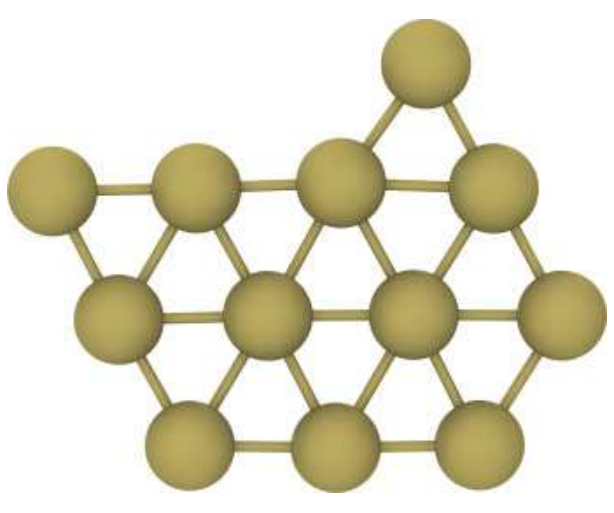

$\Delta E$=0.044 eV

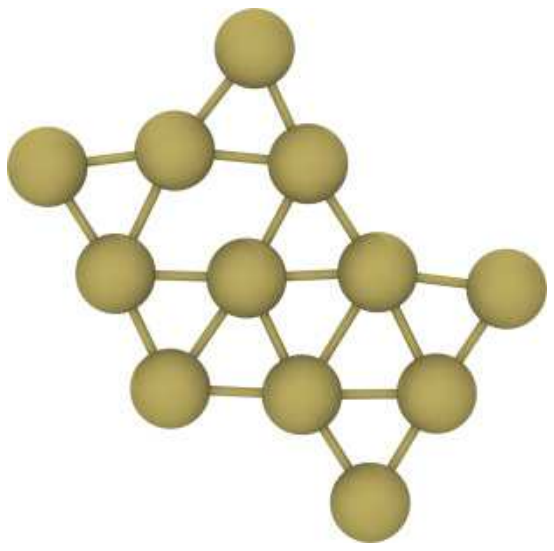

$\Delta E$=0.045 eV

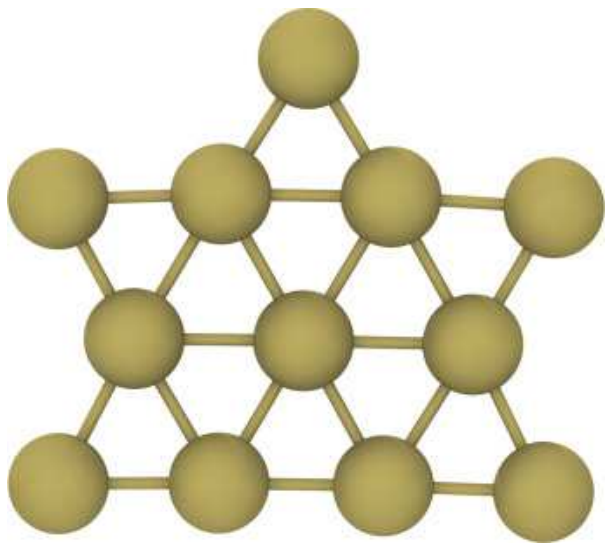

$\Delta E$=0.049 eV

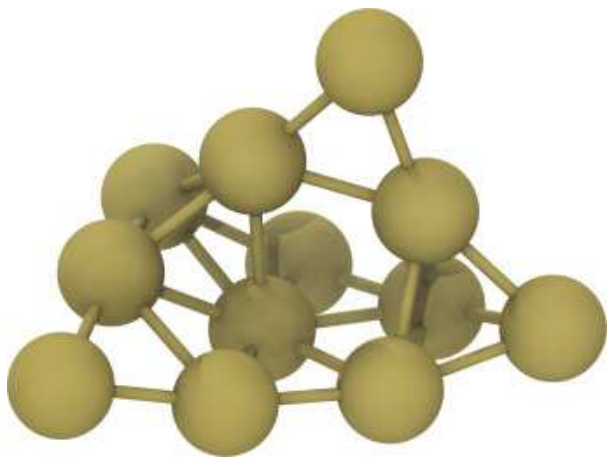

$\Delta E$=0.051 eV

$Au_{12}$(9)

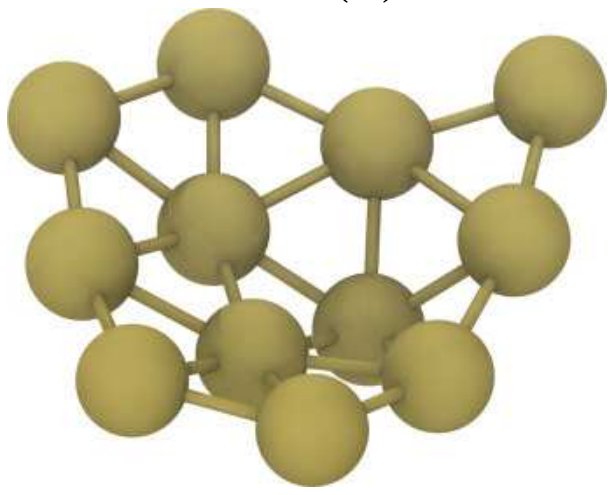

$\Delta E$=0.053 eV

**Table 2:** $Au_{13}$ clusters that are in $2nk_BT$ ($n$=13, $T$=300K) proximity of the predicted minimum energy structure. The $\Delta E$ is described as the per atom energy difference of the corresponding structure from the minimum which is given as $Au_{13}$(1). See SD file for atomic coordinates.

$Au_{13}$(1)

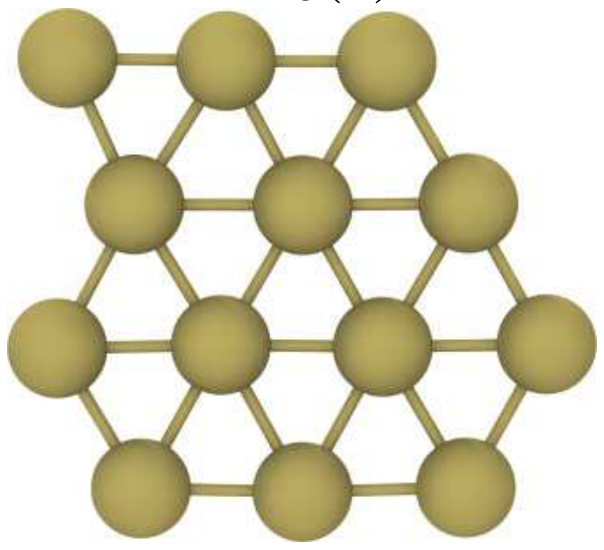

$\Delta E$=0.000 eV

$Au_{13}$(2)

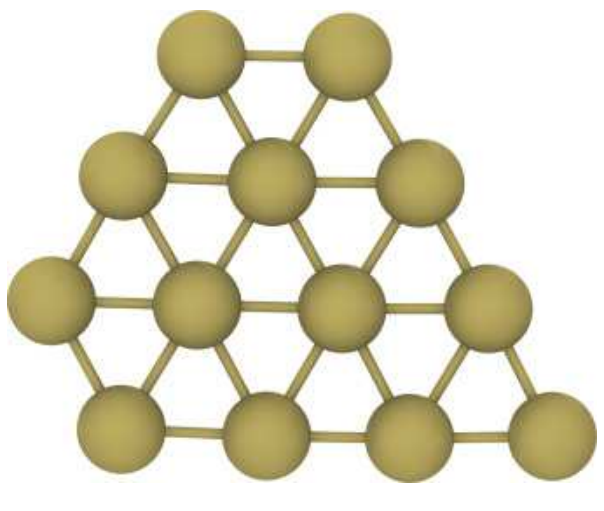

$\Delta E$=0.003 eV

$Au_{13}$(3)

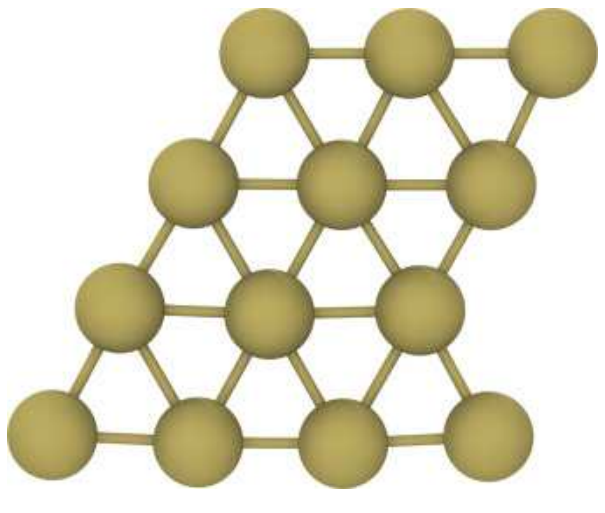

$\Delta E$=0.017 eV

$Au_{13}$(4)

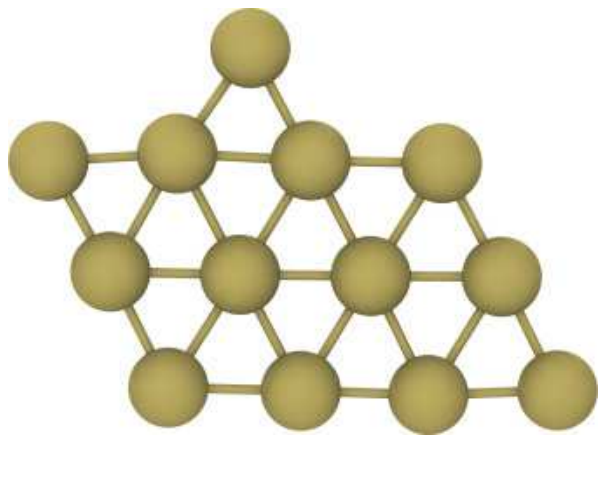

$\Delta E$=0.019 eV

$Au_{13}$(5)

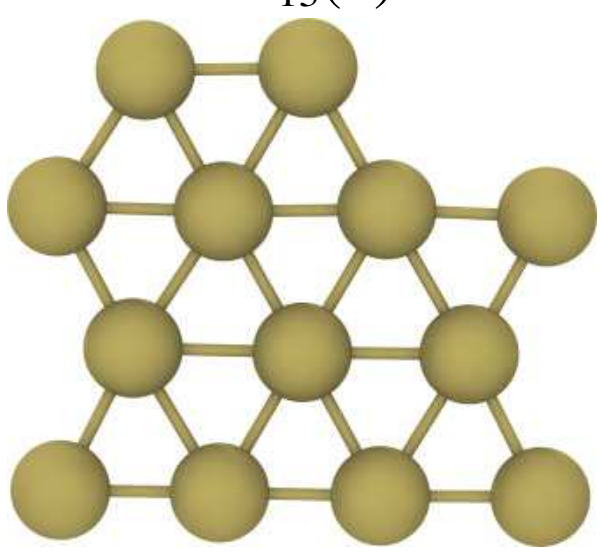

$\Delta E$=0.021 eV

$Au_{13}$(6)

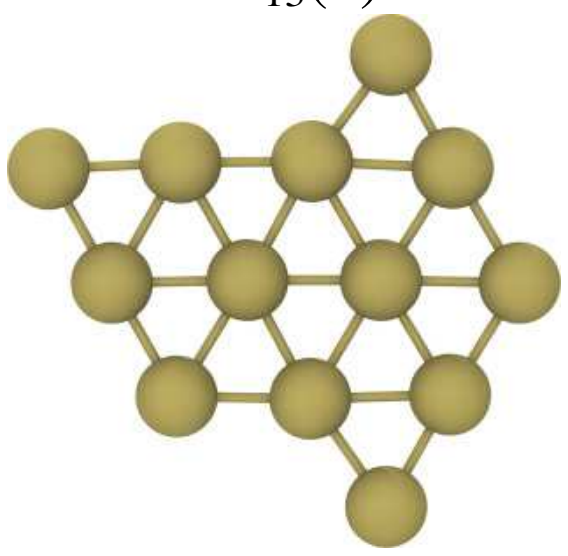

$\Delta E$=0.023 eV

$Au_{13}$(7)

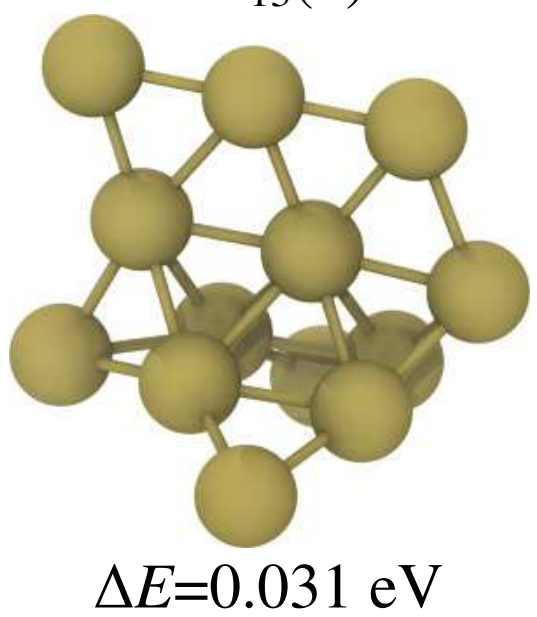

$\Delta E$=0.031 eV

$Au_{13}$(8)

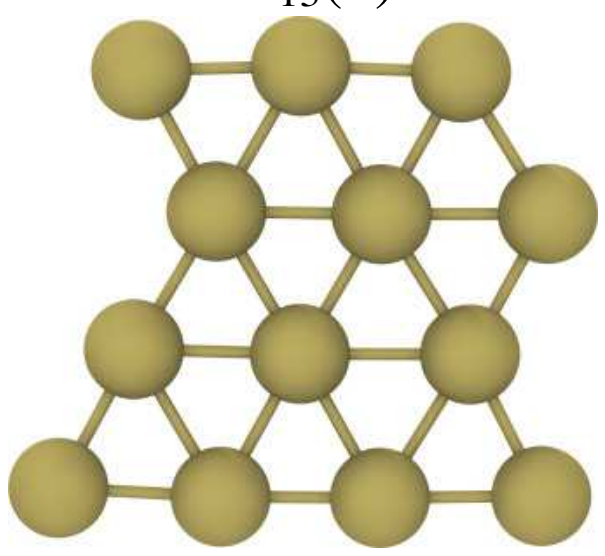

$\Delta E$=0.034 eV

$Au_{13}$(9)

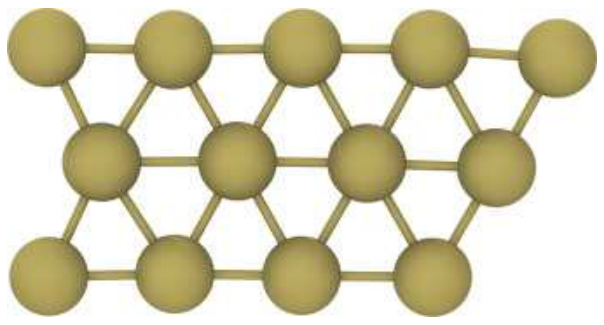

$\Delta E$=0.034 eV

$Au_{13}$(10)

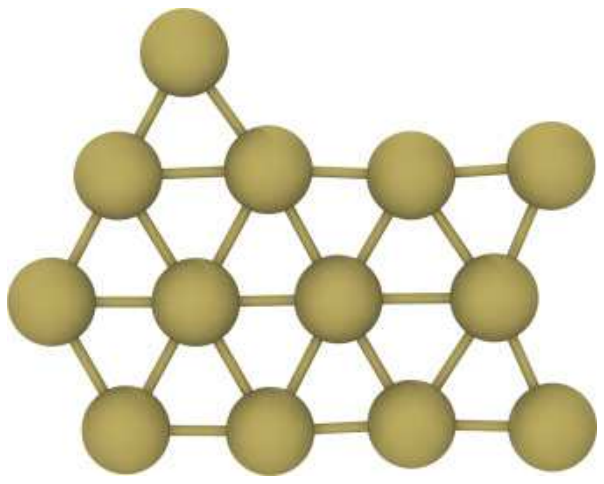

$\Delta E$=0.036 eV

$Au_{13}$(11)

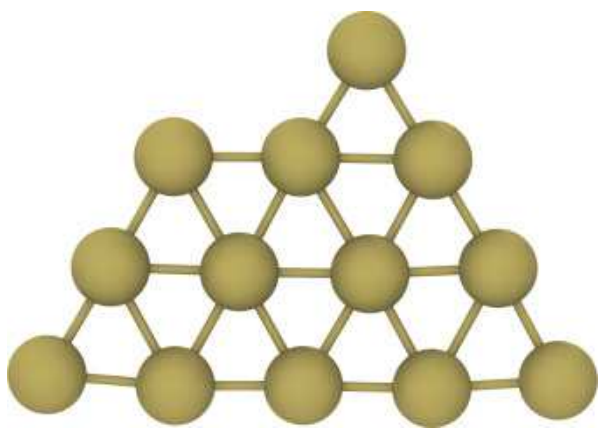

$\Delta E$=0.037 eV

$Au_{13}$(12)

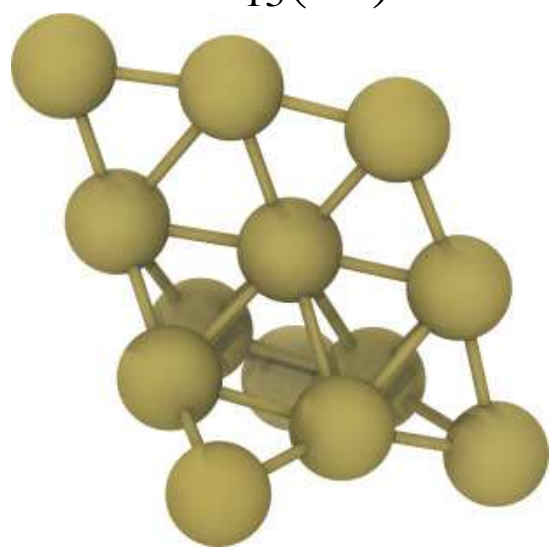

$\Delta E$=0.039 eV

$Au_{13}$(13)

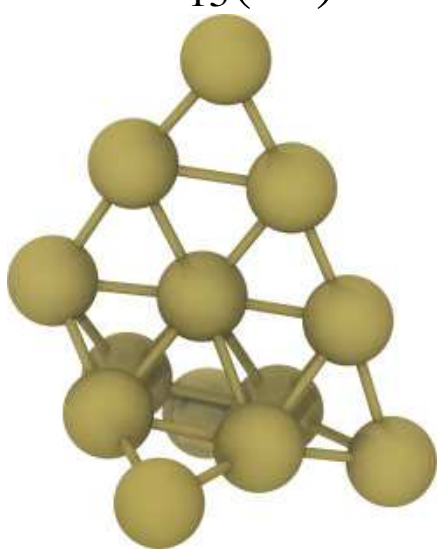

$\Delta E$=0.045 eV

$Au_{13}$(14)

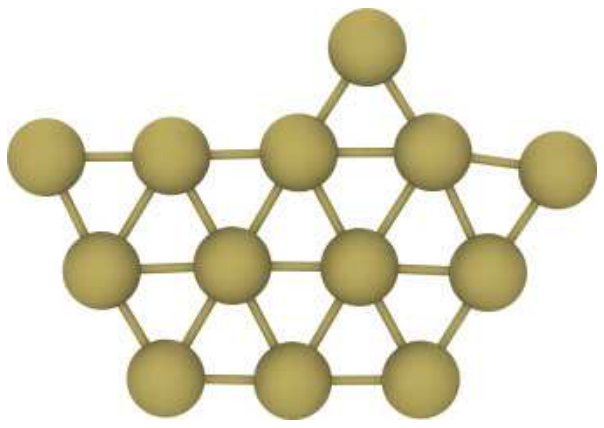

$\Delta E$=0.046 eV

$Au_{13}$(15)

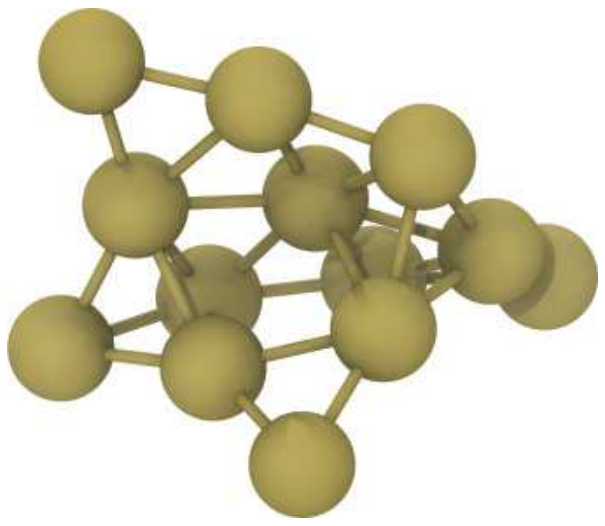

$\Delta E$=0.046 eV

$Au_{13}$(16)

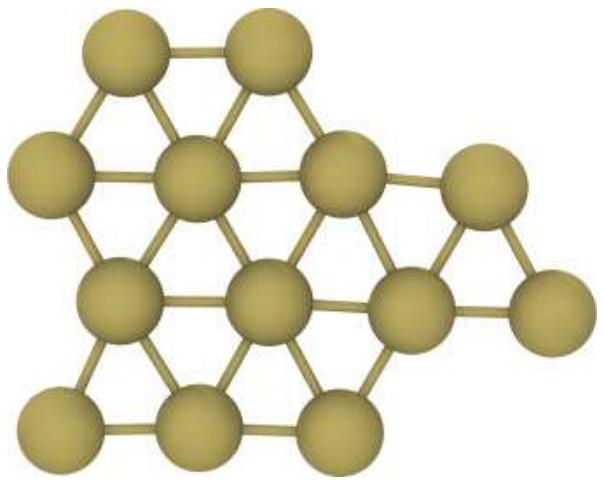

$\Delta E$=0.047 eV

$Au_{13}$(17)

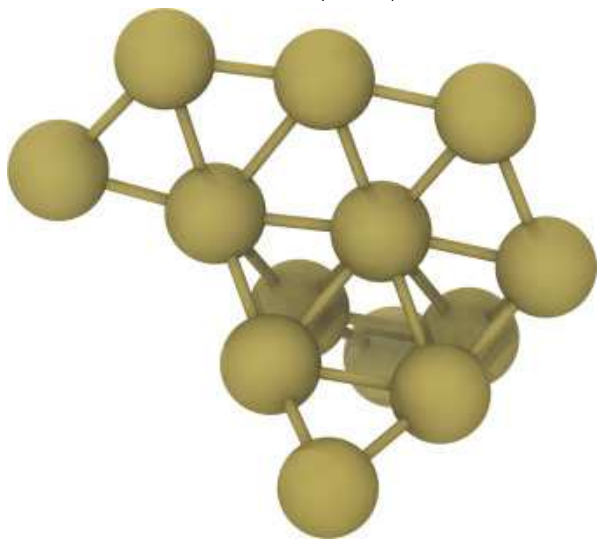

$\Delta E$=0.048 eV

$Au_{13}$(18)

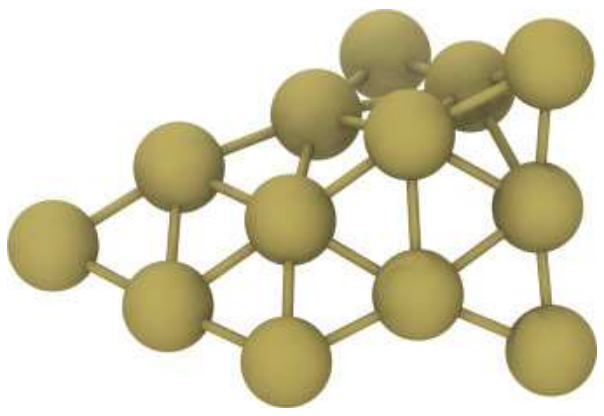

$\Delta E$=0.048 eV

$Au_{13}$(19)

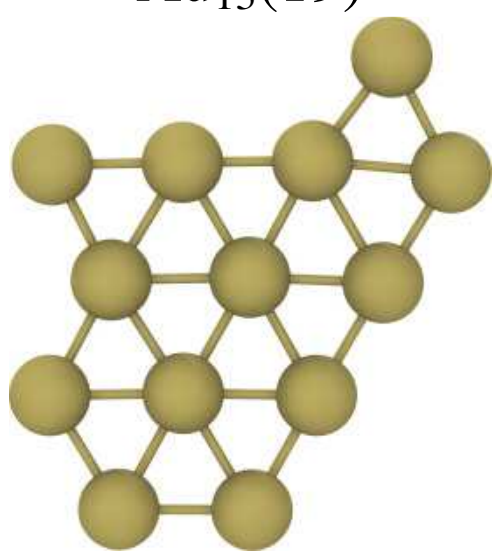

$\Delta E$=0.048 eV

$Au_{13}$(20)

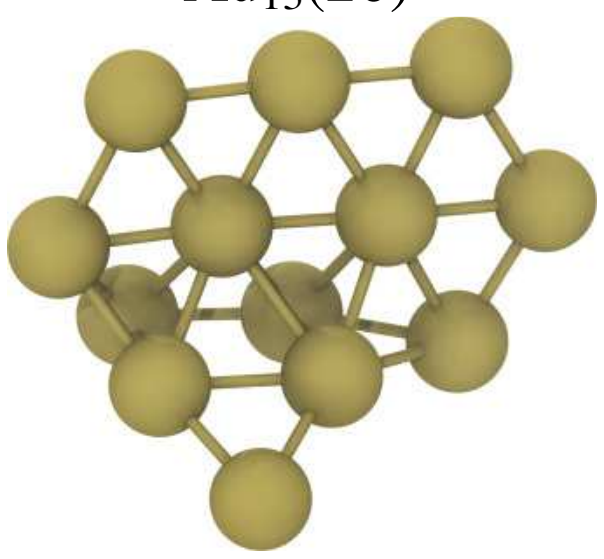

$\Delta E$=0.052 eV

$Au_{13}$(21)

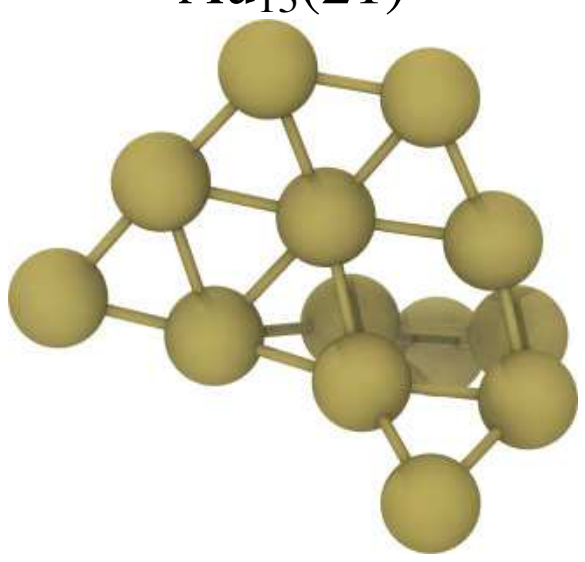

$\Delta E$=0.052 eV

**Table 3:** $Au_{14}$ clusters that are in $2nk_BT$ ($n$=14, $T$=300K) proximity of the predicted minimum energy structure. The $\Delta E$ is described as the per atom energy difference of the corresponding structure from the minimum which is given as $Au_{14}$(1). See SD file for atomic coordinates.

| | | | |
|---|---|---|---|
| $Au_{14}$(1)<br>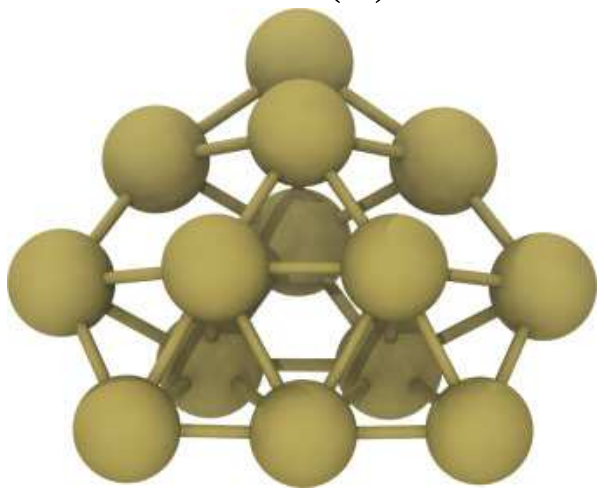<br>$\Delta E$=0.000 eV | $Au_{14}$(2)<br>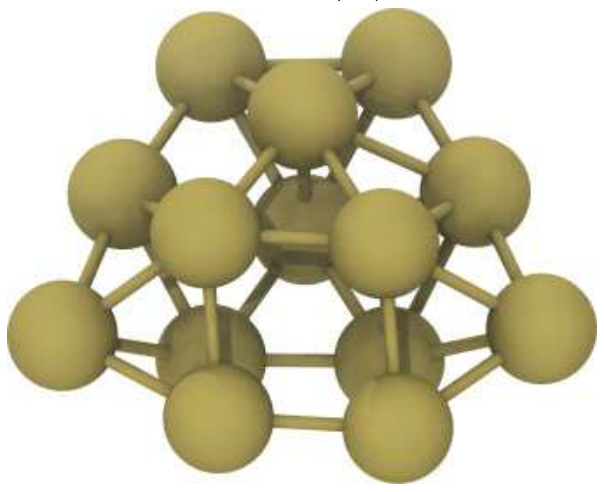<br>$\Delta E$=0.000 eV | $Au_{14}$(3)<br>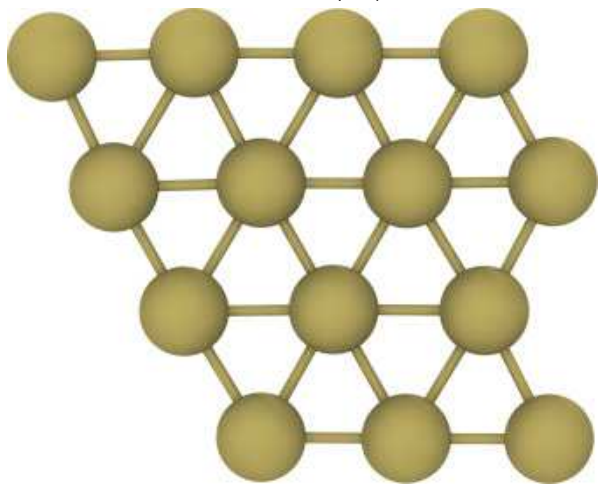<br>$\Delta E$=0.013 eV | $Au_{14}$(4)<br>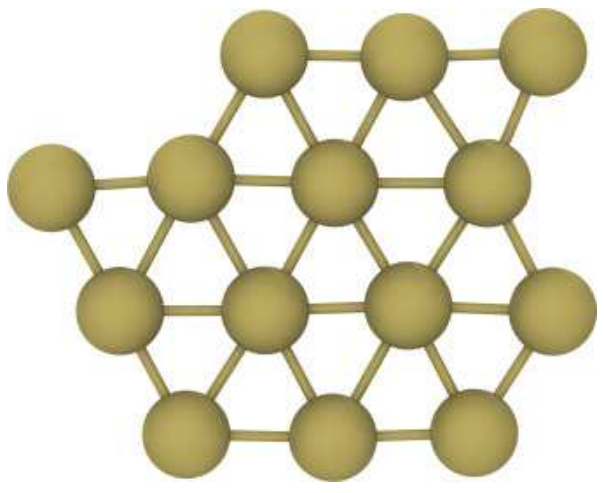<br>$\Delta E$=0.020 eV |
| $Au_{14}$(5)<br>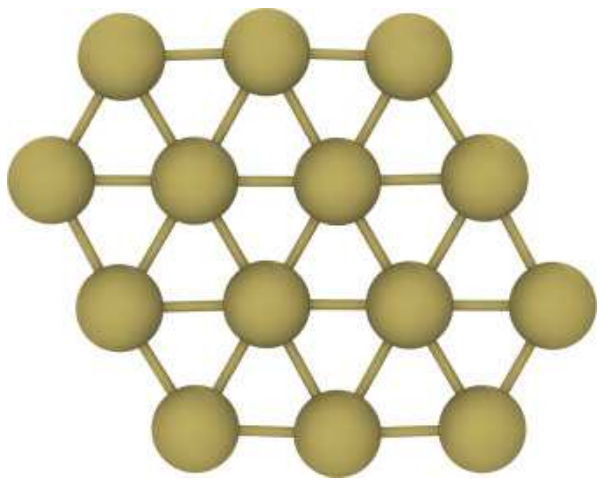<br>$\Delta E$=0.021 eV | $Au_{14}$(6)<br>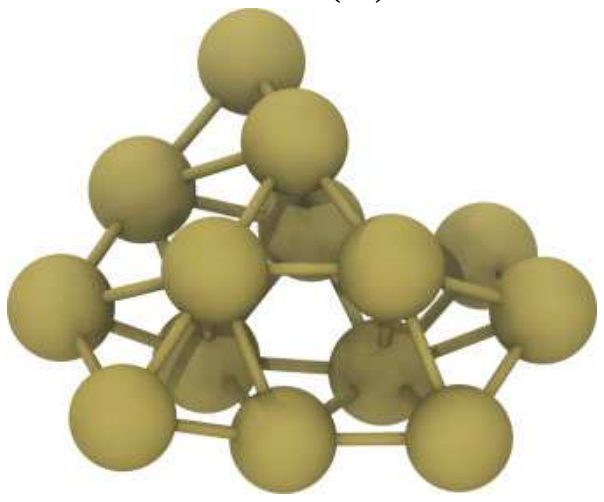<br>$\Delta E$=0.023 eV | $Au_{14}$(7)<br>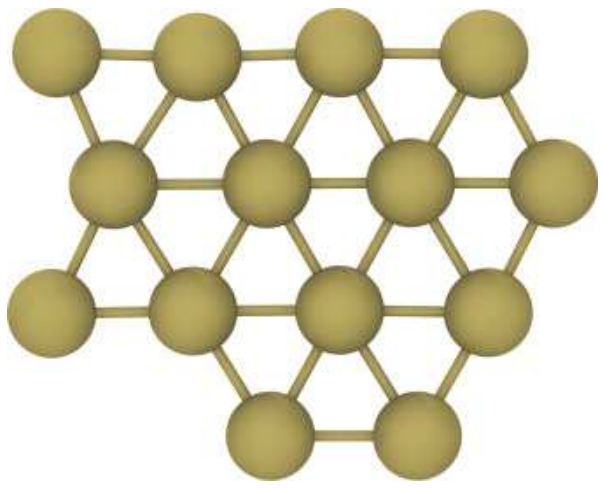<br>$\Delta E$=0.025 eV | $Au_{14}$(8)<br>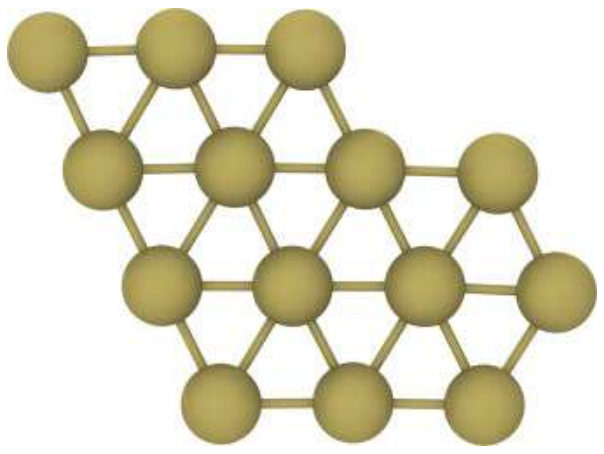<br>$\Delta E$=0.029 eV |
| $Au_{14}$(9)<br>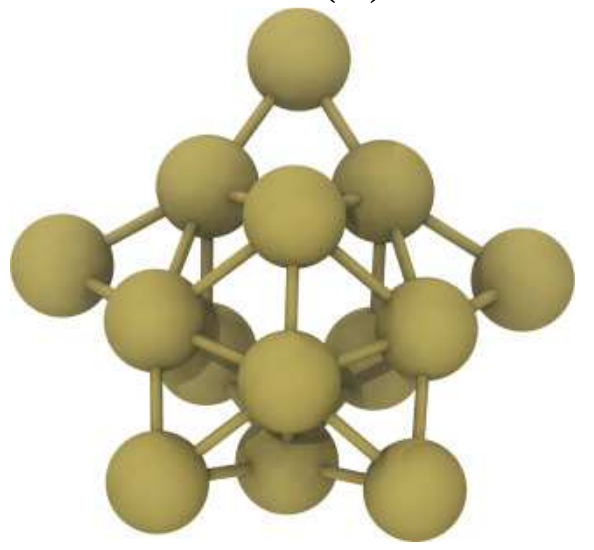<br>$\Delta E$=0.029 eV | $Au_{14}$(10)<br>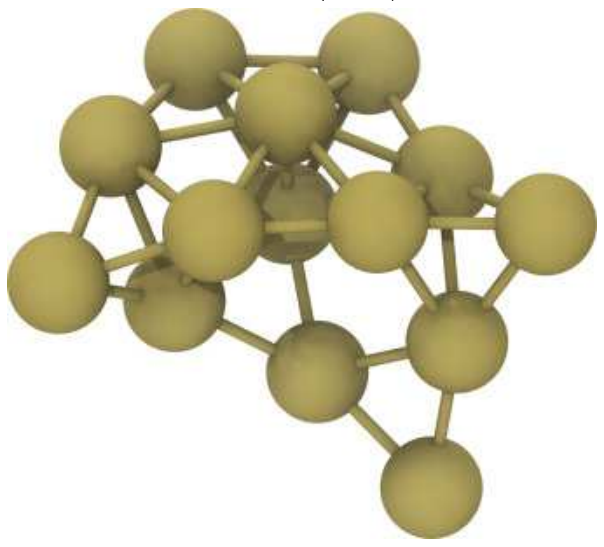<br>$\Delta E$=0.032 eV | $Au_{14}$(11)<br>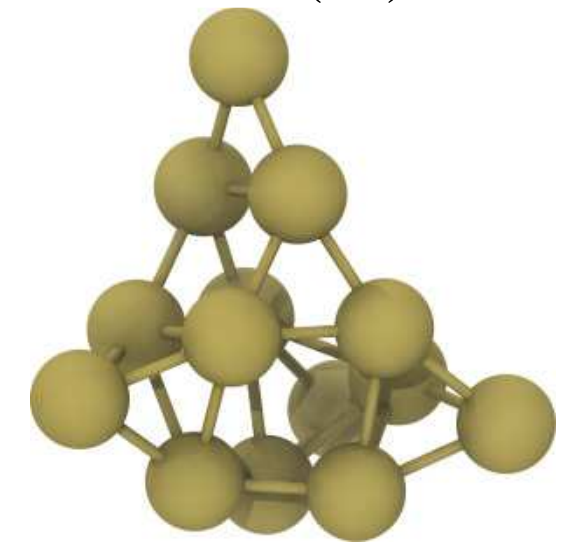<br>$\Delta E$=0.033 eV | $Au_{14}$(12)<br>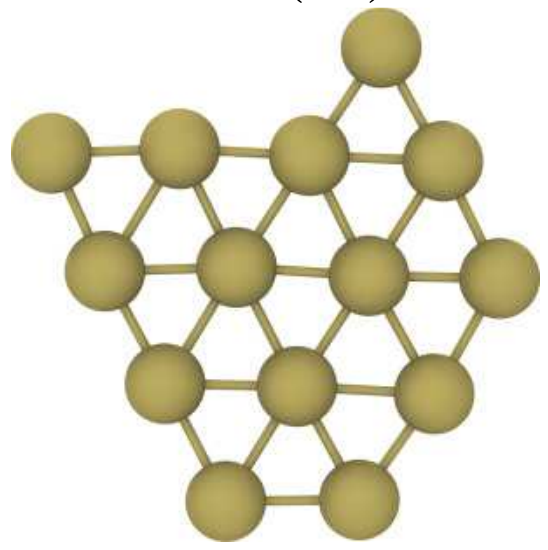<br>$\Delta E$=0.034 eV |
| $Au_{14}$(13)<br>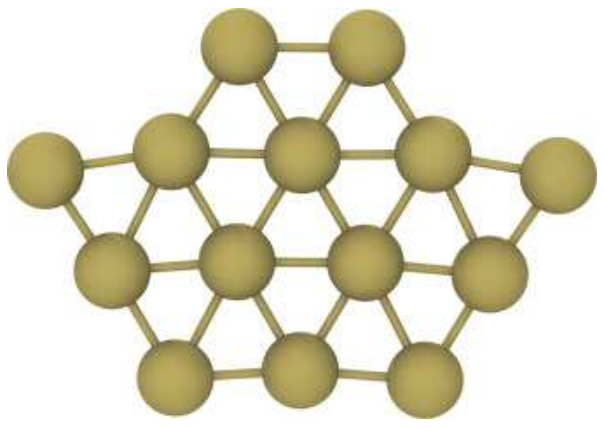<br>$\Delta E$=0.037 eV | $Au_{14}$(14)<br>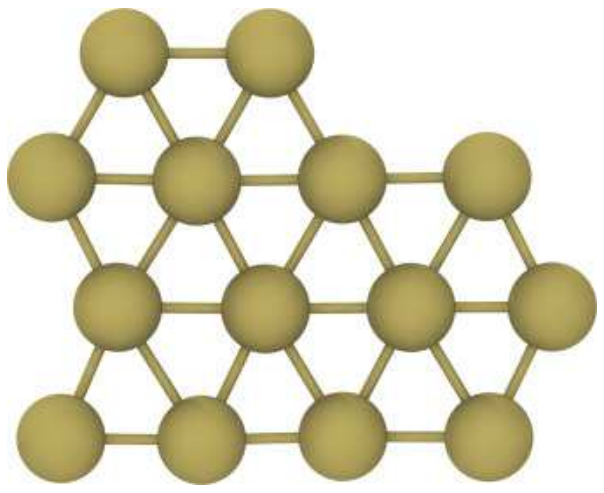<br>$\Delta E$=0.038 eV | $Au_{14}$(15)<br>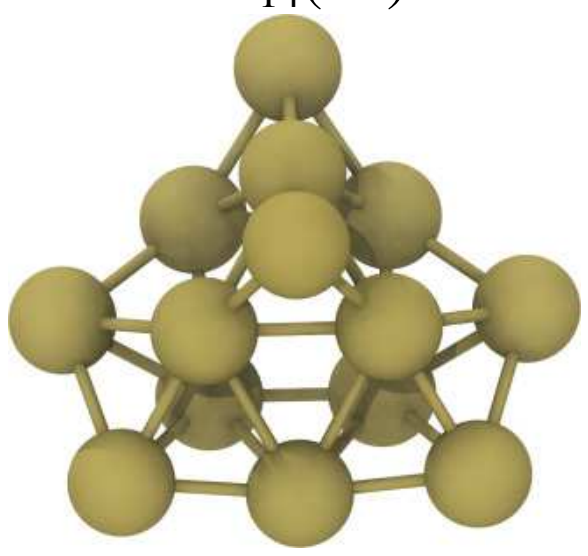<br>$\Delta E$=0.040 eV | $Au_{14}$(16)<br>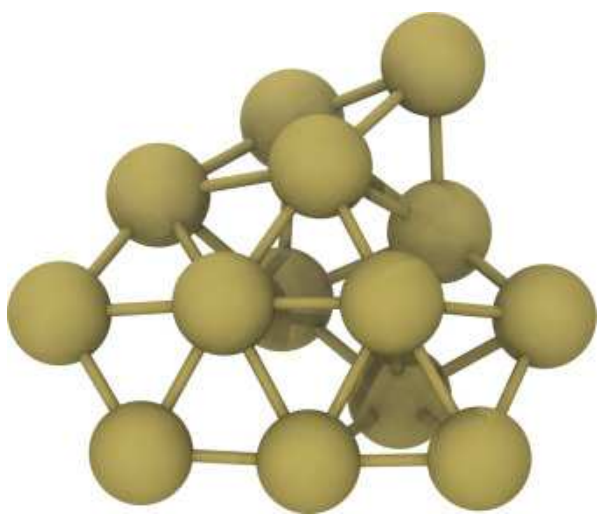<br>$\Delta E$=0.041 eV |

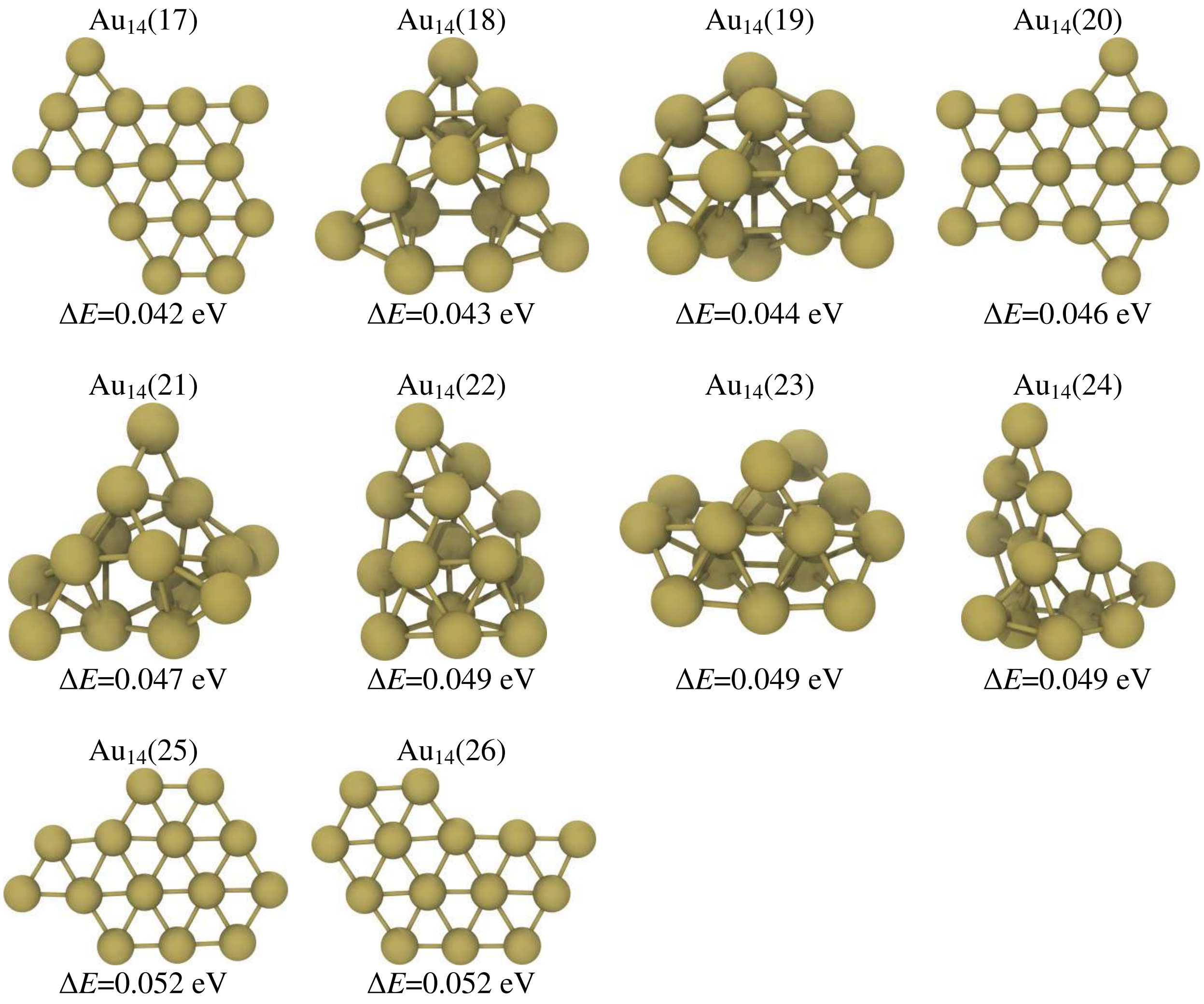

---

The global structural optimization combining GA and DFT is able to predict global minima along with many close-by structures near the 2D to 3D transition size. At the same time, we generate thousands of sample clusters from various regions of the energy and coordination landscape. Using these samples, we investigate the relationship between energy and planarity, *i.e.* whether, in general, the 2D structures are favored for $Au_{12}$ and $Au_{13}$ and 3D structures are favored for $Au_{14}$. For each cluster, we locate the best-fit plane with the least squares algorithm. The planarity of the cluster is measured by the sum of normal distances between each atom in the cluster and the fitted plane (*i.e.* residual). A lower residual means that the structure is closer to a planar geometry. In Figure 4, the formation energies of 3D $Au_{12-14}$ clusters with respect to their residual values are presented. We only include structures that have energies within 100 meV/atom of the lowest energy in this plot since the formation of clusters above this energy limit would be very unlikely. In 12-atom clusters, there seems to be a trend where the energy is lowered by decreasing the residual. Also the energies of the structures that have relatively higher residuals (1-1.5 Å) are considerably higher than suggested global minima. On the other hand, for 13 and 14-atom clusters, a number of clusters with very different degrees of planarity nonetheless have very similar energies. This illustrates the gradual rather than abrupt transition between 2D and 3D structures.

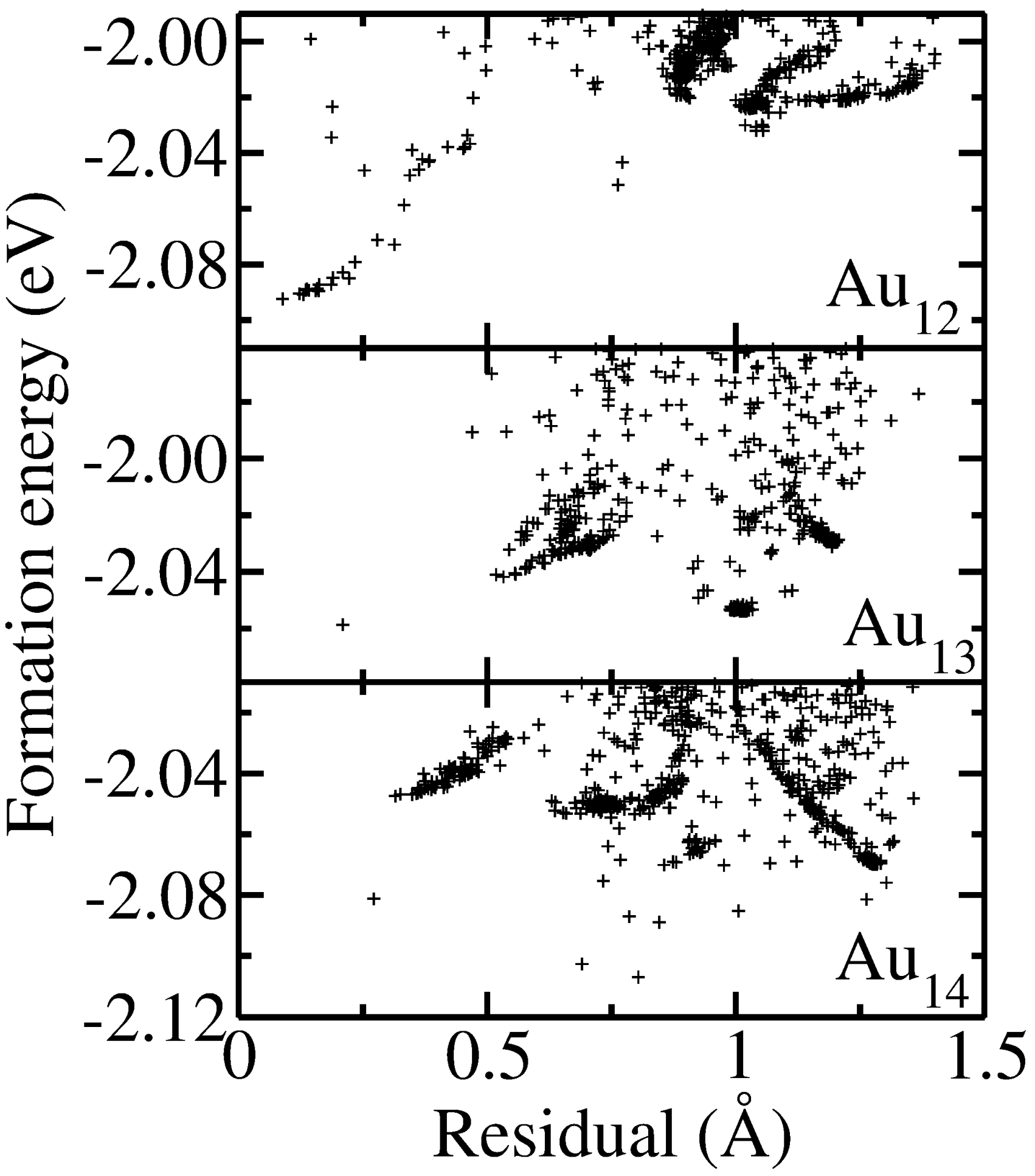

**Figure 4:** Per atom formation energies of 3D $Au_{12\text{-}14}$ clusters as a function of the total deviation of atom positions from the best plane (*i.e.* residual) that represents the corresponding cluster.

**Anionic Clusters**

We also test the consistency of energy ordering between non-spin polarized/neutral and spin-polarized/anionized ($Au_{13}^-$) calculations, since experimentally reported clusters are generally ionized. For this estimation, we use 1000 randomly generated 2D and 3D clusters. The probability of having the same energy ordering is found to be 96.9% between non-spin polarized/neutral and spin-polarized/anionized clusters. Although the formation energy order seems to be similar to a high degree for the randomly created anionic and neutral clusters, we also check the near-minima by re-evaluating the structures in Tables 1, 2 and 3, and performing four additional GA optimizations for $Au_{13}^-$. The energy comparison between neutral and anionic clusters is shown in Figure 5 for structures in Tables 1, 2 and 3. The results of GA optimizations are given in Table S1 in SI and the coordinates of the low energy structures are given in SD. The formation energies of anionic clusters ($E_f$) are calculated as given in Equation 1.

$$E_f = E_{Au_n^-} - nE_{Au} + \left[E_{MP} + (E_{Fermi} - V_{vacuum})\right]/n \qquad (1)$$

where $E_{Au_n^-}$ is the total energy of anionic cluster of $n$ atoms. $E_{Au}$ is the energy of an isolated gold atom. $E_{MP}$ is the monopole term of Makov-Payne[38] correction for the image charge interactions between the extra electron and neutralizing background charge. $E_{Fermi}$ and $V_{vacuum}$ are the Fermi level and electrostatic potential in the vacuum away from the cluster.

Re-evaluation of near-minima of neutral clusters with an excess electron does not change the energy order drastically. The minima are not altered for 12, 13 and 14-atom clusters in terms of cluster dimensionality and cluster structures. Any large drop in the energy (as in the case of the 51$^{st}$ structure in Figure 5, $Au_{14}$(21) originally in Table 3) is related to relaxation of the cluster to a new local minimum structure. From the GA optimization of $Au_{13}^-$, we obtain many structures identical to the ones found for neutral clusters. Anionic form of $Au_{13}$(1) is again the minimum energy structures of this cluster size. In addition to already determined clusters, we identify more than a dozen new 2D and 3D structures that are in $2nk_BT$ ($n$=13, $T$=300K) of the minimum as seen in Table S1.

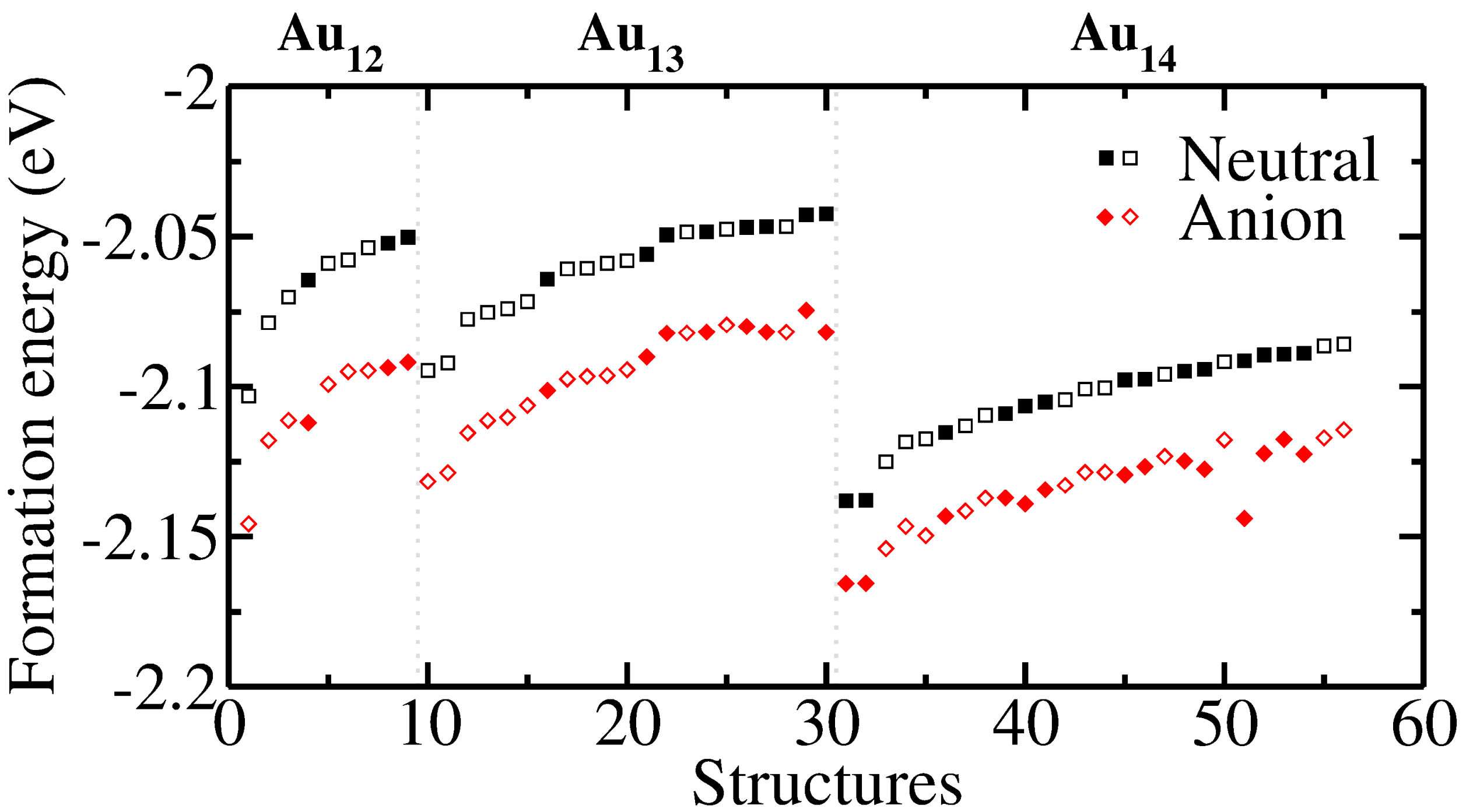

**Figure 5:** The effect of excess electron on the per atom formation energies of the low energy $Au_{12-14}$ clusters identified in Tables 1-3. Formation energies are shown for 56 structures. The clusters are given in the same order as in Tables 1, 2 and 3 for $Au_{12}$ (structures 1-9), $Au_{13}$ (structures 10-30) and $Au_{14}$ (structures 31-56). The vertical dashed lines separate one cluster size from another. Empty and filled symbols are used for 2D and 3D structures respectively.

### Relativistic Effects and *s*-*d* hybridization

One of the unusual properties of $Au_n$ clusters is that the transformation from planar to globular structures occurs at larger sizes ($n$ = 12-14) compared to other metal clusters such as $Cu_n$ and $Ag_n$ ($n$ = 6-7). A proposed reason for this observation is the enhanced 5*d*-6*s* hybridization due to relativistic effects in gold favoring the 2D structures at larger cluster sizes.[15] Our GA-DFT approach calculations also predict a 2D-3D transformation at large cluster sizes (i.e. after $Au_{13}$). In order to investigate the structure-hybridization relationship in this transition range, we calculate the *s*-*d* hybridization in both 2D and 3D clusters. In estimating the degree of hybridization ($H_{sd}$), we used two definitions: In the first, shown in Equation 2, the common area under the *s* and *d* projections of density of states (DOS), namely $g_s$ and $g_d$, are calculated up to Fermi energy ($E_F$).

$$H_{sd}{}^{a} = \int_{-\infty}^{E_F} \min\{g_s(E), g_d(E)\} dE \qquad (2)$$

In the second more rigorous approach, the degree of *s-d* hybridization is found by multiplying the local *s* and *d* orbital projection of each Kohn-Sham eigenstate ($w_s$ and $w_d$)[15], see Equation 3.

$$H_{sd}{}^{b} = \sum_{I,S}\sum_{E} W_E^2 \sum_{Q} W_Q^2 \sum_{m} w_s^{I,S,E,Q} w_{d,m}^{I,S,E,Q} \qquad (3)$$

The weights $w_s$ and $w_d$ are obtained from the projection of total wavefunction onto spherical harmonics within a sphere around each atom.[39] They are calculated at each reciprocal space point *Q*, band energy eigenvalue *E*, spin component *S* and atom *I* so summation over all these are needed. The *m* (*i.e.* projected angular momentum component) summation is only relevant for the *d* orbitals. $W_E$ and $W_Q$ are band occupancy and weight of the reciprocal space point. In Figure 6a-c, we show $H_{sd}{}^{a}$ and $H_{sd}{}^{b}$ per atom and evaluated clusters respectively. The structures for $Au_{12-14}$ are obtained from the GA optimizations. For comparison, we include the same *s-d* hybridization indices for $Au_5$, $Au_{20}$ structures taken from literature,[11, 17] a $Au_{165}$ nanocrystal created from the Wulff construction from (111), (100) and (110) surface energies,[40] and bulk Au.

In Figure 6a, $H_{sd}{}^{a}$ is presented for the 2D 5-, 12-, 13-, 14-atom and 3D 12-, 13-, 14-, 20-, 165-atom clusters and face centered cubic (bulk) gold. The planar structures assume very similar values for this measure of hybridization. 3D forms consistently have lower values of $H_{sd}{}^{a}$ compared to planar ones. $H_{sd}{}^{b}$ also show that 2D clusters possess higher number of electrons in *s-d* hybrid orbitals, see Figure 6b. Close-packed FCC gold has the smallest value in both measures. The odd numbered clusters, except for $Au_{165}$, are found to have non-zero magnetic moments and the even numbered ones are non-magnetic. $Au_{165}$ is large enough that there are many 12-fold coordinated atoms, so one may expect $Au_{165}$ to behave bulk-like. Evidently, its *s-d* hybridization index is close to that of bulk Au. In particular, we note that across the transition range of 12-14 atoms, the *s-d* hybridization indices remain constant for the 2D structures and decrease as the structures become more compact (*i.e.* going from pouch-like to face-centered close packed structures) for the 3D structures. This indicates that the energy contribution due to hybridization is more or less the same for the 2D minima. The total energy, on the other hand, increasingly favors 3D structures as the size increases. In order to investigate and quantify the energetic influence of *s-d* hybridization, we artificially change, for the lowest energy 2D and 3D structures at $n$ = 12-14 given in Figure 6, the amount of *d* electron localization by shifting the energies of the *d*-bands *via* a Hubbard $U$ correction with $U$ = 0.25 – 4 eV. We find that increasing $U$ leads to a decrease in $H_{sd}^{a}$ and $H_{sd}^{b}$ (except for $H_{sd}^{b}$ of 3D $Au_{14}$) in both planar and globular clusters as seen in Figure S4 in SI. The decrease is more pronounced for 2D than 3D structures for $n$ = 12-14. If *s-d* hybridization were responsible for stabilizing the planar structures, a more pronounced decrease in *s-d* hybridization in the 2D structures should lead them to become less stable *vis a vis* the 3D structures. But in fact, at $n$ = 12-13, 2D structures are further stabilized relative to the 3D structures about 5-6 meV/atom. The opposite trend is seen at n=14 but only marginally (~1.5 meV/atom). This evidence suggests that *s-d* hybridization is not directly correlated with the stabilization of 2D structures. This is consistent with the earlier studies on $Au_8$ clusters emphasizing that large *s-d* hybridization does not necessarily mean high stability for planar structures.[41]

For a rigorous description of hybridization, relativistic effects should be considered in the calculations.[15] The relativistic contraction of valence n*s* shells and expansion of (n-1)*d* shells cause the overlap of these states and hybridization.[16] Our DFT calculations include scalar relativistic effects (*i.e.* Darwin and mass-velocity terms). For a more complete analysis of relativistic effects, we also consider spin-orbit coupling for $Au_{12}$, $Au_{13}$ and $Au_{14}$ in both planar and 3D forms. The results of these computations are shown in Figure 6. Again, 2D structures show a higher hybridization index compared to 3D constructions. A shift in the absolute values of the $H_{sd}$ is observed in spin-orbit compared to scalar relativistic calculations. However, the differences between planar and globular structures are almost unchanged. These results point that the correct behavior in 2D-3D transition range can be captured for gold clusters with scalar relativistic calculations and addition of spin-orbit coupling does not change the hybridization behavior significantly.

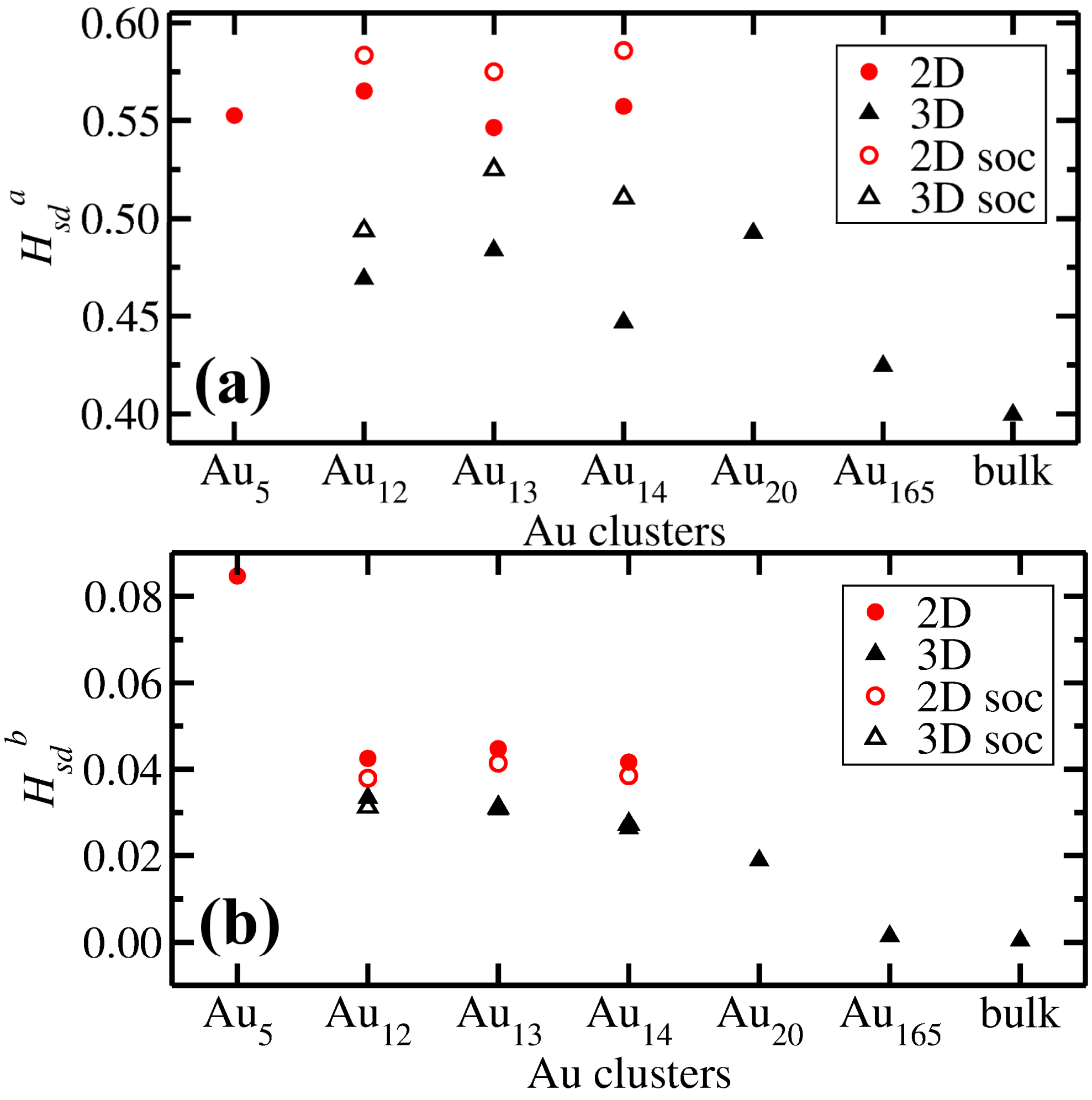

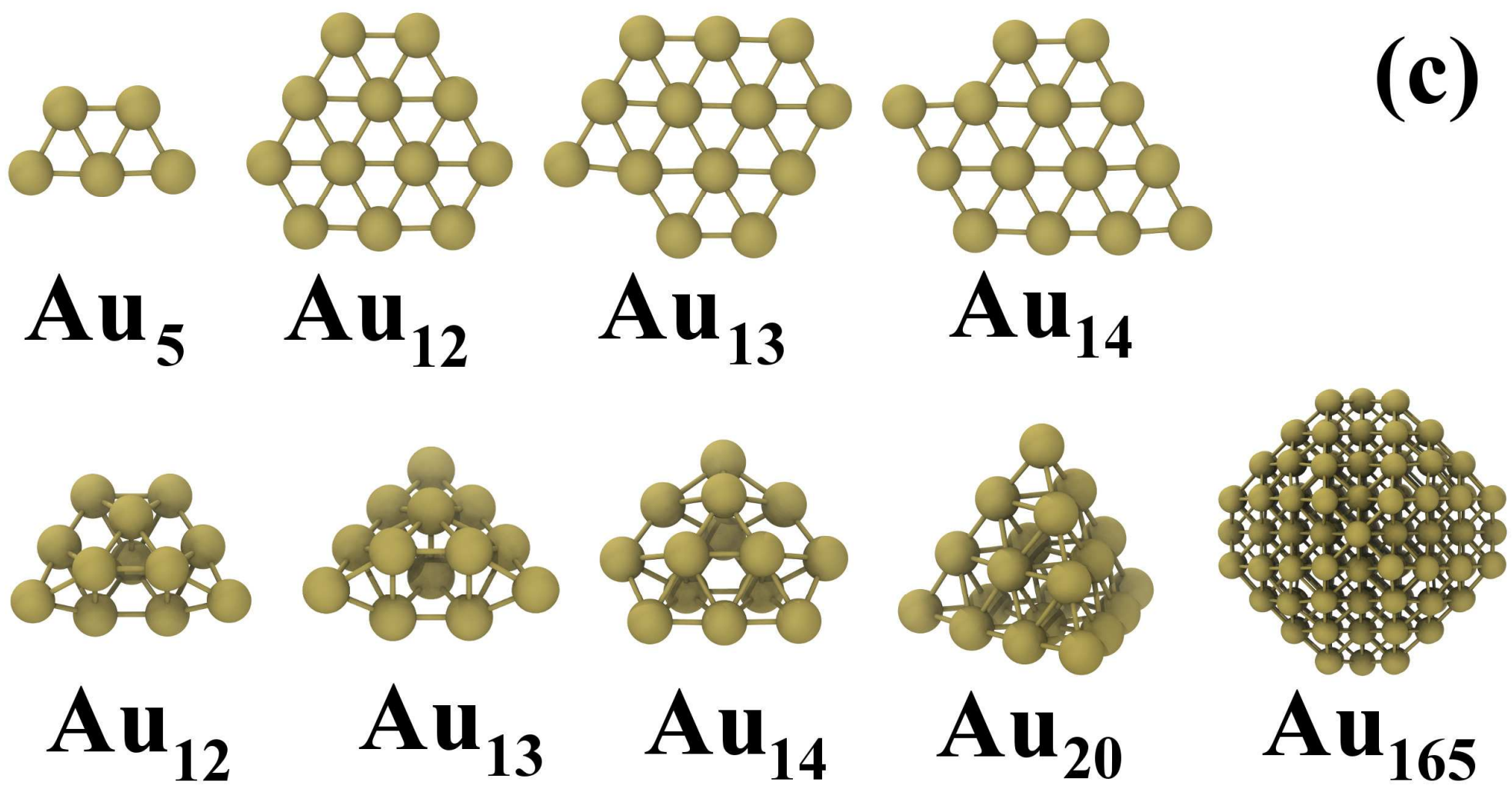

**Figure 6:** *s*-*d* band hybridization index, $H_{sd}$, defined as (a) the common area under *s* and *d* shell decomposed electronic density of states and (b) the product of *s* and *d* weights of local charges for different sized Au clusters in 2D and 3D constructions. Spin-orbit coupling (soc) is also considered in the case of $Au_{12-14}$ clusters (c) The atomic structures of the evaluated particles.

**Long Range Interactions**

Up to this point, the calculations have not involved long-range correlations. Rehr et al.,[42] using perturbation theory, estimated the contribution of dipole-dipole and higher order polarization energy to be 17% of the cohesive energy in metallic gold. These interactions are analogous to van der Waals attraction. Accordingly, we first compare different flavors of DFT van der Waals (vdW) correction methods, i.e. D2,[43, 44] D3,[45] TS,[46] and DF,[47, 48] for their contribution to the cohesive energy of bulk Au. We find that the energy contribution of the vdW corrections, defined as $|\Delta E_{vdW} - \Delta E_{PBE}|/\Delta E_{vdW}$ ($\Delta E_{vdW/PBE}$ is the formation energy of the cluster calculated with/without vdW corrections), increases in the order of TS, DF, D3 and D2, giving approximately 12%, 16%, 18% and 19% of the formation energy, when the interaction cutoffs are selected sufficiently large (see Table S2 in SI). The calculated contributions to cohesive energies for the given vdW approximations are reasonably close to earlier estimation of by Rehr et al.,[42] but the range from 12% to 19% gives us an opportunity to investigate the effect of varying vdW interaction strengths on the 2D-to-3D transition. In Figure 7a-d, the change in per atom formation energies, compared to PBE without vdW corrections, of the clusters presented in Tables 1, 2 and 3 are shown for different vdW approximations. We use two vdW interaction cutoffs, 3.8 and 14.9 Å for D2, D3 and TS. The former cutoff only includes the first nearest neighbors and the latter includes all cluster atoms. In the case of DF, a self-consistent solution is obtained for the entire cluster, thus its effect on formation energy is similar to empirical approximations at large cutoff. The calculations considering only the nearest neighbors reduce the energy of the globular structures but not enough to change the minima for 12 and 13-atom clusters. For both cutoff values, it is seen from Figure 7 that D2 gives the strongest and TS gives the weakest contribution to the formation energy as the former reduces the energy of 3D structures the most and the latter the least. For the TS calculations, the global minima of $Au_{12-14}$ have not been changed. When D2, D3 and DF methods are utilized, the energy of a globular structure, namely $Au_{12}$(4), is lowered below the planar structure (14 meV for D2, 3 meV for D3 and 3 meV for DF in per atom formation energy). In $Au_{13}$, none of the globular structures, which are shown in Table 2, is reduced in energy below planar ones. However, since the energetic order has changed in $Au_{12}$ within few meV, one might expect a similar situation for $Au_{13}$. As it turns out, one of the globular structures that are found during GA optimization of 13-atom anionic clusters is actually lower in energy than planar $Au_{13}$ when vdW interactions are considered in the neutral form. This cluster is given in Table S1 in the SI as $Au_{13}^-$(5). For the D2, D3 and DF methods, the energy difference between this structure and $Au_{13}$(1) is -10 to -17 meV per atom. When the strengths of vdW contributions to the cohesive energy of bulk gold are considered, the D3 and DF methods give the closest approximations to the estimations in Ref. [42]. The DF method is also shown to closely reproduce highly accurate results from quantum Monte Carlo and couple cluster calculations.[49] Therefore, when appropriate dispersive interactions are included, we find that the 2D-3D transformation occurs at lower values of *n*, i.e. $Au_{12}$ or $Au_{13}$ rather than $Au_{14}$. However, whether dispersive interactions are included or not, at the transition cluster size, 2D and 3D structures coexist within the vicinity of $nk_BT$ ($T$=300K), showing that the transition is still a gradual one around room temperature.

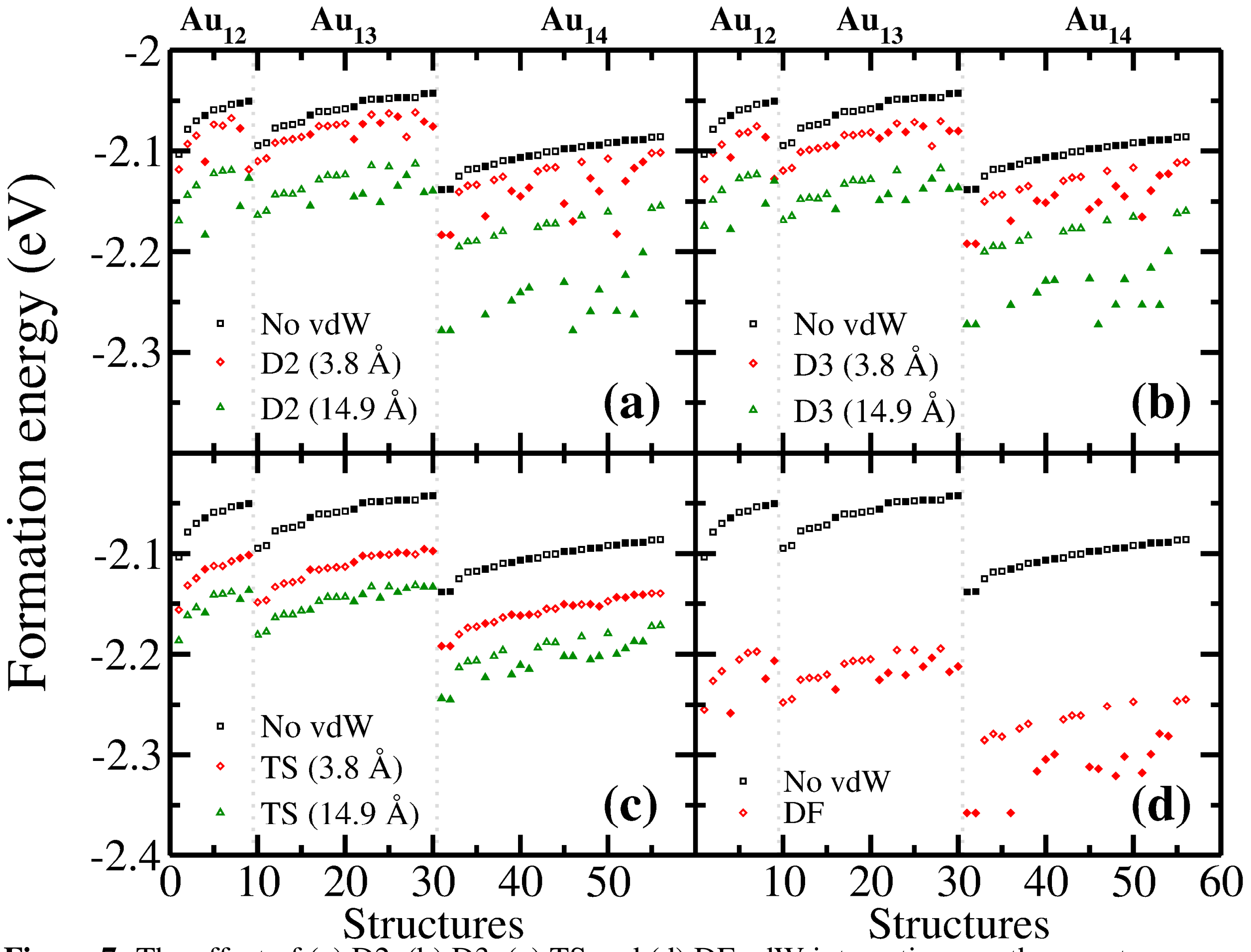

**Figure 7:** The effect of (a) D2, (b) D3, (c) TS and (d) DF vdW interactions on the per atom formation energies of the low energy $Au_{12-14}$ clusters identified in Tables 1-3. Formation energies for different interaction cutoffs, which are given in parenthesis, are shown for 56 structures. The clusters are given in the same order as in Tables 1, 2 and 3 for $Au_{12}$ (structures 1-9), $Au_{13}$ (structures 10-30) and $Au_{14}$ (structures 31-56). The vertical dashed lines separate one cluster size from another. Empty and filled symbols are used for 2D and 3D structures respectively.

## CONCLUSION

We introduce a genetic algorithm – density functional theory method for prediction of stable structures of clusters, and apply it successfully to free-standing $Au_{12\text{-}14}$ nanoclusters. The method is capable of not only finding global minimum, but numerous local minima that will assist the experimental characterization of the synthesized clusters and further computational studies concerning catalytic and photonic properties. The GA optimization is based on physically dividing and recombining clusters, as opposed to bit manipulations that are, in some cases, used to find global minimum in periodic systems. The developed GA code can utilize as many parents as possible, with the only condition that each parent should contribute at least one atom. This helps in keeping the gene pool dynamic. Considering the size of the clusters, four parents are used for crossover in this study. This method is found to outperform two parent crossover or random selection for the systems under study. For the mutations, we use a situation-dependent scheme. The lowest energy structures for 12- and 13- atom neutral clusters are found to have planar geometries whereas a 3D form is obtained as the lowest energy for $Au_{14}$ when no long-range interactions are considered. When D2, D3 and DF flavors of van der Waals interactions are included in the energy calculations, the 2D-3D transition size is reduced to below 13 atoms. The strength of dispersive interactions is found to be weaker for the TS van der Waals interaction without appreciably affecting the stability order. It is also shown that the effect of 5*d*-6*s* hybridization can be predicted by scalar relativistic calculations and the inclusion of spin-orbit coupling in the calculations did not significantly change the difference in hybridization indices between planar and globular clusters. The planar structures are found to have higher hybridization index compared to globular clusters for all sizes, and a decrease in the amount of *s*-*d* hybridization in 2D structures did not consistently correlate with energetic destabilization, leading us to conclude that *s*-*d* hybridization is not a significant factor in 2D-3D transition. When the clusters are ionized, the minima configurations are not altered and formation energy order between clusters is mostly unchanged. In all these calculations, we find several dozen clusters with energies that are in close proximity to the lowest energy structures. We predict that these lowest energy and near-lowest energy structures are likely to coexist at room temperature and above due to thermal excitations. The existence of many structures within a small energy interval may explain the long standing debate on the transition size and the global minima of the gold clusters around the studied size range.

## METHODS

The atomic configurations of gold clusters are optimized by combining GA and DFT calculations. The GA provides a non-local sampling of cluster structures over the phase space using genetic operations, and the DFT calculations are used for local optimization and total energy calculations.

**Genetic Algorithm:** We developed a GA code that is distinct from existing GA codes primarily in that it uses a spatial decomposition scheme for crossover with 4 parents, as shown in Figure 8. Furthermore, the mutation scheme and rate are adjusted automatically during the optimization. Spatial decomposition is realized by dividing each cluster (*i.e.* parent) into four parts, each of which is a connected subcluster of atoms, using either planar or irregular cutting of the parent structure. The mating process involves taking these subclusters from four parents and recombining them into a new cluster. Other genetic operations such as mutations, parent exchange (*i.e.* introduction of randomly generated parents in place of existing ones), and parent cloning (*i.e.* fit parents may be copied to next generation subject to only mutation) are included to increase the optimization speed and prevent premature convergence of the process. For mutations, two operations are considered: (1) A bulk mutation where a randomly selected atom is moved along a random vector within a certain sphere around the mass center of the cluster, (2) an edge/surface mutation where an atom (at position $\vec{r}_o$ ), chosen among the ones having the lowest coordination numbers, is moved along the edge/surface. A surface mutation vector is determined on a plane with normal ( $\vec{N}$ ) defined by adding the vectors from nearest neighbors to the mutating atom, weighted by the inverse distance, *i.e.* $\vec{N} = \sum_i \frac{1}{|\vec{r}_i - \vec{r}_o|}(\vec{r}_i - \vec{r}_o)$, where the index *i* runs over all nearest neighbors. The surface mutation vector is selected inside a radius of 1.5 bond lengths on the plane defined by $\vec{N}$. Consequently, the mutating atom moves to a lower density region at the surface. Bulk mutations are applied during the initial generations where the energy differences between structures are large. Edge/surface mutations, on the other hand, are considered when similar structures start to dominate the population. The mutation rate is set to 20% initially but adjusted on the fly depending on the stagnancy of the lowest energy during the evolution. The population size is selected as 20 and at each generation, 10 structures with the lowest DFT energies are selected for mating operations. For the initial generation, structures are generated randomly subject only to limits on the largest (20 Å) and smallest (2.4 Å) distance between any two gold atoms, in order to give reasonable and sufficiently varied initial guesses for cluster size and atomic bond length.

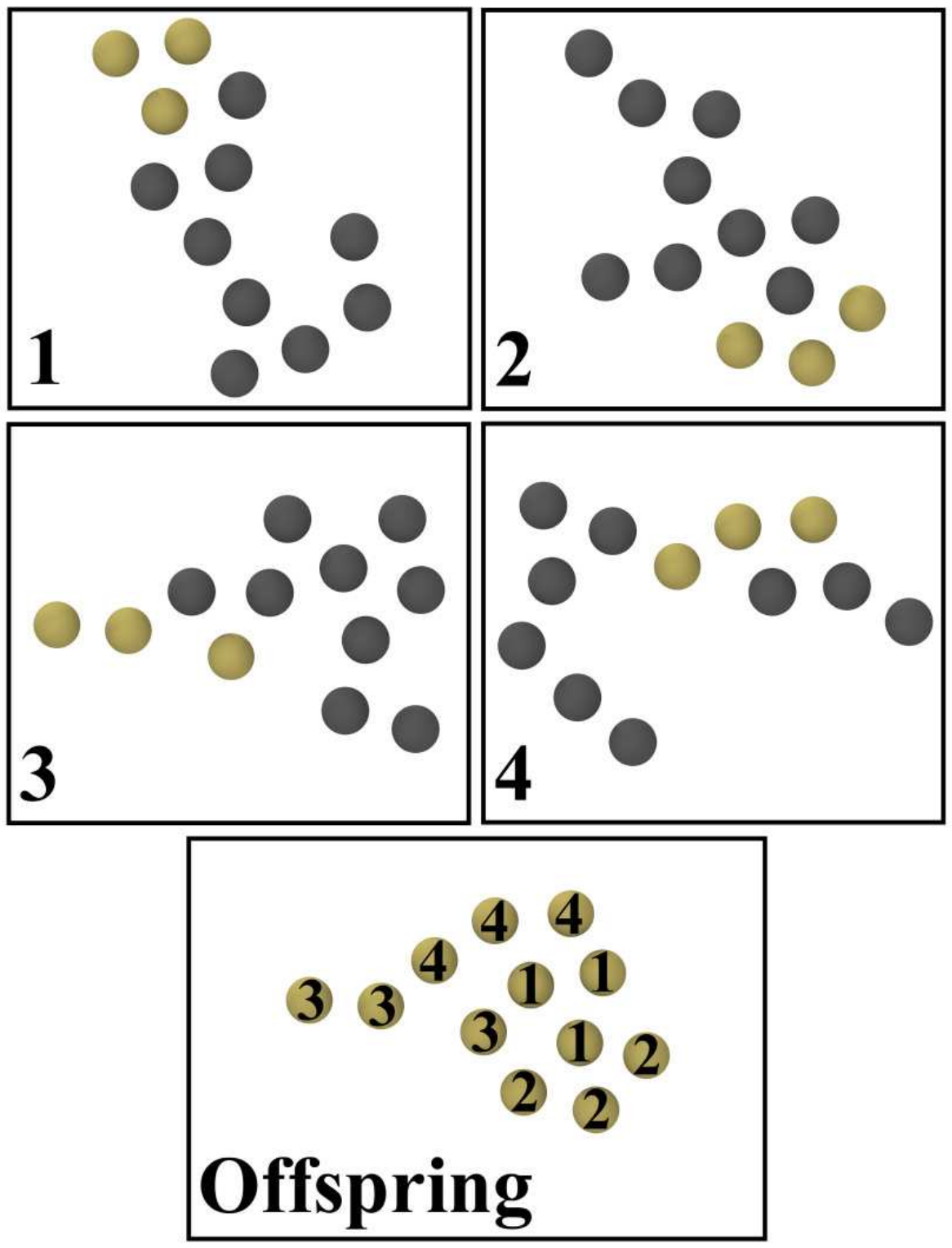

**Figure 8:** 4-parent crossover scheme for a 2D gold cluster. The first 4 panes represent the parents in which the contributed genes are yellow colored. The last pane shows the offspring which is formed from the combination of the selected atom groups from the parents.

We evaluate the energy distribution of structures obtained from GA sampling to report its merits in adequately covering the phase space. In this respect, we first compare the 4-parent versus 2-parent crossover GA schemes. For each type of crossover, we perform 4 independent GA optimizations with a population of 20, lasting 40 generations. When the entire sampling from these calculations is investigated, it is seen that 4-parent crossover scheme samples a larger energy space and produces lower energy structures, as seen in Figure 9a. We also compare the obtained energy distribution from GA with that obtained from random sampling. The same number of ionic relaxations (*i.e.* 18 conjugate gradient steps) is considered for structures created by both sampling methods. In Figure 9b, we show the energy distribution of random sampling and in Figure 9c, the energy distribution from a GA optimization run. It is seen that the randomly generated clusters produce a Gaussian-like distribution with certain energies sampled much more than the rest. The results in Figure 9c show that GA samples a larger energy space extending both to lower and higher energies in a more uniform fashion. This is practically important if the DFT results are to be used as training set for the parameterization of empirical potentials.[50] It is true that the random sampling can be further extended to higher energies by widening the range of accepted atom-atom separations and the distribution can be balanced by careful selection of structures from the randomly generated samples. However, it is much harder to create sufficient number of low energy structures with random generation and any parameterization of empirical methods with a randomly generated training set will be deficient in predictions around ground state.

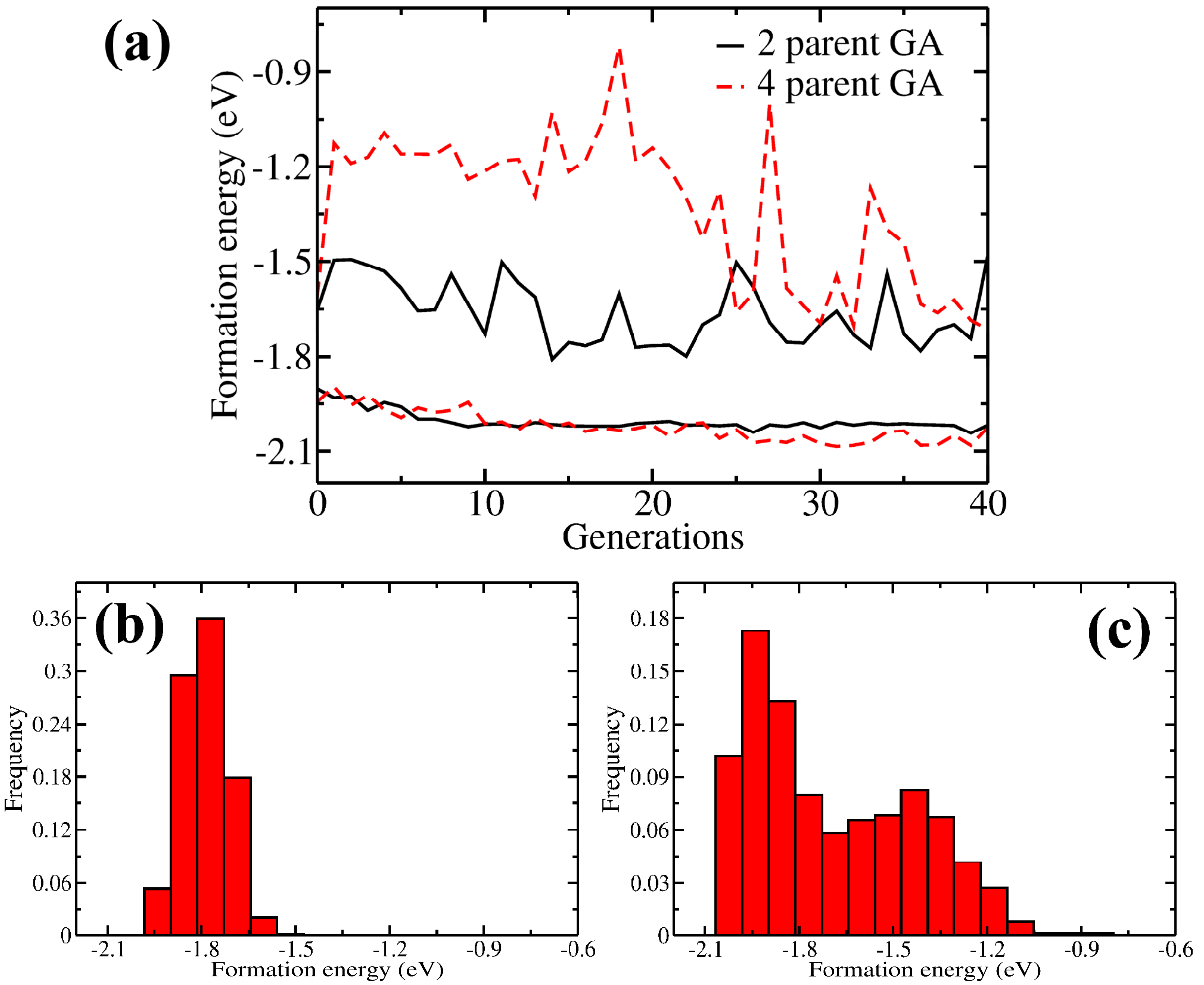

**Figure 9:** (a) The upper and lower limits of the $Au_{13}$ formation energies at each generation during GA optimization with 2-parent (black solid lines) and 4-parent (red dashed lines) crossover schemes. A population of 20 is used for both schemes. Comparison between the energy sampling of (b) randomly-generated and (c) GA-generated $Au_{13}$ nanoclusters. The data is based on 1000 randomly generated clusters for (b) and the first 1000 structures obtained from a 4-parent GA optimization (i.e. 20 clusters over 50 generations) for (c). Per atom formation energies are given for all graphs.

**DFT:** First principles calculations on clusters are performed with a plane wave basis as implemented in the DFT code VASP.[51, 52] The projector augmented wave (PAW) method is used for efficient description of valence states near the cores. The Perdew-Burke-Ernzerhof (PBE)[53] parameterization of the generalized gradient approximation (GGA) is selected for electron exchange-correlation since this functional is found to perform better for nanoclusters.[54, 55] When the vdW-DF method is used for the dispersion interactions, the optB86b-vdW functional is employed. After testing the energy convergence and relative stability of clusters (see Figure S5 in SI and the discussion therein), we determine that a plane-wave energy cutoff of 230 eV converges the energy to 7.5 meV per atom with little to no change in relative stability of gold clusters. Therefore this cutoff is adequate for the DFT calculations during the GA optimization. For 2D clusters, we use a simulation box with the dimensions 30×30×15 $Å^3$, whereas for 3D clusters a 30×30×30 $Å^3$ box is used, in order to minimize spurious cluster-cluster interactions through periodic boundary conditions. Additional information on the convergence of energy and simulation box size is given in SI. In all DFT calculations, only the Γ-point is used in reciprocal space. DFT ionic relaxations during the GA run are performed partially (*i.e.* only 18 ionic steps) using conjugate gradient minimization. Partial relaxation is useful in balancing the number of local and global minimization steps and reducing the computational cost. Also, partially relaxed structures ensure a more inclusive sampling around local minima.

To further reduce computational cost, non-spin-polarized DFT computations are performed during GA optimization. We compare the results of spin-polarized and non-spin-polarized calculations on 1000 randomly-generated $Au_{13}$ structures. The total energies of spin-polarized evaluations are found to be on the average 15 meV lower than the non-spin-polarized results, but the energy order between different structures is largely unaffected. The probability of two clusters having the same order in terms of energy with spin-polarized and non-spin-polarized calculations is found to be 99.7%.

One of our aims is to capture $Au_{12\text{-}14}$ clusters that are energetically relevant under typical synthesis conditions. Small cluster production techniques such as laser vaporization[56] are highly non-equilibrium and expose nanoclusters to a range of temperatures. We report all $Au_n$ clusters that are within $2nk_BT$ ($n$ = number of atoms, $T$ = 300K) from the global minimum energy structures. To identify these clusters, after the GA calculations are completed, we further relax the structures within $3nk_BT$ of the predicted minima, using an increased energy cutoff (300 eV) and spin-polarization, until the energy difference between two ionic relaxation steps are converged to $10^{-4}$ eV/atom. We select structures from $3nk_BT$ proximity because during GA optimization the structures are only partially relaxed and full relaxation can bring some structures below the $2nk_BT$ cap. For these calculations we use a 30×30×30 $Å^3$ simulation box regardless of the dimensionality of the cluster.

***Supplementary Information***

Details of the DFT calculations for gold clusters; plots of genetic algorithm optimization of 2D $Au_{12}$, 3D $Au_{13}$ and 2D $Au_{14}$; low energy structures for anionic $Au_{13}$; the plots showing the variation of energy and *s-d* hybridization in $Au_{12\text{-}14}$ clusters via changing Hubbard correction term U;  the contribution of vdW interactions to the cohesive energy of bulk gold. The structures of all the low energy $Au_{12}$, $Au_{13}$, $Au_{14}$ and $Au_{13}^-$ in Cartesian coordinates are given in a text file in Supplementary Dataset.

## ACKNOWLEDGEMENTS

Use of the Center for Nanoscale Materials, an Office of Science user facility, was supported by the U. S. Department of Energy, Office of Science, Office of Basic Energy Sciences, under Contract No. DE-AC02-06CH11357. We gratefully acknowledge the computing resources provided on Blues and Fusion, high-performance computing clusters operated by the Laboratory Computing Resource Center at Argonne National Laboratory. This work used the Extreme

Science and Engineering Discovery Environment (XSEDE), which is supported by National Science Foundation grant number ACI-1053575.[57]

## REFERENCES

1. Haruta M., Kobayashi T., Sano H. & Yamada N. Novel gold catalysts for the oxidation of carbon monoxide at a temperature far below 0°C. *Chem. Lett.* **16**, 405-408 (1987).
2. Hashmi A. S. Gold-catalyzed organic reactions. *Chem. Rev.* **107**, 3180-3211 (2007).
3. Zheng J., Petty J. T. & Dickson R. M. High quantum yield blue emission from water-soluble au8 nanodots. *J. Am. Chem. Soc.* **125**, 7780-7781 (2003).
4. Daniel M. C. & Astruc D. Gold nanoparticles: Assembly, supramolecular chemistry, quantum-size-related properties, and applications toward biology, catalysis, and nanotechnology. *Chem. Rev.* **104**, 293-346 (2004).
5. Saha K., Agasti S. S., Kim C., Li X. & Rotello V. M. Gold nanoparticles in chemical and biological sensing. *Chem. Rev.* **112**, 2739-2779 (2012).
6. Zheng J., Zhang C. & Dickson R. M. Highly fluorescent, water-soluble, size-tunable gold quantum dots. *Phys. Rev. Lett.* **93**, 077402 (2004).
7. Haruta M. Size- and support-dependency in the catalysis of gold. *Catal. Today* **36**, 153-166 (1997).
8. Fernández E. M., Soler J. M., Garzón I. L. & Balbás L. C. Trends in the structure and bonding of noble metal clusters. *Phys. Rev. B* **70**, 165403 (2004).
9. Öğüt S., Idrobo J. C., Jellinek J. & Wang J. Structural, electronic, and optical properties of noble metal clusters from first principles. *J. Clust. Sci.* **17**, 609-626 (2006).
10. Gilb S., Weis P., Furche F., Ahlrichs R. & Kappes M. M. Structures of small gold cluster cations ($Au_n^+$, n<14): Ion mobility measurements versus density functional calculations. *J. Chem. Phys.* **116**, 4094 (2002).
11. Häkkinen H., *et al.* On the electronic and atomic structures of small $Au_n^-$ (n = 4–14) clusters: A photoelectron spectroscopy and density-functional study. *J. Phys. Chem. A* **107**, 6168-6175 (2003).
12. Xiao L., Tollberg B., Hu X. & Wang L. Structural study of gold clusters. *J. Chem. Phys.* **124**, 114309 (2006).
13. Xing X., Yoon B., Landman U. & Parks J. H. Structural evolution of au nanoclusters: From planar to cage to tubular motifs. *Phys. Rev. B* **74**, 165423 (2006).
14. Grönbeck H. & Andreoni W. Gold and platinum microclusters and their anions: Comparison of structural and electronic properties. *Chem. Phys.* **262**, 1-14 (2000).
15. Hakkinen H., Moseler M. & Landman U. Bonding in Cu, Ag, and Au clusters: Relativistic effects, trends, and surprises. *Phys. Rev. Lett.* **89**, 033401 (2002).
16. Pyykko P. Theoretical chemistry of gold. *Angew. Chem. Int. Ed. Engl.* **43**, 4412-4456 (2004).
17. Gruene P., *et al.* Structures of neutral $Au_7$, $Au_{19}$, and $Au_{20}$ clusters in the gas phase. *Science* **321**, 674-676 (2008).
18. Bulusu S., Li X., Wang L. S. & Zeng X. C. Evidence of hollow golden cages. *P. Natl. Acad. Sci. U S A* **103**, 8326-8330 (2006).

19. Wang J., Ning H., Ma Q. M., Liu Y. & Li Y. C. $Au_{42}$: A possible ground-state noble metallic nanotube. *J. Chem. Phys.* **129**, 134705 (2008).
20. Wang J.*, et al.* Hollow cages versus space-filling structures for medium-sized gold clusters: The spherical aromaticity of the $Au_{50}$ cage. *J. Phys. Chem. A* **109**, 9265-9269 (2005).
21. Wales D. J. & Doye J. P. K. Global optimization by basin-hopping and the lowest energy structures of Lennard-Jones clusters containing up to 110 atoms. *J. Phys. Chem. A* **101**, 5111-5116 (1997).
22. Lee H. M., Ge M., Sahu B. R., Tarakeshwar P. & Kim K. S. Geometrical and electronic structures of gold, silver, and gold−silver binary clusters: Ori□ins of ductility of gold and gold−silver alloy formation. *J. Phys. Chem. B* **107**, 9994-10005 (2003).
23. Kirkpatrick S., Gelatt C. D. & Vecchi M. P. Optimization by simulated annealing. *Science* **220**, 671-680 (1983).
24. Eiben A. E. & Smith J. E. *Introduction to Evolutionary Computing*. Springer (2003).
25. Goedecker S. Minima hopping: An efficient search method for the global minimum of the potential energy surface of complex molecular systems. *J. Chem. Phys.* **120**, 9911-9917 (2004).
26. Kennedy J. & Eberhart R. Particle swarm optimization. In: *IEEE International Conference on Neural Networks Proceedings.* pp 1942-1948, Perth, WA, Australia, (1995).
27. Heiles S., Logsdail A. J., Schafer R. & Johnston R. L. Dopant-induced 2D-3D transition in small Au-containing clusters: DFT-global optimisation of 8-atom Au-Ag nanoalloys. *Nanoscale* **4**, 1109-1115 (2012).
28. Alexandrova A. N. & Boldyrev A. I. Search for the $Li_n^{0/+1/-1}$(n= 5−7) lowest-energy structures using the ab initio gradient embedded genetic algorithm (gega). Elucidation of the chemical bonding in the lithium clusters. *J. Chem. Theory Comput.* **1**, 566-580 (2005).
29. Jóhannesson G. H.*, et al.* Combined electronic structure and evolutionary search approach to materials design. *Phys. Rev. Lett.* **88**, 255506 (2002).
30. Shao N.*, et al.* Probing the structural evolution of medium-sized gold clusters: $Au_n^-$ (n = 27-35). *J. Am. Chem. Soc.* **132**, 6596-6605 (2010).
31. Sen F. G.*, et al.* Towards accurate prediction of catalytic activity in $IrO_2$ nanoclusters via first principles-based variable charge force field. *J. Mater. Chem. A* **3**, 18970-18982 (2015).
32. Yildirim H., Kinaci A., Zhao Z. J., Chan M. K. & Greeley J. P. First-principles analysis of defect-mediated Li adsorption on graphene. *ACS Appl. Mater. Interfaces* **6**, 21141-21150 (2014).
33. Xiao Y. L. & Williams D. E. Genetic algorithm - a new approach to the prediction of the structure of molecular clusters. *Chem. Phys. Lett.* **215**, 17-24 (1993).
34. Deaven D. M. & Ho K. M. Molecular geometry optimization with a genetic algorithm. *Phys. Rev. Lett.* **75**, 288-291 (1995).
35. Vilhelmsen L. B. & Hammer B. A genetic algorithm for first principles global structure optimization of supported nano structures. *J. Chem. Phys.* **141**, 044711 (2014).
36. Do H. & Besley N. A. Structural optimization of molecular clusters with density functional theory combined with basin hopping. *J. Chem. Phys.* **137**, 134106 (2012).

37. Gehrke R. & Reuter K. Assessing the efficiency of first-principles basin-hopping sampling. *Phys. Rev. B* **79**, 085412 (2009).
38. Makov G. & Payne M. C. Periodic boundary conditions in *ab initio* calculations. *Phys. Rev. B* **51**, 4014-4022 (1995).
39. Kresse G. & Joubert D. From ultrasoft pseudopotentials to the projector augmented-wave method. *Phys. Rev. B* **59**, 1758-1775 (1999).
40. Barmparis G. D. & Remediakis I. N. Dependence on Co adsorption of the shapes of multifaceted gold nanoparticles: A density functional theory. *Phys. Rev. B* **86**, 085457 (2012).
41. Grönbeck H. & Broqvist P. Comparison of the bonding in $Au_8$ and $Cu_8$: A density functional theory study. *Phys. Rev. B* **71**, 073408 (2005).
42. Rehr J. J., Zaremba E. & Kohn W. Van der Waals forces in the noble metals. *Phys. Rev. B* **12**, 2062-2066 (1975).
43. Grimme S. Semiempirical GGA-type density functional constructed with a long-range dispersion correction. *J. Comput. Chem.* **27**, 1787-1799 (2006).
44. Amft M., Lebegue S., Eriksson O. & Skorodumova N. V. Adsorption of Cu, Ag, and Au atoms on graphene including van der Waals interactions. *J. Phys. Condens. Mat.* **23**, 395001 (2011).
45. Grimme S., Antony J., Ehrlich S. & Krieg H. A consistent and accurate ab initio parametrization of density functional dispersion correction (DFT-D) for the 94 elements H-Pu. *J. Chem. Phys.* **132**, 154104 (2010).
46. Tkatchenko A. & Scheffler M. Accurate molecular van der Waals interactions from ground-state electron density and free-atom reference data. *Phys. Rev. Lett.* **102**, 073005 (2009).
47. Klimeš J., Bowler D. R. & Michaelides A. Van der Waals density functionals applied to solids. *Phy. Rev. B* **83**, 195131 (2011).
48. Klimes J., Bowler D. R. & Michaelides A. Chemical accuracy for the van der Waals density functional. *J. Phys. Condens. Mat.* **22**, 022201 (2010).
49. Berland K., *et al.* Van der Waals forces in density functional theory: A review of the vdW-DF method. *Rep. Prog. Phys.* **78**, 066501 (2015).
50. Narayanan B., *et al.* Describing the diverse geometries of gold nanoclusters - a first-principles based bond order potential. *To be published* (http://arxiv.org/abs/1511.03319), (2015).
51. Hohenberg P. & Kohn W. Inhomogeneous electron gas. *Phys. Rev.* **136**, B864-B871 (1964).
52. Kresse G. & Furthmuller J. Efficient iterative schemes for ab initio total-energy calculations using a plane-wave basis set. *Phys. Rev. B* **54**, 11169-11186 (1996).
53. Perdew J. P., Burke K. & Ernzerhof M. Generalized gradient approximation made simple. *Phys. Rev. Lett.* **77**, 3865-3868 (1996).
54. Serapian S. A., Bearpark M. J. & Bresme F. The shape of $Au_8$: Gold leaf or gold nugget? *Nanoscale* **5**, 6445-6457 (2013).
55. Järvi T. T., *et al.* Development of a ReaxFF description for gold. *Eur. Phys. J. B* **66**, 75-79 (2008).
56. Lee S., *et al.* Oxidative decomposition of methanol on subnanometer palladium clusters: The effect of catalyst size and support composition. *J. Phys. Chem. C* **114**, 10342-10348 (2010).

57. Towns J., *et al.* XSEDE: Accelerating scientific discovery. *Comput. Sci. Eng.* **16**, 62-74 (2014).

# Supplementary Information

# Unraveling the Planar-Globular Transition in Gold Nanoclusters through Evolutionary Search

*Alper Kinaci,[1] Badri Narayanan,[1] Fatih G. Sen,[1] Michael J. Davis,[2] Stephen K. Gray,[1] Subramanian K. R. S. Sankaranarayanan,[1] Maria K. Y. Chan[1*]*

[1]Center for Nanoscale Materials, Argonne National Laboratory, Lemont, IL 60439 USA

[2]Chemical Sciences Division, Argonne National Laboratory, Lemont, IL 60439 USA

**Correspondence and requests for materials should be addressed to M.C. (email: mchan@anl.gov)*

## GA optimization of neutral 2D $Au_{12}$, 3D $Au_{13}$ and 2D $Au_{14}$

Structural evolution of 2D $Au_{12}$, 3D $Au_{13}$ and 2D $Au_{14}$ are presented in Figures S1, S2 and S3 respectively. These figures include the per atom formation energies of all evaluated members at each generation and some selected structures that have the lowest energies at the corresponding generations from the evolution path. In calculation of formation energies, single gold atom in vacuum is used as the reference state.

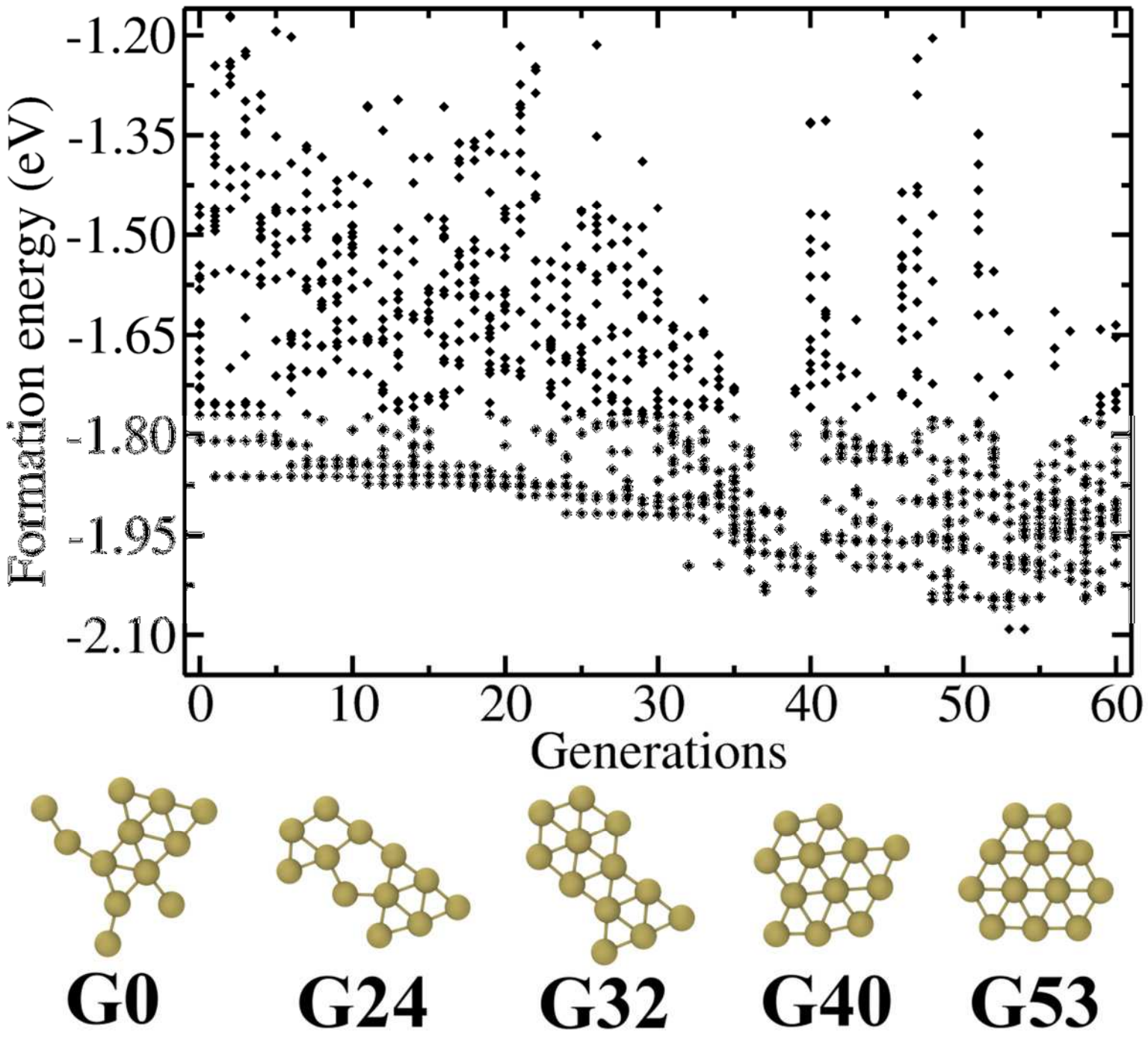

**Figure S1:** Per atom formation energies and structures of 2D $Au_{12}$ clusters as a function of generation number through genetic algorithm optimization. The lowest energy clusters at selected generations (given by G#) are also presented in the bottom panel.

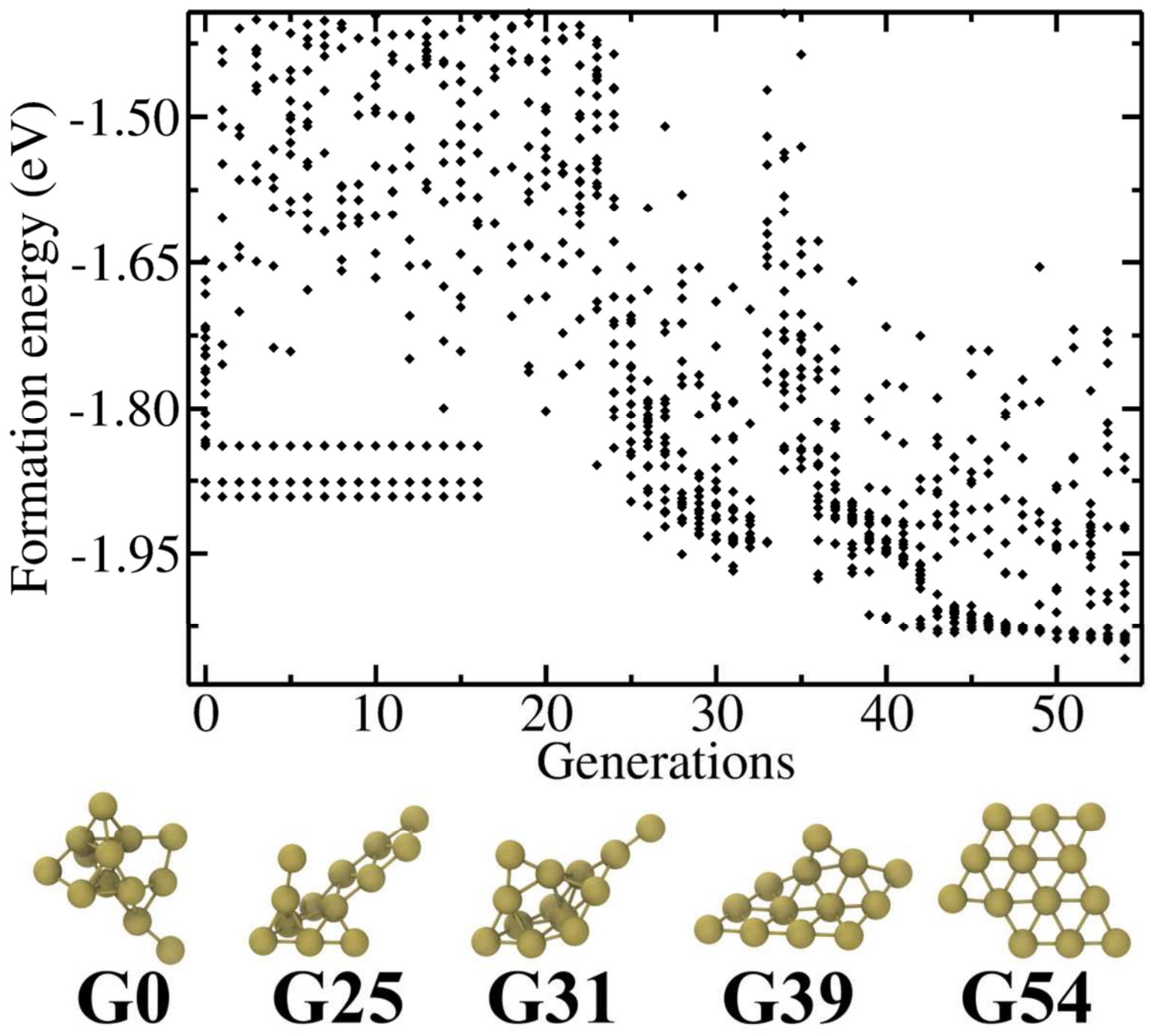

**Figure S2:** Per atom formation energies and structures of 3D $Au_{13}$ clusters as a function of generation number through genetic algorithm optimization. The lowest energy clusters at selected generations (given by G#) are also presented in the bottom panel.

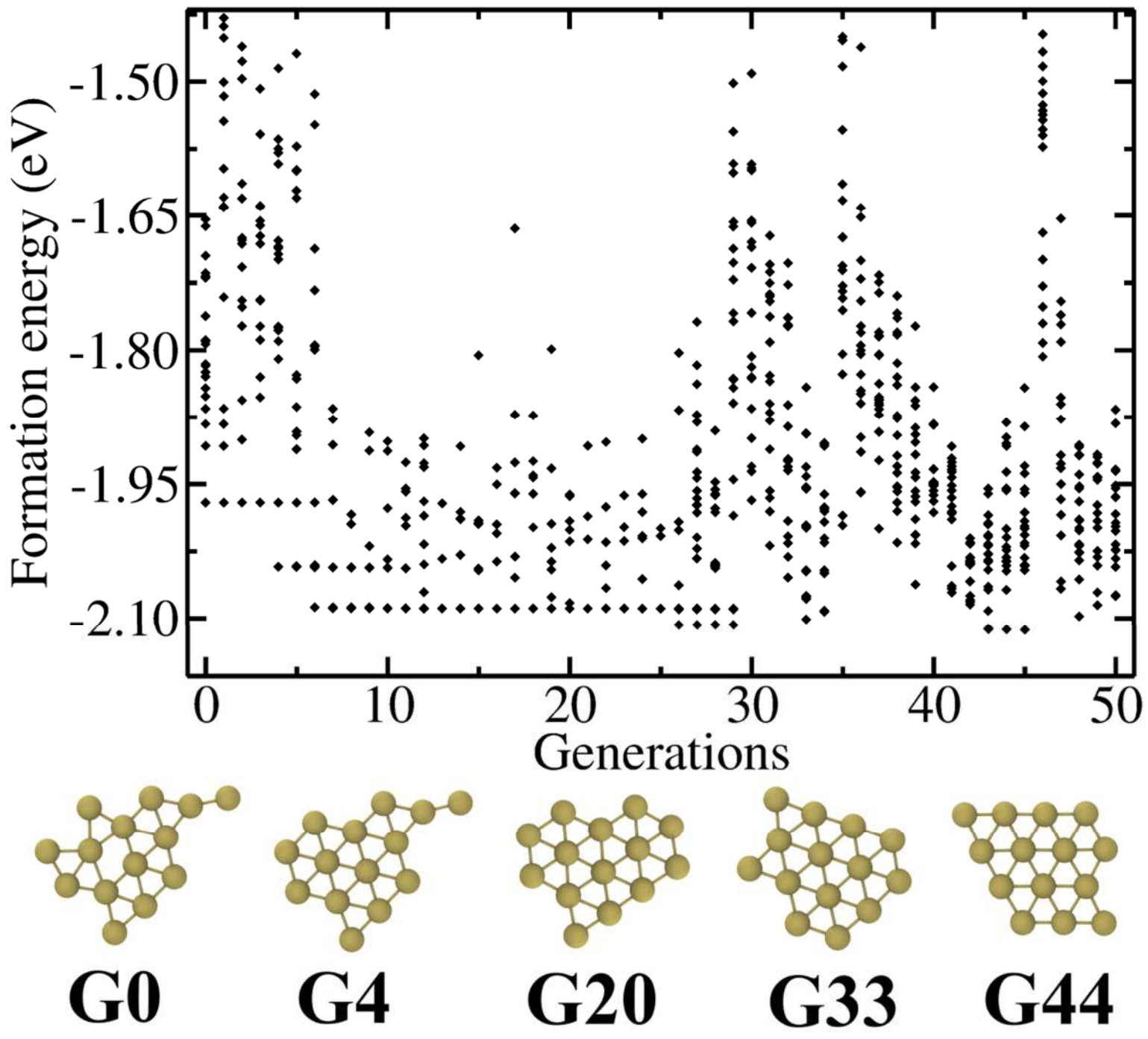

**Figure S3:** Per atom formation energies and structures of 2D $Au_{14}$ clusters as a function of generation number through genetic algorithm optimization. The lowest energy clusters at selected generations (given by G#) are also presented in the bottom panel.

## Structures of neutral clusters in xyz format

The Supplementary Dataset file Au_nanoclusters.doc contains the Cartesian coordinates of the clusters presented in Tables 1, 2 and 3 in the main text. The file may be opened as a text file to read the coordinates. The coordinates are presented in .xyz format. The first line denotes the number of atoms in the cluster; the second line is the name of the cluster i.e. the name given in Tables 1-3; following lines contain the elements symbol, x, y and z coordinates in Angstroms for each atom in the cluster. The structures are placed directly end-to-end in the file following the order in Tables 1, 2 and 3.

## GA optimization of anionic $Au_{13}$

In Table S1 we present the anionic 13-atom gold clusters that have energies within $2k_BT$ of the identified minimum for this system. These structures are obtained as a result of 4 independent GA optimizations with Au atoms having freedom of motion in three dimensions. The DFT and GA calculations have similar setups as the ones for neutral clusters except the clusters have an excess electron.

**Table S1:** $Au_{13}^-$ that are in $2nk_BT$ ($n$=13, $T$=300K) proximity of the predicted minimum energy structure for the anionic clusters. The $\Delta E$ is described as the per atom energy difference of the corresponding structure from the minimum which is given as $Au_{13}^-$(1). If a structure is same as one of the neutral clusters, the name of the neutral cluster (i.e. shown in Table 2) is given in brackets.

| $Au_{13}^-$(1) [$Au_{13}$(1)] | $Au_{13}^-$(2) [$Au_{13}$(2)] | $Au_{13}^-$(3) [$Au_{13}$(3)] | $Au_{13}^-$(4) [$Au_{13}$(4)] |
|---|---|---|---|
| 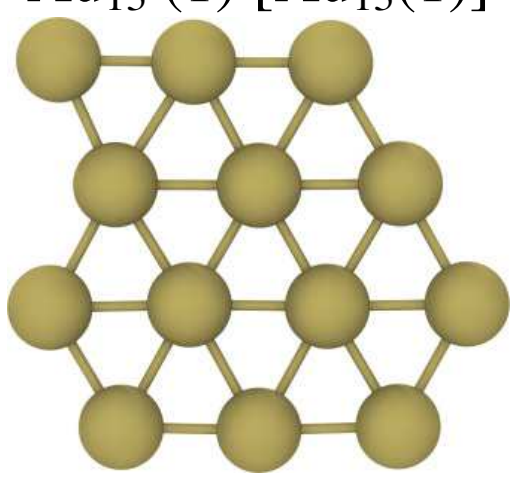 | 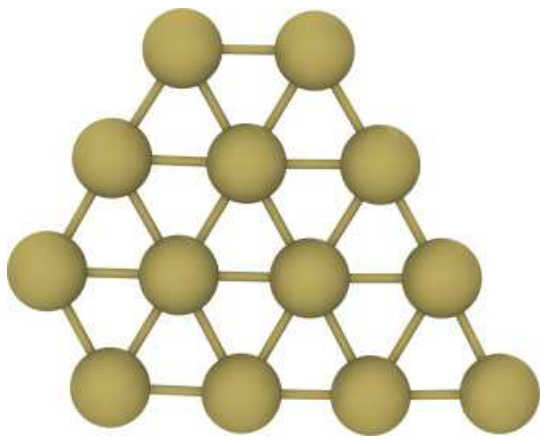 | 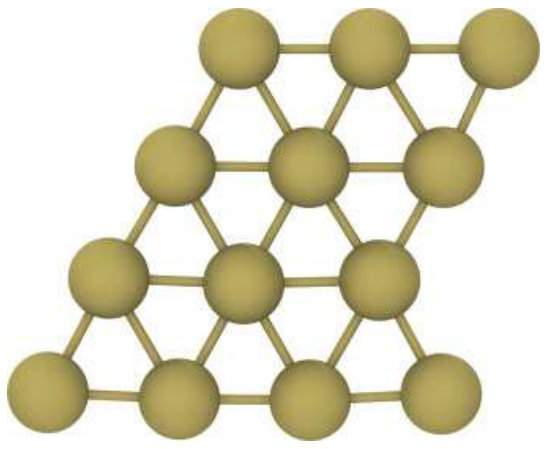 | 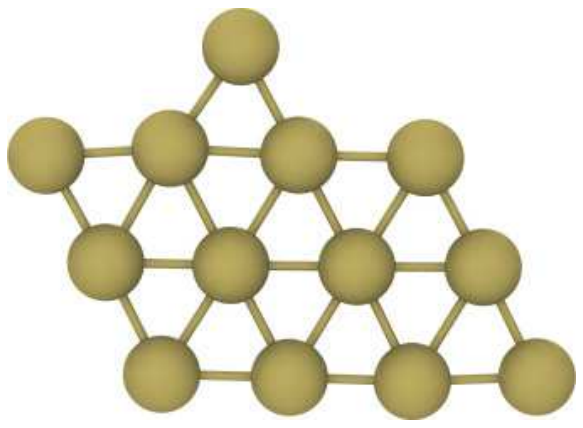 |
| $\Delta E$=0.000 eV | $\Delta E$=0.003 eV | $\Delta E$=0.014 eV | $\Delta E$=0.019 eV |
| $Au_{13}^-$(5) | $Au_{13}^-$(6) | $Au_{13}^-$(7) [$Au_{13}$(6)] | $Au_{13}^-$(8) |
| 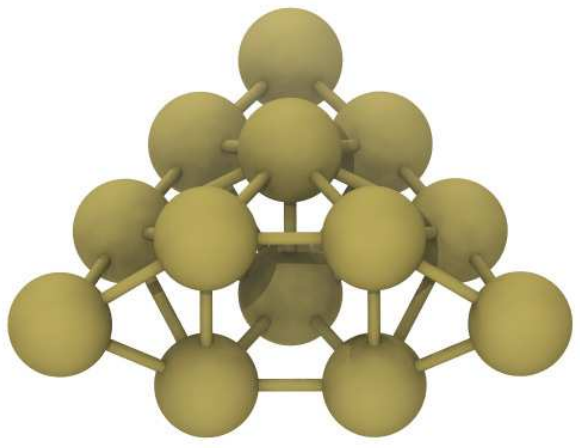 | 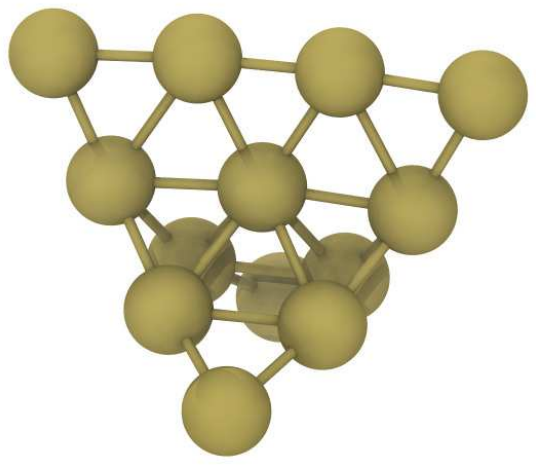 | 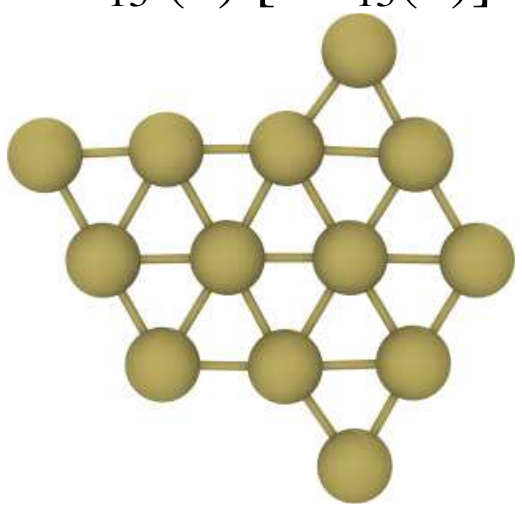 | 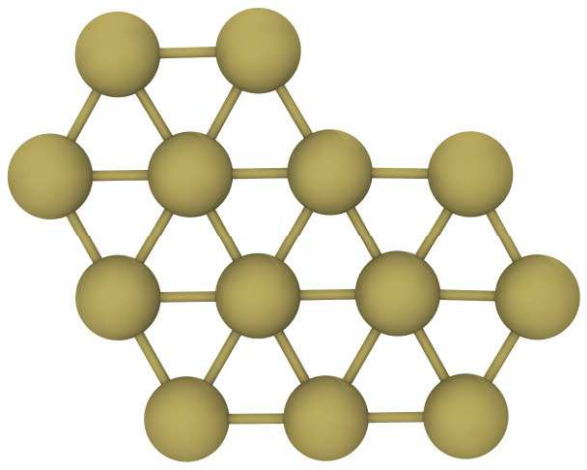 |
| $\Delta E$=0.021 eV | $\Delta E$=0.024 eV | $\Delta E$=0.025 eV | $\Delta E$=0.028 eV |
| $Au_{13}^-$(9) | $Au_{13}^-$(10) [$Au_{13}$(7)] | $Au_{13}^-$(11) | $Au_{13}^-$(12) |
| 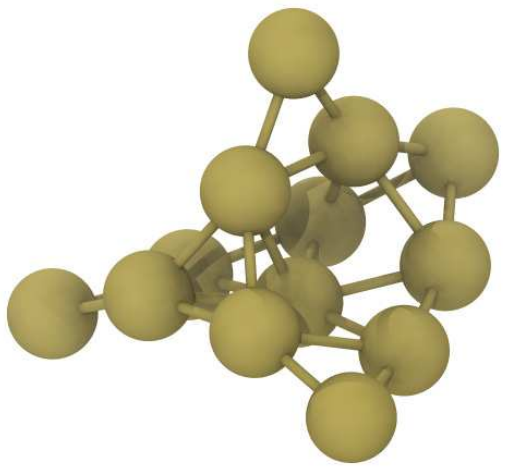 | 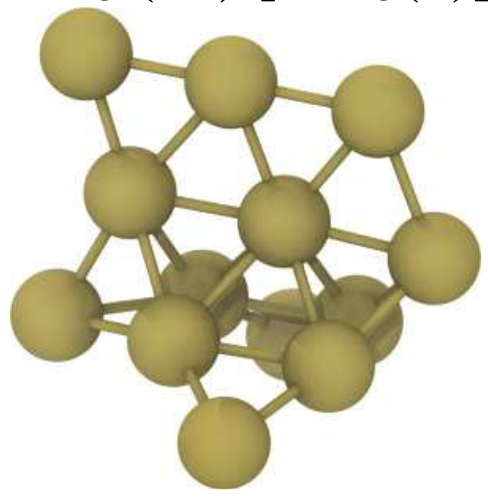 | 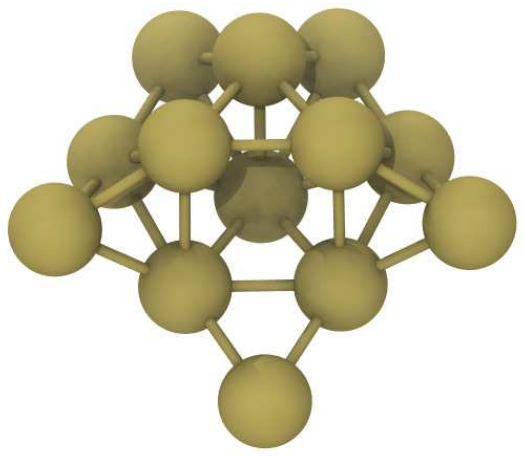 | 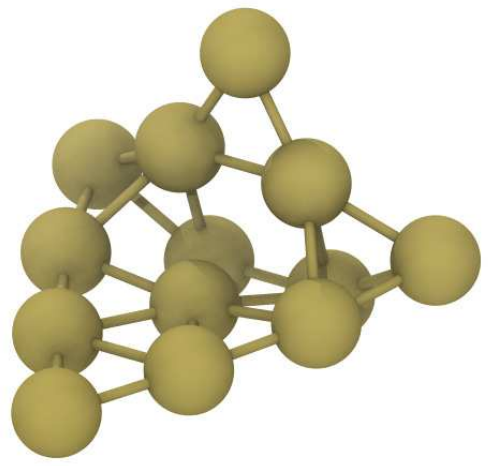 |
| $\Delta E$=0.029 eV | $\Delta E$=0.031 eV | $\Delta E$=0.032 eV | $\Delta E$=0.033 eV |
| $Au_{13}^-$(13) [$Au_{13}$(8)] | $Au_{13}^-$(14) | $Au_{13}^-$(15) [$Au_{13}$(10)] | $Au_{13}^-$(16) |

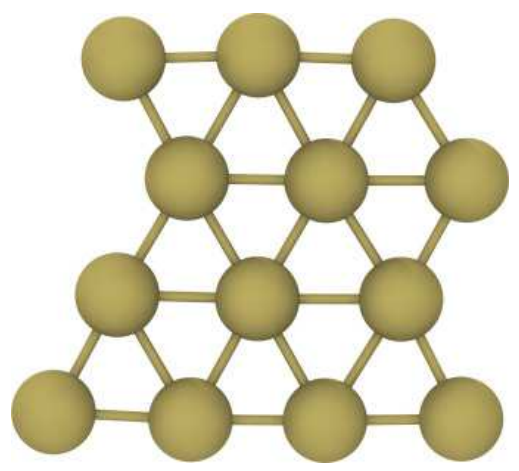 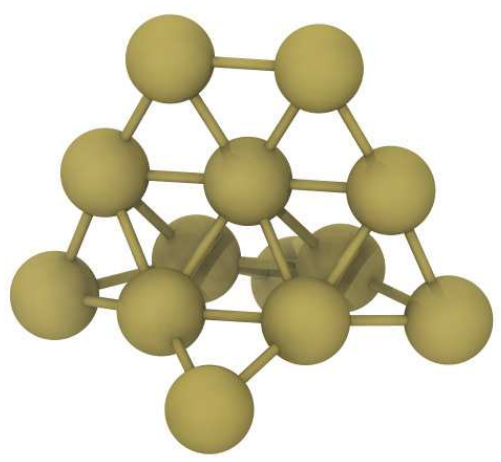 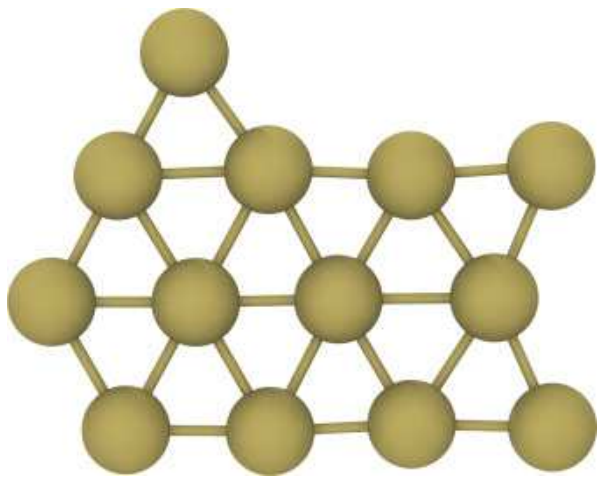 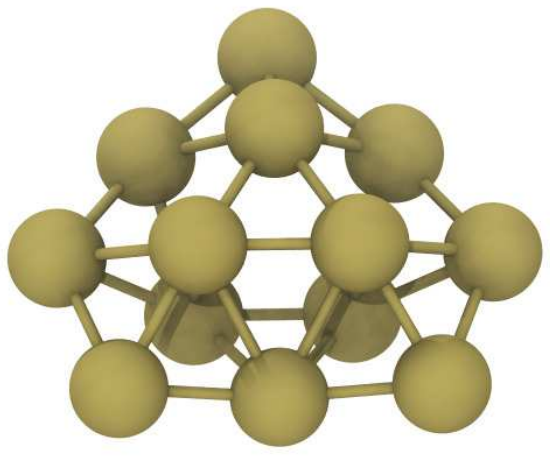

Δ*E*=0.033 eV | Δ*E*=0.034 eV | Δ*E*=0.035 eV | Δ*E*=0.035 eV

$Au_{13}^{-}$(17) [$Au_{13}$(11)] | $Au_{13}^{-}$(18) | $Au_{13}^{-}$(19) | $Au_{13}^{-}$(20) [$Au_{13}$(12)]

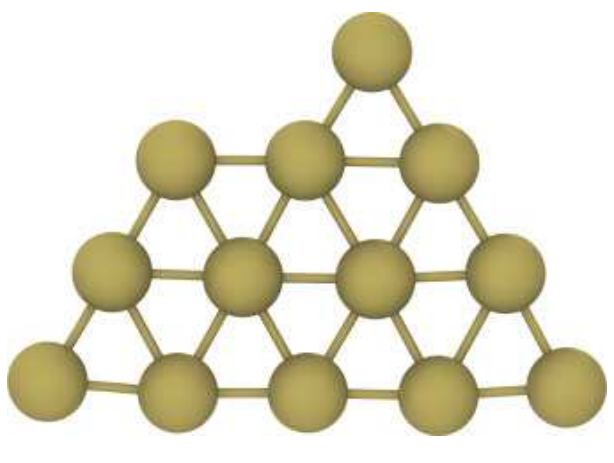 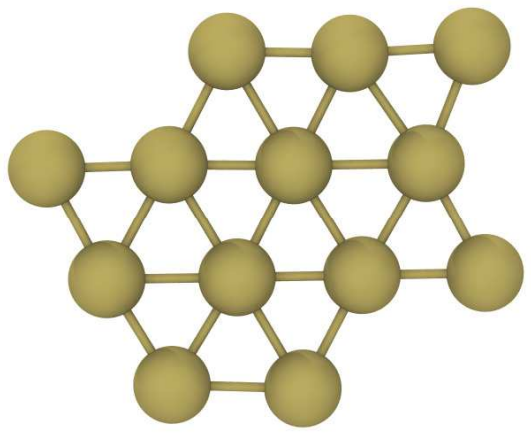 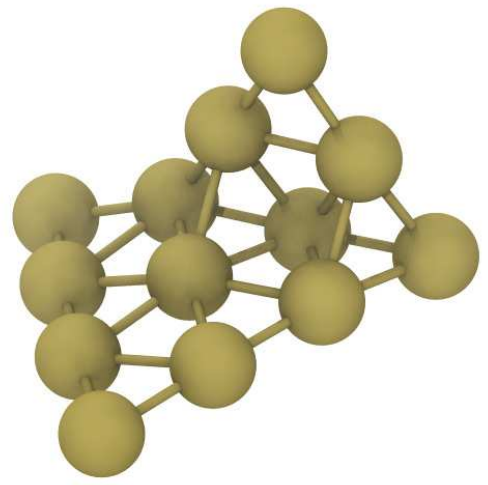 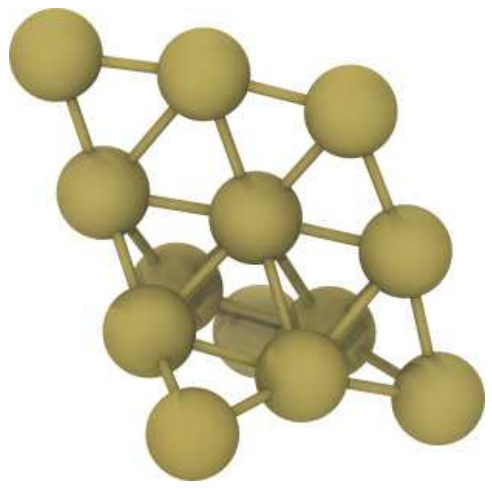

Δ*E*=0.036 eV | Δ*E*=0.038 eV | Δ*E*=0.041 eV | Δ*E*=0.042 eV

$Au_{13}^{-}$(21) | $Au_{13}^{-}$(22) | $Au_{13}^{-}$(23) | $Au_{13}^{-}$(24)

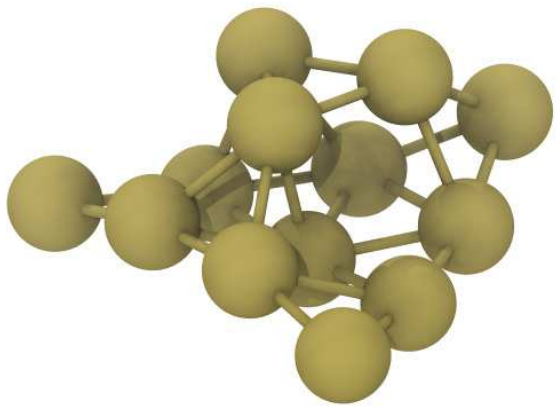 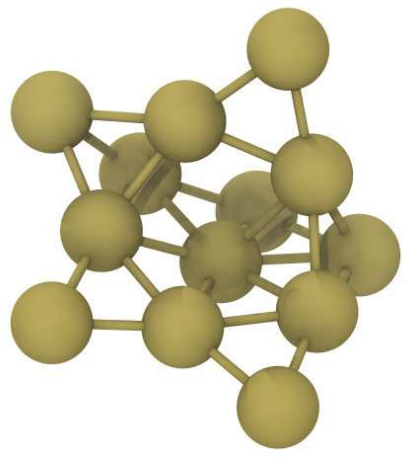 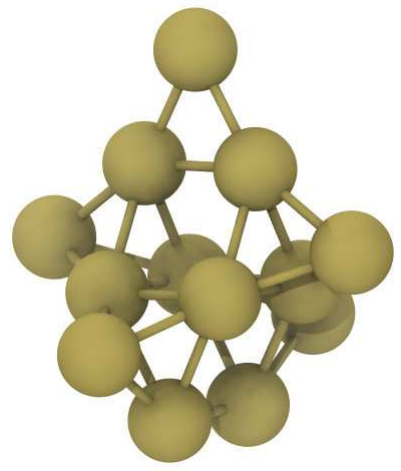 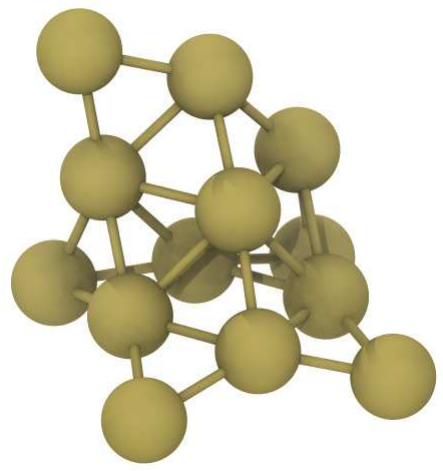

Δ*E*=0.042 eV | Δ*E*=0.043 eV | Δ*E*=0.043 eV | Δ*E*=0.043 eV

$Au_{13}^{-}$(25) | $Au_{13}^{-}$(26) | $Au_{13}^{-}$(27) | $Au_{13}^{-}$(28)

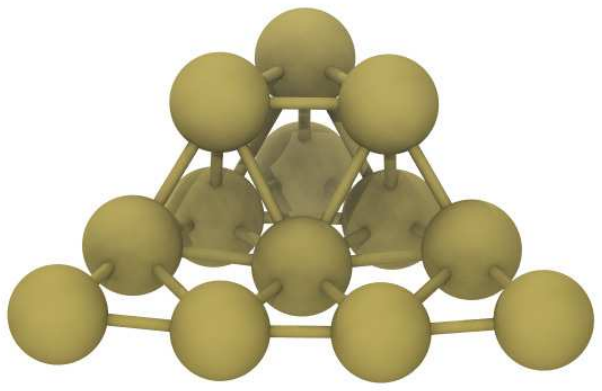 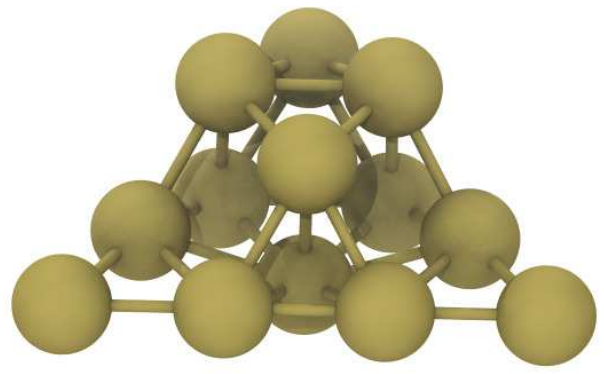 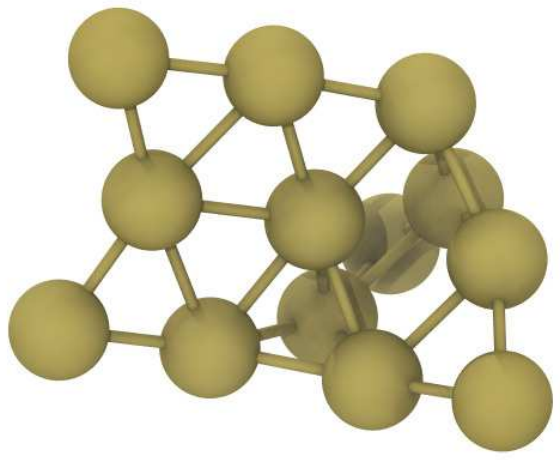 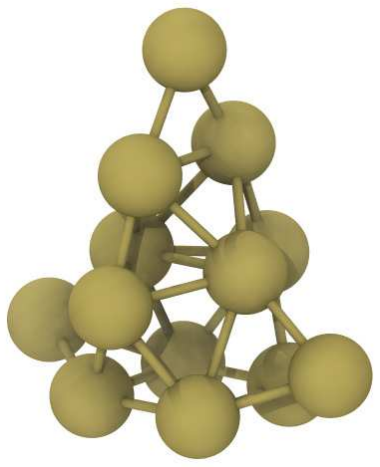

Δ*E*=0.045 eV | Δ*E*=0.046 eV | Δ*E*=0.048 eV | Δ*E*=0.048 eV

$Au_{13}^{-}$(29) | $Au_{13}^{-}$(30) [$Au_{13}$(13)] | $Au_{13}^{-}$(31) [$Au_{13}$(14)] | $Au_{13}^{-}$(32)

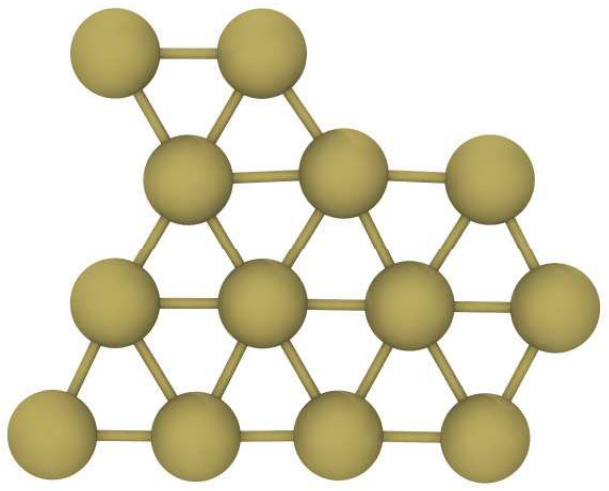 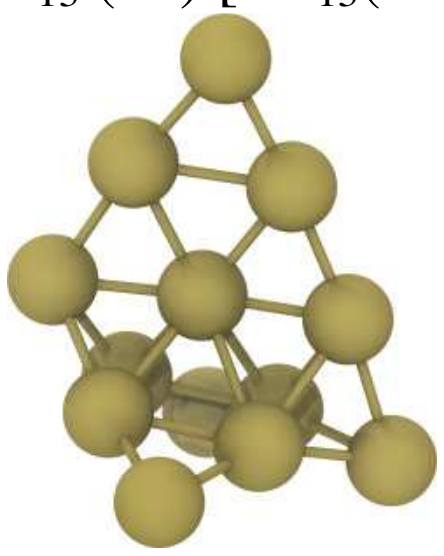 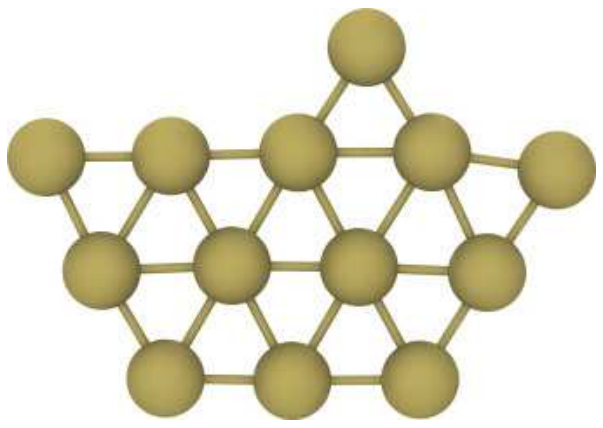 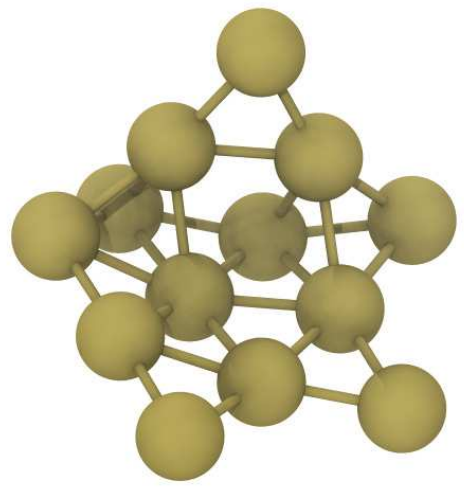

ΔE=0.049 eV | ΔE=0.049 eV | ΔE=0.049 eV | ΔE=0.050 eV

$Au_{13}^{-}$(33)

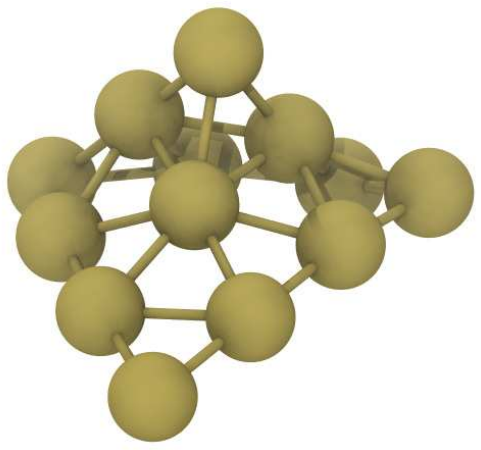

ΔE=0.050 eV

$Au_{13}^{-}$(34)

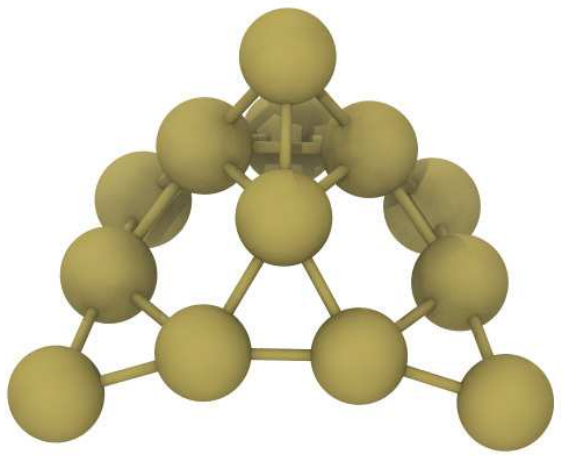

ΔE=0.051 eV

$Au_{13}^{-}$(35)

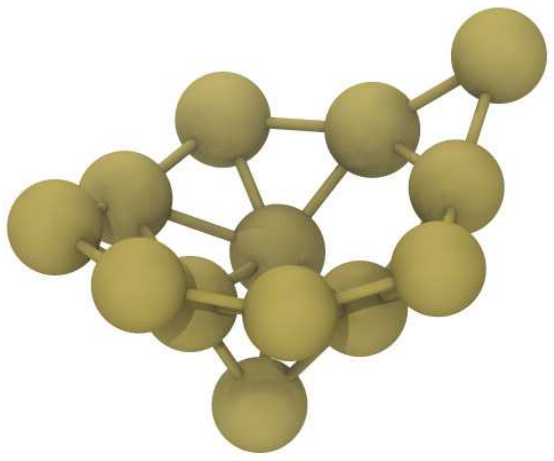

ΔE=0.051 eV

$Au_{13}^{-}$(36)

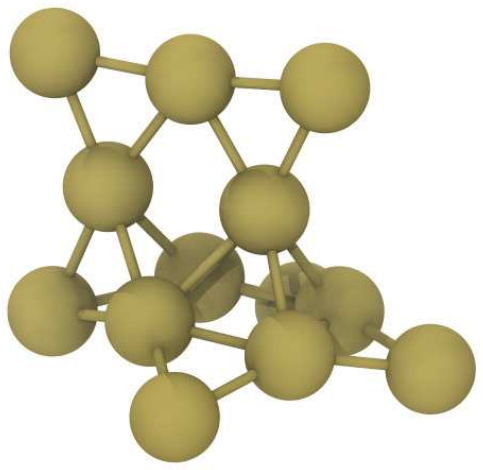

ΔE=0.051 eV

---

**Structures of anionic clusters in xyz format**

The Cartesian coordinates of the clusters in Table S1 are included in the Supplementary Dataset file Au_nanoclusters.doc as just after the neutral clusters. The format and the naming scheme are same as the ones for neutral clusters. The structures are placed directly end-to-end in the file with the order given in Table S1 following the structures given for neutral cluster in Tables 1, 2 and 3.

## Hybridization in *s-d Orbitals*

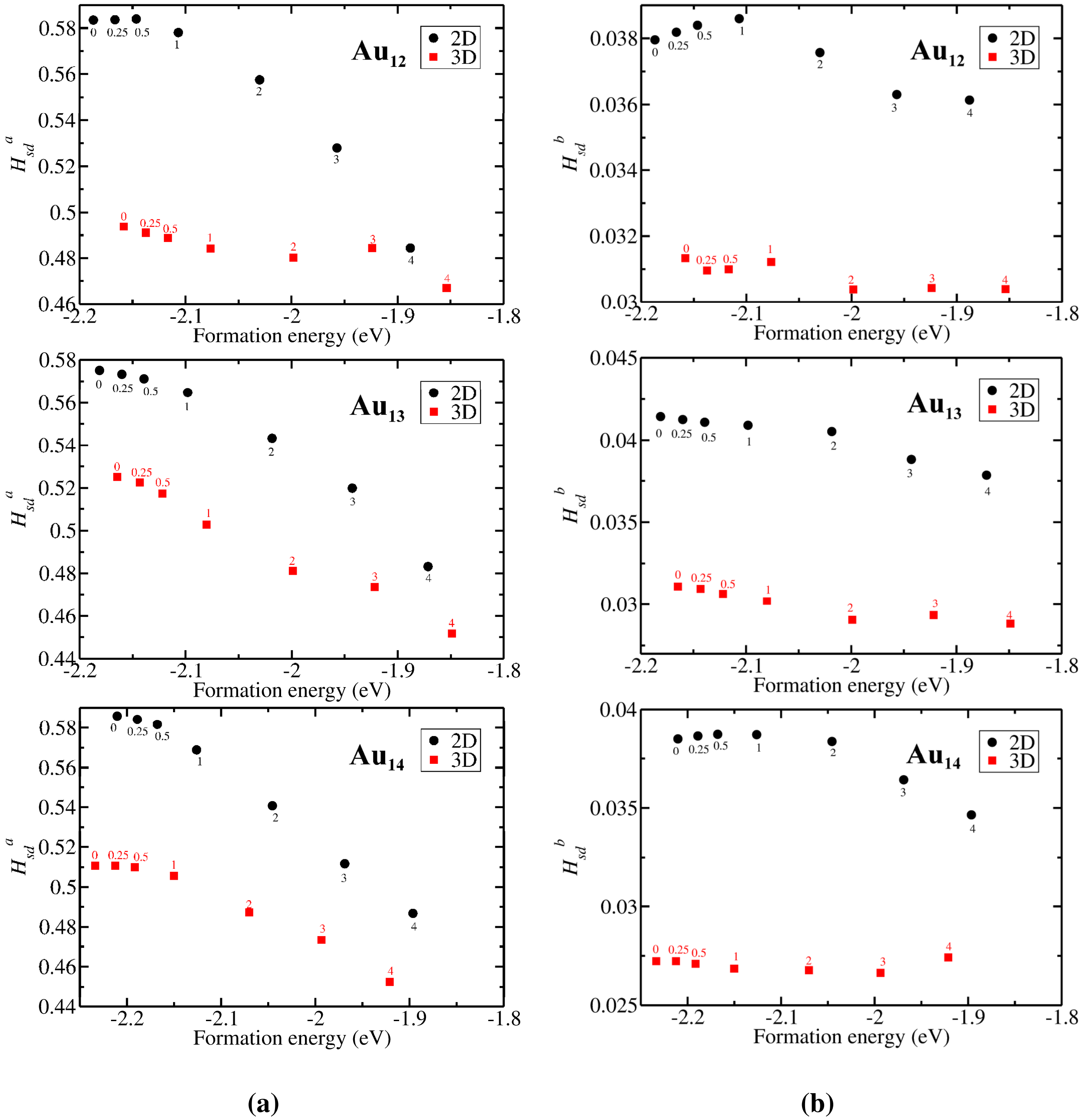

**Figure S4:** The variation of *s*-*d* band hybridization index, $H_{sd}$, with respect to energy for different values of Hubbard correction term U. (a) $H_{sd}^{a}$ is the common area under *s* and *d* shell decomposed electronic density of states while (b) $H_{sd}^{b}$ is the product of *s* and *d* weights of local charges for different sized Au clusters in 2D and 3D constructions. The U values corresponding to each data point is indicated next to it.

## Long Range Interactions

In Table S2, cohesive energies calculated from PBE+vdW are given for bulk gold. D2, D3, TS and DF vdW approximations are given with respect to different vdW interaction cutoff values from 3.8 to 21 Å. The cohesive energy does not change after ~15 Å appreciably.

**Table S2:** The cohesive energy of FCC gold from PBE calculations including the contribution of different flavors (D2, D3, TS and DF) of van der Waals interactions. The contribution of vdW is calculated for considering several interaction cutoffs. In parenthesis the percent contribution of the vdW to the cohesive energy is given.

| vdW cutoff (Å) | Cohesive energy/atom (eV) with vdW | | | | |
|---|---|---|---|---|---|
| | No vdW | D2 | D3 | TS | DF |
| 21 | | -3.684 (18.9%) | -3.631 (17.7%) | -3.410 (12.4%) | |
| 19 | | -3.682 (18.9%) | -3.630 (17.7%) | -3.409 (12.4%) | |
| 17 | | -3.680 (18.8%) | -3.629 (17.7%) | -3.408 (12.4%) | |
| 14.9 | | -3.676 (18.7%) | -3.626 (17.6%) | -3.405 (12.3%) | |
| 3.8 | | -3.088 (3.3%) | -3.125 (4.4%) | -3.052 (2.1%) | |
| Not applicable | -2.987 (0%) | | | | -3.550 (15.9%) |

## Details of DFT calculations

During the genetic algorithm optimization, we use an energy cutoff of 230 eV for DFT calculations. To test the adequacy of this cutoff energy, we randomly generate 1000 $Au_{13}$ clusters and calculated the cohesive energies using 230 and 520 eV cutoff. Between these two sets, we find that %99.4 of all binary comparisons (i.e. between any two clusters within a set) in terms of energy order is same. We also investigate the changes in the order and the relative differences in energy for $Au_{13}$ clusters around the identified minimum with respect changes in cutoff energy, see Figure S5. It is seen that the order and the relative differences in the formation energy for these low energy structures are virtually unchanged.

In cluster simulations, the size of the surrounding vacuum should be large enough to minimize the spurious periodic image interactions. The smallest cluster-cluster distances observed when the structure is constrained to 2 dimensions since these clusters have the largest diameter. We searched all our structures generated during GA optimization and the largest Au-Au separation in a cluster is found to be 21.5 Å for a 2D structure as expected. This leaves at least 8.5 Å between the atoms of the periodic image clusters since the simulation box has an xy plane with dimension 30×30 $Å^2$. In order to quantify the error introduced by the image interactions, we model the same cluster configuration in a 60×60×15 $Å^3$ cell. The energy difference between the calculations utilizing 30×30×15 $Å^3$ and 60×60×15 $Å^3$ simulation boxes is found to be 1.65 meV/atom. Note that such open structures have very high energies and around global minima, the clusters are much more compact. Thus distances between the image clusters are much larger in compact clusters. In this case the errors for the low energy structures are expected to be insignificant.

After GA optimization, clusters within $3nk_BT$ of the identified minima are revaluated by fully relaxing their atomic coordinates. For these simulations, we use 30×30×30 $Å^3$ boxes. We also increase the planewave cutoff energy to 300 eV and consider spin polarization. The ionic relaxation is continued until the energy is converged within $10^{-4}$ eV/atom.

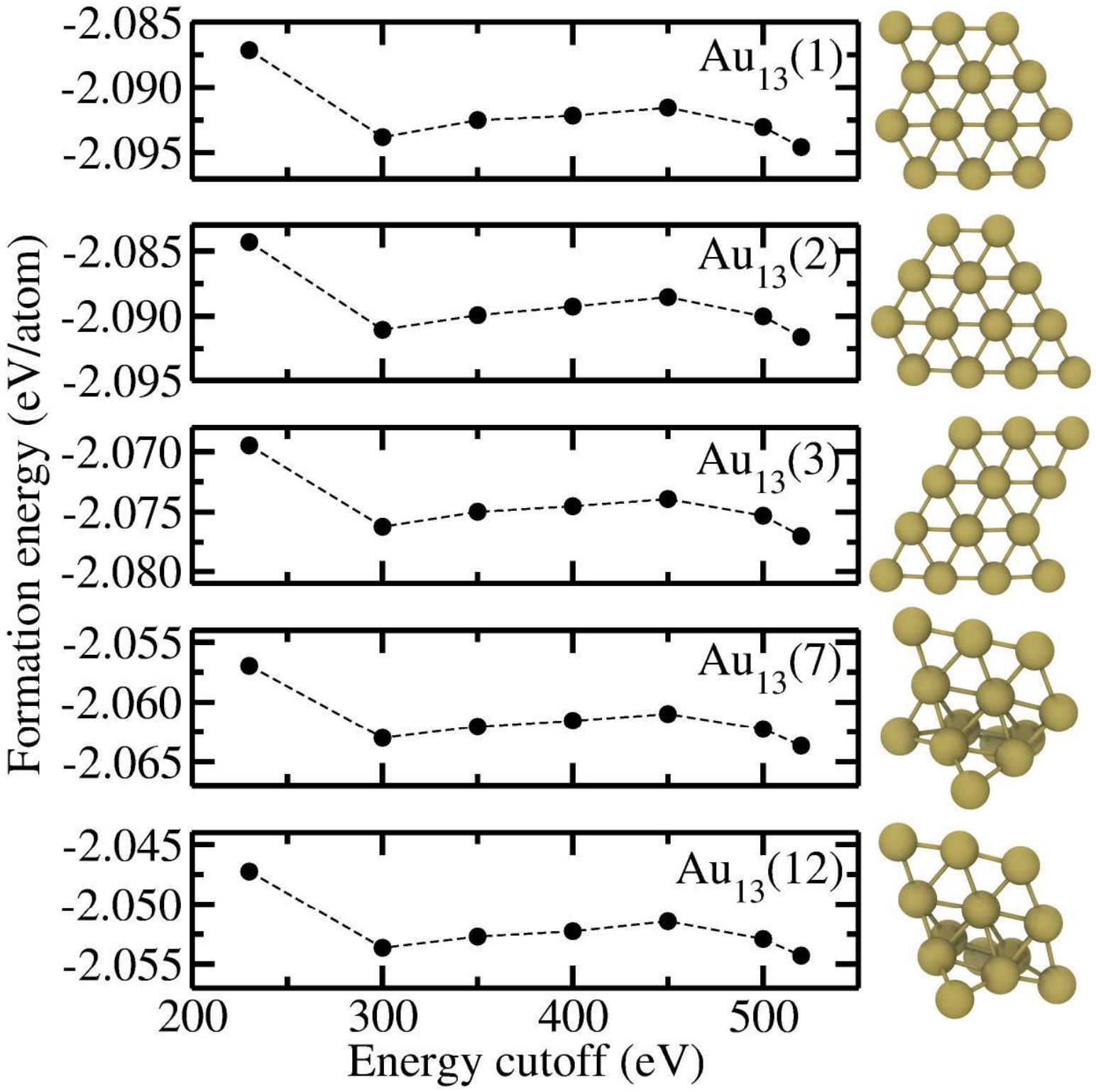

**Figure S5:** Per atom formation energies of low energy $Au_{13}$ clusters as a function of planewave kinetic energy cutoff. The reference for the formation energy calculation is the isolated Au atom. The atomic structures corresponding to these clusters are given on the right side of each plot.

12
Au12(1)

| | | | |
|---|---|---|---|
| Au | 16.497139013771381144 | 18.462017401583317167 | 15.426895931497931258 |
| Au | 18.439446682160561153 | 18.011305723626318098 | 17.256157562520520798 |
| Au | 16.926784710509416243 | 15.914937706098559289 | 16.411880156563793065 |
| Au | 18.888412666379007732 | 15.519178440530357221 | 18.253385253397699017 |
| Au | 17.415866633665874730 | 13.398185948379548904 | 17.447059650252654706 |
| Au | 15.881190169632153797 | 11.350041788473737014 | 16.570544729624426594 |
| Au | 14.527887224670438471 | 18.796844375937414640 | 13.601708607853106869 |
| Au | 13.015686709644215924 | 16.698419230667109758 | 12.755178262152746527 |
| Au | 14.981790833185124612 | 16.305473367290087339 | 14.592968536335138907 |
| Au | 13.426164119224418769 | 14.199158893318223562 | 13.704688734146067475 |
| Au | 15.429646409191839496 | 13.840668669359519782 | 15.575702433777181000 |
| Au | 13.915261556965846879 | 11.745340339735836466 | 14.731325006878773110 |

12
Au12(2)

| | | | |
|---|---|---|---|
| Au | 16.203052229406445406 | 18.572348678105495168 | 15.960034590716265512 |
| Au | 18.425967736211461556 | 18.020962978324565285 | 17.388695352244262438 |
| Au | 16.982507890174520782 | 15.967535103847415812 | 16.341500860211525037 |
| Au | 19.247440378113566339 | 15.486112829666719648 | 17.817148706885753029 |
| Au | 17.854554249699301494 | 13.416385447238027950 | 16.792522880472624536 |
| Au | 16.292769880910128677 | 11.362307226448313813 | 15.659994208770191548 |
| Au | 12.507369984469105972 | 16.820465894718584110 | 13.413865148204489941 |
| Au | 14.747674973741057514 | 16.571252543912631694 | 14.886971063267301574 |
| Au | 18.437302633210197911 | 10.799566171978963069 | 17.043724981644668759 |
| Au | 13.330061850524817402 | 14.375997627373807930 | 13.835269470110389989 |
| Au | 15.557153365074821494 | 13.937628192850755937 | 15.302417638343047912 |
| Au | 14.079430593464286048 | 11.827233153534738364 | 14.215165198130073065 |

12
Au12(3)

| | | | |
|---|---|---|---|
| Au | 19.582387777176510468 | 19.957912344511903768 | 7.500000000000000000 |
| Au | 21.171791681448134170 | 17.781465251645020231 | 7.500000000000000000 |
| Au | 22.577153668221264127 | 15.532708905919497155 | 7.500000000000000000 |
| Au | 23.963188615361211475 | 13.251047142289211322 | 7.500000000000000000 |
| Au | 21.340209790063493500 | 13.022902786532117858 | 7.500000000000000000 |
| Au | 20.022207305729985194 | 10.710233833814271520 | 7.500000000000000000 |
| Au | 18.292863054725167160 | 17.674457100784444208 | 7.500000000000000000 |
| Au | 19.855665922098388876 | 15.436763755433428713 | 7.500000000000000000 |
| Au | 17.084790515131192734 | 15.297912632638565711 | 7.500000000000000000 |
| Au | 18.590914729822102913 | 13.052626279587446589 | 7.500000000000000000 |
| Au | 15.792518854332527667 | 12.935749858925172262 | 7.500000000000000000 |
| Au | 17.304506193890272669 | 10.700122258919073204 | 7.500000000000000000 |

12
Au12(4)

| | | | |
|---|---|---|---|
| Au | 16.248370659225177093 | 15.898326411067985831 | 13.945898591973701741 |
| Au | 17.867229611664459554 | 15.402504071344630177 | 16.022275869261296322 |
| Au | 18.250250282053318784 | 15.617513616431347501 | 18.743845886423205371 |
| Au | 17.084408372281821187 | 13.357417961958816477 | 17.855115067301095877 |
| Au | 16.479889252904282415 | 13.327337643341797602 | 14.879130944836257200 |
| Au | 15.805637013417166870 | 11.320609468552401466 | 16.772186114486522968 |
| Au | 13.850544118064677335 | 16.964676710077295496 | 13.099971341429204941 |
| Au | 14.031777073829799107 | 16.324183310123991930 | 15.775059623639156925 |
| Au | 15.650275711758785846 | 15.827306292380903585 | 17.851706839616930722 |
| Au | 13.862584440484598147 | 14.344134028036874540 | 13.725014586987271059 |
| Au | 14.319888904328447410 | 13.741641858833906653 | 16.662142770540580017 |
| Au | 14.073171658986931121 | 11.849759082849887548 | 14.555882337504710833 |

12

Au12(5)

| | | | |
|---|---|---|---|
| Au | 19.450544817212708892 | 19.929815573246202121 | 7.500000000000000000 |
| Au | 20.988709856903639661 | 17.777598607710057621 | 7.500000000000000000 |
| Au | 22.486547181488106162 | 15.544918622708808087 | 7.500000000000000000 |
| Au | 19.700872679889119610 | 15.366672836345502517 | 7.500000000000000000 |
| Au | 21.315297021684418866 | 13.149427824490874883 | 7.500000000000000000 |
| Au | 20.063560091518475303 | 10.799711877395063553 | 7.500000000000000000 |
| Au | 16.622041849024999038 | 19.820939138948173053 | 7.500000000000000000 |
| Au | 18.158873536238001378 | 17.566130665622086582 | 7.500000000000000000 |
| Au | 15.407689586229400547 | 17.455319702605432752 | 7.500000000000000000 |
| Au | 16.919562988795505021 | 15.086818602374552967 | 7.500000000000000000 |
| Au | 14.225176830023807284 | 15.128234822989849562 | 7.500000000000000000 |
| Au | 18.510496252992126642 | 12.946643448563746048 | 7.500000000000000000 |

12

Au12(6)

| | | | |
|---|---|---|---|
| Au | 20.691829899018500782 | 17.973336385863117215 | 7.500000000000000000 |
| Au | 22.163464894808040384 | 15.794322494976025695 | 7.500000000000000000 |
| Au | 19.471167091892752410 | 15.481915984667875108 | 7.500000000000000000 |
| Au | 23.884846717738991373 | 13.823456277260147473 | 7.500000000000000000 |
| Au | 21.327739492278229960 | 13.099536215630322999 | 7.500000000000000000 |
| Au | 20.010552855812743900 | 10.772101420320598919 | 7.500000000000000000 |
| Au | 16.254869292595842012 | 19.683900446261048955 | 7.500000000000000000 |
| Au | 17.980431036019162150 | 17.693189667008066834 | 7.500000000000000000 |
| Au | 15.229794715939760863 | 17.254864746087726246 | 7.500000000000000000 |
| Au | 16.809439867760655574 | 14.960745617272683816 | 7.500000000000000000 |
| Au | 14.119304284987151732 | 14.905836932808053774 | 7.500000000000000000 |
| Au | 18.563007190148237413 | 12.959884361844624223 | 7.500000000000000000 |

12

Au12(7)

| | | | |
|---|---|---|---|
| Au | 20.690593429652356150 | 17.920959782926814796 | 7.500000000000000000 |
| Au | 22.329634558188903526 | 15.772048938515094818 | 7.500000000000000000 |
| Au | 19.634907791445851188 | 15.415625565372220507 | 7.500000000000000000 |
| Au | 23.973726964059572708 | 13.710642457278698458 | 7.500000000000000000 |
| Au | 21.334069538648179787 | 13.210647093160105214 | 7.500000000000000000 |
| Au | 20.312798010952111127 | 10.805306195223428389 | 7.500000000000000000 |
| Au | 19.136205484634757568 | 20.051323306003776992 | 7.500000000000000000 |
| Au | 17.957850929752034119 | 17.636979835100472513 | 7.500000000000000000 |
| Au | 15.367879447728661546 | 17.290735303547108970 | 7.500000000000000000 |
| Au | 16.891023334457219818 | 15.079934484644690329 | 7.500000000000000000 |
| Au | 18.584739102995250448 | 12.859204044017166879 | 7.500000000000000000 |
| Au | 15.951132425485109678 | 12.606867480210333099 | 7.500000000000000000 |

12
Au12(8)

| | | | |
|---|---|---|---|
| Au | 17.839006217901765439 | 15.244012317671982615 | 15.694471518393550369 |
| Au | 19.465297189513119491 | 14.806251503762691613 | 17.773677681411300000 |
| Au | 15.899779676380966009 | 15.766430102496780563 | 13.956355407952930392 |
| Au | 17.137643546481225343 | 13.571821023219539981 | 17.900622450728107538 |
| Au | 15.873629556914593763 | 13.376555658489710510 | 15.336232379200339437 |
| Au | 15.109824851211678620 | 11.822771246299046410 | 17.521260770163944898 |
| Au | 12.696286166135241302 | 16.366906722540608854 | 17.920596897789238255 |
| Au | 13.520498509365998174 | 15.901261816490141143 | 15.464646415847248306 |
| Au | 13.455024032356762476 | 16.332461506261889639 | 12.777787020669336471 |
| Au | 14.510440793817473093 | 14.468518513358134214 | 17.638700988352240984 |
| Au | 13.725383242125936079 | 13.788260883272467794 | 13.609505699061692496 |
| Au | 13.993481140794861162 | 11.460931462137669001 | 14.870118317430586430 |

12
Au12(9)

| | | | |
|---|---|---|---|
| Au | 16.083087406318096413 | 18.015281208584148942 | 15.230200302560662706 |
| Au | 18.271737150310315201 | 17.944465581095059292 | 16.833765791368449527 |
| Au | 16.520982798114747681 | 15.916747844962022995 | 16.995160635107779967 |
| Au | 19.117911645501273199 | 15.490086849135698444 | 17.768454649333051520 |
| Au | 17.506941769247806207 | 13.375068237655208137 | 17.400604135279319706 |
| Au | 16.040131062572712040 | 11.369766928910165049 | 16.313352821657783664 |
| Au | 15.322898125209441389 | 19.249600961197160132 | 13.000740626768015673 |
| Au | 14.091944479150919634 | 16.944918482390939118 | 13.330326637242951193 |
| Au | 14.035625068252565839 | 16.302071120338858634 | 15.984615320026632546 |
| Au | 13.334719476711224928 | 14.450897986554867458 | 13.996422425532074385 |
| Au | 15.029328052692358497 | 13.809634724224981284 | 16.105796296123855882 |
| Au | 14.041181740918583287 | 11.856625011950612247 | 14.470053497999352032 |

13
Au13(1)

| | | | |
|---|---|---|---|
| Au | 18.342703168298825744 | 17.122768394358743649 | 7.500000000000000000 |
| Au | 16.503885149045196101 | 19.009044635794730027 | 7.500000000000000000 |
| Au | 20.227177257832597235 | 15.235647410358749809 | 7.500000000000000000 |
| Au | 17.557112012914469545 | 14.520511706619206649 | 7.500000000000000000 |

Au 19.530140665671098077 12.605282365249252052 7.500000000000000000
Au 18.800157104357232640 10.026268928765228239 7.500000000000000000
Au 15.578785758228310243 16.502628594618975910 7.500000000000000000
Au 12.984541446089847838 15.706348762669414043 7.500000000000000000
Au 14.951197978996194848 13.831956771325833344 7.500000000000000000
Au 12.328906371517209806 13.075474997311239633 7.500000000000000000
Au 16.839676036541863624 11.913194146550763364 7.500000000000000000
Au 14.192427466576040374 11.142652577382131440 7.500000000000000000
Au 16.168033137931264065 9.280959789995787190 7.500000000000000000
13
Au13(2)
Au 14.189427529496967395 14.255496899600522909 7.500000000000000000
Au 16.571345007689039619 12.980237444348928122 7.500000000000000000
Au 18.973013559049093146 11.710846551852768016 7.500000000000000000
Au 14.240619777079896124 11.496200755376595026 7.500000000000000000
Au 16.658666388298957628 10.261959098385791123 7.500000000000000000
Au 11.898386809654970975 12.769093296574256513 7.500000000000000000
Au 16.489986641812006241 15.671213593632270289 7.500000000000000000
Au 14.072571335309547180 17.013209063378617714 7.500000000000000000
Au 9.598382424197769325 14.116893355648857167 7.500000000000000000
Au 18.958170390573442887 14.399510425359510535 7.500000000000000000
Au 11.812848871481381963 15.599960304622351615 7.500000000000000000
Au 16.413554120711939532 18.388026797426888947 7.500000000000000000
Au 18.812224644645240801 17.083754913793729457 7.500000000000000000
13
Au13(3)
Au 17.152114242006042844 16.944827556968842686 7.500000000000000000
Au 19.664915566596246066 17.727502298631069522 7.500000000000000000
Au 19.306289438806565784 15.085213556585769012 7.500000000000000000
Au 18.746143477464968896 12.498266357248754233 7.500000000000000000
Au 16.125251373988923831 11.610127852181644670 7.500000000000000000
Au 20.631866072048218541 10.656260071680861756 7.500000000000000000
Au 18.122317427423592306 9.742870012911005517 7.500000000000000000
Au 16.687529354864285125 14.271177822734701479 7.500000000000000000
Au 14.618866185260355195 16.159193008846411743 7.500000000000000000
Au 14.083886529404409771 13.490327540168237874 7.500000000000000000
Au 12.979109358420222620 8.308371467353218520 7.500000000000000000
Au 13.470333902448510699 10.902850251709894636 7.500000000000000000
Au 15.572435274267670380 8.974719023979300658 7.500000000000000000
13
Au13(4)
Au 18.536233742606075481 17.948443951908515714 15.054978839346023989
Au 16.125865774427335708 17.763111781178007931 13.756667488320101000
Au 19.740120427462777286 15.670644170185942912 15.807943685783673260

| | | | |
|---|---|---|---|
| Au | 17.371814795164645773 | 15.483784410777163387 | 14.549549939569301316 |
| Au | 21.010203246800493559 | 13.506796815152840807 | 16.587314035686546276 |
| Au | 18.673036771138601608 | 13.107786882940217765 | 15.362863195755981138 |
| Au | 17.470597309944896125 | 10.793454685011438343 | 14.834666962817726343 |
| Au | 13.810029977302818338 | 17.601237339346003097 | 12.495791180041626589 |
| Au | 11.480362729770781982 | 17.259854372223415453 | 11.224939139168384372 |
| Au | 14.995347632970389640 | 15.261644542856970475 | 13.278044833602152153 |
| Au | 12.620463505875816779 | 15.023612168549165702 | 11.985909568368342093 |
| Au | 13.842257893168810057 | 12.798543099939701406 | 12.775691684297463269 |
| Au | 16.233541136367051649 | 12.994565884930441513 | 14.060759502242309082 |

13
Au13(5)

| | | | |
|---|---|---|---|
| Au | 16.338087014420725751 | 18.017606828409473252 | 7.500000000000000000 |
| Au | 18.311456040599900064 | 16.170505335724442375 | 7.500000000000000000 |
| Au | 15.645230715612010286 | 15.422787110300214763 | 7.500000000000000000 |
| Au | 17.626100321925292747 | 13.559630034409355659 | 7.500000000000000000 |
| Au | 19.467985707862744960 | 11.625433498369121210 | 7.500000000000000000 |
| Au | 16.944522678926745130 | 10.871102004156741216 | 7.500000000000000000 |
| Au | 16.154754790184821189 | 8.336901146139302554 | 7.500000000000000000 |
| Au | 13.706730615111549554 | 17.402144484480363218 | 7.500000000000000000 |
| Au | 12.982909405514144296 | 14.824914149364287752 | 7.500000000000000000 |
| Au | 10.472975431079847297 | 13.958473709520992756 | 7.500000000000000000 |
| Au | 14.965169049417557545 | 12.821460566947179061 | 7.500000000000000000 |
| Au | 12.352589765947111999 | 12.090516583644077997 | 7.500000000000000000 |
| Au | 14.245329848397638273 | 10.196840425534263730 | 7.500000000000000000 |

13
Au13(6)

| | | | |
|---|---|---|---|
| Au | 18.221831437057524283 | 19.115045207833503582 | 15.045574647651152489 |
| Au | 19.061879214684708472 | 16.702094370452261529 | 15.588463086952945247 |
| Au | 17.564491060159362235 | 14.593758348721713958 | 14.822793642488109711 |
| Au | 19.992594038132963874 | 14.240636969656508271 | 16.178091374750710685 |
| Au | 18.533802297008151072 | 12.136531321091212021 | 15.430869176939495446 |
| Au | 17.169988033856657950 | 10.011364270306373214 | 14.744486454440105305 |
| Au | 14.290761794362571990 | 17.396677683277768978 | 12.935372662743326089 |
| Au | 11.956906208930819346 | 17.554347327214554042 | 11.632634651924025349 |
| Au | 16.605897358933319197 | 17.137294347555481977 | 14.216788089224207781 |
| Au | 12.829202830040120631 | 15.118367572248471831 | 12.183551605289583364 |
| Au | 15.216950432592003750 | 14.882012276535183659 | 13.512271280494250192 |
| Au | 13.728461103228804419 | 12.687689922281483845 | 12.746019233471930221 |
| Au | 16.072141752012694127 | 12.353681445825257512 | 14.056703086630088606 |

13
Au13(7)

| | | | |
|---|---|---|---|
| Au | 19.604420313582782143 | 14.777604312653060958 | 14.662681227048167898 |
| Au | 17.504085115277774776 | 15.517517422098377367 | 16.456577223743103389 |
| Au | 17.174859164699174841 | 13.560196687596006271 | 14.528619895129356365 |
| Au | 15.956116602070773425 | 13.345617013827565600 | 17.069490519878950607 |
| Au | 15.048056861411403418 | 15.260260901030317626 | 15.271185420194761662 |
| Au | 13.115253177957930220 | 15.483423814527336404 | 13.472921712024827912 |
| Au | 19.164216146653615169 | 12.840751288873379465 | 12.844173515060264279 |
| Au | 18.724422287799395548 | 10.930022576478624075 | 11.060200566375048226 |
| Au | 16.684227418406614163 | 11.523456286506860380 | 12.643468348789170363 |
| Au | 15.402278890667538747 | 11.316428341075885911 | 15.192915877625322807 |
| Au | 15.310669319198726868 | 10.884343133382371249 | 17.804123933547678860 |
| Au | 14.502113213620535959 | 13.227705693516517371 | 13.391801793671881171 |
| Au | 14.177480804653551871 | 10.555117667433176365 | 12.853922279911031978 |

13

Au13(8)

| | | | |
|---|---|---|---|
| Au | 14.970495427001198507 | 14.243791310666301442 | 7.500000000000000000 |
| Au | 12.448966319707260197 | 15.509653655513677251 | 7.500000000000000000 |
| Au | 15.157175462208126504 | 11.510014430544833530 | 7.500000000000000000 |
| Au | 12.723397517641544141 | 12.774429044713860648 | 7.500000000000000000 |
| Au | 12.438214036308746557 | 18.140161680717884707 | 7.500000000000000000 |
| Au | 17.401689228504718443 | 12.947549122277663969 | 7.500000000000000000 |
| Au | 10.495423931469611389 | 11.252961820663280434 | 7.500000000000000000 |
| Au | 14.847386765454341884 | 16.973435375963596528 | 7.500000000000000000 |
| Au | 12.825807231689690724 | 9.976080579079976474 | 7.500000000000000000 |
| Au | 19.547778392962158023 | 14.449968863512490103 | 7.500000000000000000 |
| Au | 10.254289829288341096 | 13.994941434683443759 | 7.500000000000000000 |
| Au | 15.228158872668926094 | 8.856724182731895212 | 7.500000000000000000 |
| Au | 17.202659185094798744 | 15.780439898931451381 | 7.500000000000000000 |

13

Au13(9)

| | | | |
|---|---|---|---|
| Au | 18.305585357952629977 | 19.151417531853894616 | 7.500000000000000000 |
| Au | 21.887832543243327166 | 15.949913071052112201 | 7.500000000000000000 |
| Au | 19.336418242808587564 | 16.714323822173589917 | 7.500000000000000000 |
| Au | 17.432775366193542510 | 14.591522775821124114 | 7.500000000000000000 |
| Au | 20.069529435485200963 | 14.016196318928734854 | 7.500000000000000000 |
| Au | 18.268959355126607846 | 11.988037055972583289 | 7.500000000000000000 |
| Au | 16.562780249729598836 | 17.159829030404630146 | 7.500000000000000000 |
| Au | 14.746020246264206932 | 15.210564225647877024 | 7.500000000000000000 |
| Au | 15.568065976067259726 | 12.605630764144706291 | 7.500000000000000000 |
| Au | 12.871232973908627528 | 13.196193673576134486 | 7.500000000000000000 |
| Au | 13.667553743198595839 | 10.656521839065460355 | 7.500000000000000000 |
| Au | 16.447083393457660350 | 10.017387182373491683 | 7.500000000000000000 |
| Au | 14.508591081564826908 | 8.164807098985496836 | 7.500000000000000000 |

13

Au13(10)

| | | | |
|---|---|---|---|
| Au | 15.221901769973731788 | 15.292807525740311547 | 7.500000000000000000 |
| Au | 17.720114363221263432 | 16.795680618296607634 | 7.500000000000000000 |
| Au | 19.990699272537447939 | 18.194251010853385253 | 7.500000000000000000 |
| Au | 22.442321342939060003 | 16.894520984790581508 | 7.500000000000000000 |
| Au | 17.567475345765096506 | 19.569360531485081367 | 7.500000000000000000 |
| Au | 17.722334577102294872 | 14.112691838820012435 | 7.500000000000000000 |
| Au | 19.904493185367481800 | 20.916736665584139132 | 7.500000000000000000 |
| Au | 22.538733259946518217 | 14.284479564727002199 | 7.500000000000000000 |
| Au | 20.078789609823761708 | 15.371858422600841010 | 7.500000000000000000 |
| Au | 15.355712703534306840 | 18.113812256498043496 | 7.500000000000000000 |
| Au | 22.315850254690346333 | 19.557249398244334060 | 7.500000000000000000 |
| Au | 13.054020891465810905 | 16.827483089961155827 | 7.500000000000000000 |
| Au | 15.557868023633217192 | 12.629881592398472634 | 7.500000000000000000 |

13

Au13(11)

| | | | |
|---|---|---|---|
| Au | 15.805678307950879358 | 19.808379654001971204 | 7.500000000000000000 |
| Au | 17.165130966070933027 | 17.498191675280665436 | 7.500000000000000000 |
| Au | 18.505489934281847297 | 15.205487594610305635 | 7.500000000000000000 |
| Au | 15.819720175131559259 | 15.119421969336904965 | 7.500000000000000000 |
| Au | 19.734525529316332637 | 12.835137309436868591 | 7.500000000000000000 |
| Au | 17.087639281901001453 | 12.698573385998455976 | 7.500000000000000000 |
| Au | 14.397582576123330256 | 17.586950038371263361 | 7.500000000000000000 |
| Au | 11.734035345405327888 | 17.626155483802563140 | 7.500000000000000000 |
| Au | 10.390229717689130950 | 15.326938216437222096 | 7.500000000000000000 |
| Au | 13.088461429499179900 | 15.194187893974044457 | 7.500000000000000000 |
| Au | 9.031267995936188342 | 13.081558600481562493 | 7.500000000000000000 |
| Au | 11.719232193824565158 | 12.817702432519984868 | 7.500000000000000000 |
| Au | 14.396678554869492217 | 12.790177928747946012 | 7.500000000000000000 |

13

Au13(12)

| | | | |
|---|---|---|---|
| Au | 18.675097125987463187 | 14.532395117988368582 | 15.142800931117722740 |
| Au | 17.375281343896908481 | 16.020092181409122389 | 17.051874180568749040 |
| Au | 16.979760576459653976 | 12.847362471593907429 | 13.802526089169008472 |
| Au | 16.426226211878763195 | 13.509336426015671861 | 16.697249331696454533 |
| Au | 15.956841233228345800 | 15.517665065803406321 | 14.798968961530647448 |
| Au | 14.354648704867592812 | 13.711910206294227521 | 13.329244346729240078 |
| Au | 14.034643177789000390 | 16.338289821104435617 | 13.187789226592856906 |
| Au | 19.581751270034246204 | 12.903976839755795680 | 13.165380155061935596 |
| Au | 17.970570629934293549 | 11.225668787060055109 | 11.789812222716701484 |
| Au | 16.136583657511557988 | 9.679191833074838769 | 10.618101709362200324 |
| Au | 14.815921686383388689 | 11.737114454499558036 | 15.194565757146897056 |
| Au | 15.168044370147766742 | 11.472132763478398942 | 17.807955860746641719 |
| Au | 15.151507927880574655 | 11.217029520922128683 | 12.525809668560791366 |

13

Au13(13)

| | | | |
|---|---|---|---|
| Au | 17.875133894191947093 | 15.428979098627360500 | 16.550806176972102435 |
| Au | 20.041976162209987677 | 14.395617959029067379 | 15.309489990085092259 |
| Au | 15.828141078835296085 | 13.567619509568936920 | 16.877649012170984122 |
| Au | 17.505410286963059008 | 13.516569414644921565 | 14.616887907051351547 |
| Au | 19.517552656412860301 | 12.342661257374571093 | 13.287944272885452435 |
| Au | 17.004018524946154400 | 11.594671144223987369 | 12.778914904258840579 |
| Au | 21.960926104610884124 | 13.192549672882856626 | 13.942312548335534572 |
| Au | 14.193415293878231864 | 16.112200145744889568 | 12.784669931522046227 |
| Au | 15.577966221576598826 | 15.525128653920809541 | 14.972858729947301271 |
| Au | 14.896679056242362549 | 13.579035802780115461 | 13.047708094392115541 |
| Au | 15.134818504915223514 | 11.590118262644624636 | 14.995494987062237158 |
| Au | 14.812811679654082297 | 11.222242453401722884 | 17.591583970343929622 |
| Au | 14.340739623563035110 | 10.990557689155682652 | 12.474947527973487382 |

13

Au13(14)

| | | | |
|---|---|---|---|
| Au | 15.911597371767502551 | 19.926366497680895407 | 7.500000000000000000 |
| Au | 17.117341911095781626 | 17.594493645586499753 | 7.500000000000000000 |
| Au | 18.439205818374652779 | 15.263696683151902533 | 7.500000000000000000 |
| Au | 15.722872366251936072 | 15.256677409666135503 | 7.500000000000000000 |
| Au | 14.416399979260294373 | 12.906462147778036353 | 7.500000000000000000 |
| Au | 17.170609588690417979 | 12.868576157776395519 | 7.500000000000000000 |
| Au | 15.846703152761449473 | 10.529878184396954666 | 7.500000000000000000 |
| Au | 14.315462819776639591 | 17.779577642676951399 | 7.500000000000000000 |
| Au | 11.703500197035928920 | 17.616139111752669777 | 7.500000000000000000 |
| Au | 12.976255608628900262 | 15.283352326271282706 | 7.500000000000000000 |
| Au | 11.755378428366459076 | 12.872748165766283890 | 7.500000000000000000 |
| Au | 10.523221879552824731 | 10.533735201439164086 | 7.500000000000000000 |
| Au | 13.192485140437982594 | 10.422914351056670412 | 7.500000000000000000 |

13

Au13(15)

| | | | |
|---|---|---|---|
| Au | 16.447820496989418615 | 15.607242591571424128 | 14.342860426951821395 |
| Au | 17.648701419335875329 | 15.278628199568732526 | 16.770594083511287664 |
| Au | 19.176525118588834573 | 15.163958095510141177 | 14.460897983034914915 |
| Au | 16.146031843868851041 | 12.856469945756979101 | 17.349772328104776875 |
| Au | 17.550368145485204963 | 13.010874304272505242 | 15.041948172047979071 |
| Au | 15.244308814037738742 | 11.559925807412909649 | 15.198777494162152735 |
| Au | 16.943293673009822697 | 14.581848435106287809 | 19.225644989517284955 |
| Au | 14.340640738008834987 | 16.097566712995199367 | 12.831352866951855063 |
| Au | 14.893418589662015705 | 13.554064669738245641 | 13.229516572168456889 |
| Au | 19.187839391914803144 | 12.981804703520756661 | 12.863659496566500096 |
| Au | 16.758459733172649919 | 11.502407531478290537 | 12.699054300730868050 |
| Au | 14.115655391476181535 | 10.992463861233552791 | 12.744315690510566341 |
| Au | 18.895139125449489370 | 10.838757679834811753 | 11.304895528741282007 |

13

Au13(16)

| | | | |
|---|---|---|---|
| Au | 15.943862349808640033 | 19.780119204863527926 | 7.500000000000000000 |
| Au | 17.205839304238519816 | 17.499059428369879043 | 7.500000000000000000 |
| Au | 18.615744175650878134 | 15.250608369971082823 | 7.500000000000000000 |
| Au | 15.817577717612209653 | 15.167099975101955067 | 7.500000000000000000 |
| Au | 17.298865160666029084 | 12.921410311596817522 | 7.500000000000000000 |
| Au | 15.975745180000494372 | 10.570744437238870361 | 7.500000000000000000 |
| Au | 14.389758574714017314 | 17.584281340330068844 | 7.500000000000000000 |
| Au | 11.689167455810999030 | 17.515061165843626156 | 7.500000000000000000 |
| Au | 10.406419519121227424 | 15.142222505622338957 | 7.500000000000000000 |
| Au | 13.127065255622811790 | 15.168258037450108588 | 7.500000000000000000 |
| Au | 18.679670373841819497 | 10.653485736936566752 | 7.500000000000000000 |
| Au | 11.729329024357252109 | 12.806069405673831341 | 7.500000000000000000 |
| Au | 14.430074960554964747 | 12.749123267001081672 | 7.500000000000000000 |

13

Au13(17)

| | | | |
|---|---|---|---|
| Au | 19.939763260163410052 | 14.556492802044317614 | 14.913447980195561726 |
| Au | 17.801473578542690746 | 15.468208329465364059 | 16.427010369346042751 |
| Au | 15.924944021232816382 | 13.548853101604461102 | 16.902403832553151375 |
| Au | 15.415379044143978859 | 11.641170383190431181 | 14.876847512440557253 |
| Au | 15.425113796864394544 | 15.576499287226456048 | 15.093139344624406206 |
| Au | 14.810704610945061077 | 11.206601080294996464 | 17.425988873809814095 |
| Au | 19.578039748941638010 | 12.731616891087758958 | 12.956435393289607916 |
| Au | 18.998412827342722409 | 10.934974206470521096 | 11.097313103163719106 |
| Au | 17.439070081769180831 | 13.591117037930970923 | 14.419294362020398381 |
| Au | 16.787472865552199863 | 11.726365947544524104 | 12.505468881157101535 |
| Au | 16.591674009508246002 | 9.932162090059028614 | 10.535962760011022610 |
| Au | 14.927460168789989936 | 13.653196391901392559 | 13.077615419466898317 |
| Au | 13.616957617203672370 | 15.960904971179141754 | 13.199274477921552062 |

13

Au13(18)

| | | | |
|---|---|---|---|
| Au | 19.234572390134825781 | 17.629799080069993522 | 12.980229741585260328 |
| Au | 19.210704344748435801 | 15.647897747826640114 | 14.674191182249495924 |
| Au | 16.804858402809571771 | 14.766586066304915548 | 15.783378435389355587 |
| Au | 19.311760818911778870 | 14.396812791632321193 | 17.045172721717349873 |
| Au | 18.594530026119542754 | 12.762655185496967647 | 15.057986945912976040 |
| Au | 17.790074271080339230 | 10.585855468987256600 | 13.846049837768349633 |
| Au | 14.343117959427367225 | 17.318780929292849180 | 12.752661388669826792 |
| Au | 12.005013772104087266 | 17.470073830093152623 | 11.435781750653228883 |
| Au | 16.796857550284489236 | 16.779843379604756137 | 13.702031711079365905 |
| Au | 14.801842395016844733 | 14.943144682978482862 | 14.000194720917226832 |
| Au | 12.648434353321826862 | 15.071341074437089702 | 12.334414582592126308 |
| Au | 13.361688497595471503 | 12.698273414196837194 | 13.315705619607246746 |
| Au | 15.947017753445889454 | 12.493272709078572902 | 14.195551898859196882 |

13
Au13(19)

| | | | |
|---|---|---|---|
| Au | 18.997824219957575309 | 20.588859477494427352 | 7.500000000000000000 |
| Au | 20.836256131472442377 | 18.587513656210983726 | 7.500000000000000000 |
| Au | 18.173232040811793553 | 18.008922996782096959 | 7.500000000000000000 |
| Au | 20.067487650774577190 | 16.029870005917477016 | 7.500000000000000000 |
| Au | 17.307615295973800329 | 15.421736382956192202 | 7.500000000000000000 |
| Au | 21.710624244438438524 | 13.962361461363483883 | 7.500000000000000000 |
| Au | 19.114196654006207154 | 13.393630880033411401 | 7.500000000000000000 |
| Au | 16.329851153139298248 | 20.067450064025781131 | 7.500000000000000000 |
| Au | 15.498689557883407630 | 17.517154160933479545 | 7.500000000000000000 |
| Au | 14.618798851985415510 | 14.995745702320222748 | 7.500000000000000000 |
| Au | 16.514961248307070463 | 12.841926497946284158 | 7.500000000000000000 |
| Au | 13.698672197522935079 | 12.521495838141648349 | 7.500000000000000000 |
| Au | 15.348465573726782551 | 10.447387610874264041 | 7.500000000000000000 |

13
Au13(20)

| | | | |
|---|---|---|---|
| Au | 19.963875200365123419 | 14.555732870067835449 | 14.905961272112767446 |
| Au | 17.768580892771094426 | 15.578332482944016846 | 16.303087224729413407 |
| Au | 17.425385632543083148 | 13.558402466566095157 | 14.550280679384712812 |
| Au | 15.819027559570940866 | 13.687652095888230619 | 16.861966363892040732 |
| Au | 15.355084532979304868 | 15.512717791027055725 | 14.809497526887467345 |
| Au | 15.346842070316261797 | 11.471540308636877015 | 15.204528090622632774 |
| Au | 14.696018973486326686 | 11.471037403446651126 | 17.779412419274194690 |
| Au | 19.551385072890443695 | 12.566210074342146541 | 13.120716497216983853 |
| Au | 18.983008506244381408 | 10.651103734466229156 | 11.282832307121726245 |
| Au | 17.034307912945095609 | 11.417846029633839322 | 12.984843207011330435 |
| Au | 16.437018415354685885 | 9.578917835079996834 | 11.118817584793166731 |
| Au | 14.818368442260096529 | 13.389039873874800435 | 13.265520347207631957 |
| Au | 14.428336223273376504 | 10.699670933026265374 | 12.653333626745775220 |

13
Au13(21)

| | | | |
|---|---|---|---|
| Au | 20.063077032580277859 | 14.660300533104425824 | 14.616116793408131969 |
| Au | 17.989625547177858778 | 15.932338772480186151 | 15.858001509107172211 |
| Au | 17.506702189994470586 | 13.673890771977617931 | 14.400910294828930702 |
| Au | 15.450830365855829029 | 15.126838770374542165 | 15.627021918343880102 |
| Au | 13.090888916454550994 | 14.686655429288958175 | 14.364394172227544999 |
| Au | 12.896131611685536456 | 14.063278141081209327 | 11.774552102932666742 |
| Au | 19.713545626831251667 | 12.501362363347496398 | 13.120567254508539889 |
| Au | 19.268377411737617422 | 10.388035982469871144 | 11.604966585377448851 |
| Au | 17.125610515304295234 | 11.445245553572252817 | 12.799969621795764141 |
| Au | 15.717468591131009248 | 11.612282988659480054 | 15.162070098051572131 |
| Au | 15.062893372046849194 | 13.231824957035648538 | 13.071468918538121429 |
| Au | 14.330415616029418047 | 10.641482703972398483 | 17.202245786832655483 |

Au 13.806456369171071330 13.151540011635987071 16.497294082047140762
14
Au14(1)
Au 14.041857300670661957 13.578520956689100530 15.064948341065232285
Au 13.859150199130034764 16.291514116327796557 14.454377691188092570
Au 11.629384851273641743 14.660423311992843765 14.049778130608112647
Au 12.642583235783135365 12.048603475470164526 13.314126428797971258
Au 13.977060481685439797 14.431368960349463038 12.368346282077517628
Au 12.512874944470770089 12.965543598583831653 10.557560238950621567
Au 12.782702470538620076 15.635295185343112578 10.279824347842390964
Au 12.454044608005688133 17.365830329980237678 12.433083460302787771
Au 11.213237216754798453 10.630818838767687851 11.501319316764803347
Au 10.133975093352177055 15.852112625402000390 12.152602896219516282
Au 10.183202332921442945 13.160346693882081226 12.252368704493065366
Au 10.843255806901902716 17.606379327287669412 10.131065760128519671
Au 10.193019549541292079 14.934181186113747231 9.523970592202061880
Au 9.979246968970592491 12.258220997809923247 9.609117237358919539
14
Au14(2)
Au 13.413524432988383595 16.276895761195369516 13.533779056704799970
Au 14.133829551067295327 14.046957533847107769 11.994520401600087922
Au 12.747172397771805663 15.864930423272959104 10.533004263667615774
Au 11.744704566585790317 14.073202168505865473 14.020092673889699597
Au 10.571959607590404318 15.921103861328402829 12.407804355240706684
Au 10.458173600435907602 17.454918392855830689 10.029059748584051803
Au 10.116326810768768496 14.804261181756070087 9.684353072439938259
Au 14.421586669752443299 13.979477335277916694 14.756163460953576561
Au 13.226769879581789624 11.818280906297266242 13.380856293866475681
Au 12.404485175394261987 13.172632558200758623 10.148830093484672688
Au 9.899082302947329381 12.070440980599572711 9.523048714638695955
Au 11.530012596829548599 10.908918884309880326 11.494961618255450375
Au 10.124211826793258595 13.243074164806049708 12.060438188814357474
Au 12.217963913494022776 18.175302534747743266 12.054702904858720203
14
Au14(3)
Au 15.735740924173862609 20.038934216788543807 7.500000000000000000
Au 18.342160605652260585 20.682320018668431061 7.500000000000000000
Au 20.156372653367906622 18.677559586216869292 7.500000000000000000
Au 17.552884598896206114 18.000421159977857855 7.500000000000000000
Au 19.497264677716053427 16.052661261396803383 7.500000000000000000
Au 16.785250924660235938 15.405696746612386860 7.500000000000000000
Au 21.228889660665892336 14.026545681447894154 7.500000000000000000
Au 18.659097528052924986 13.402785974641599154 7.500000000000000000
Au 16.044328386134573350 12.785418426885781784 7.500000000000000000

| | | | |
|---|---|---|---|
| Au | 14.104361535448965981 | 14.712225650098091023 | 7.500000000000000000 |
| Au | 13.140844074641329087 | 19.412008600653365420 | 7.500000000000000000 |
| Au | 10.606564640018993018 | 18.710212539024556833 | 7.500000000000000000 |
| Au | 14.920846239866358829 | 17.366364201882639406 | 7.500000000000000000 |
| Au | 12.296410428704644957 | 16.637975847704439047 | 7.500000000000000000 |

14

Au14(4)

| | | | |
|---|---|---|---|
| Au | 17.539071181195446059 | 15.326655419472208663 | 7.500000000000000000 |
| Au | 14.877106656389891270 | 14.713258710379932381 | 7.500000000000000000 |
| Au | 19.309027320065979438 | 13.352386390795121685 | 7.500000000000000000 |
| Au | 16.784700209577430030 | 12.601992524884842339 | 7.500000000000000000 |
| Au | 14.127770879567936646 | 12.086055446220687060 | 7.500000000000000000 |
| Au | 15.810450967005941436 | 17.319144753692260252 | 7.500000000000000000 |
| Au | 15.956280491007206024 | 10.013656266914868098 | 7.500000000000000000 |
| Au | 13.373541405264884219 | 9.487961750963739505 | 7.500000000000000000 |
| Au | 13.105174431929095036 | 16.768385935770005801 | 7.500000000000000000 |
| Au | 10.534201197217290868 | 16.095927827530069010 | 7.500000000000000000 |
| Au | 12.272242468522714276 | 14.033176239709479916 | 7.500000000000000000 |
| Au | 9.592863783754628670 | 13.540127846610380047 | 7.500000000000000000 |
| Au | 11.327347323091625952 | 11.459216698249349520 | 7.500000000000000000 |
| Au | 10.846188685409979158 | 8.825354784807139197 | 7.500000000000000000 |

14

Au14(5)

| | | | |
|---|---|---|---|
| Au | 17.528977162577273674 | 15.466997940640766274 | 7.500000000000000000 |
| Au | 14.843613894450392365 | 14.746142374028448074 | 7.500000000000000000 |
| Au | 19.348188368124684899 | 13.421093244288927693 | 7.500000000000000000 |
| Au | 16.714720487850776465 | 12.799014090483744610 | 7.500000000000000000 |
| Au | 14.085550395949923086 | 12.182159387259000383 | 7.500000000000000000 |
| Au | 18.599220385397362776 | 10.846047586986896860 | 7.500000000000000000 |
| Au | 15.958747897664547821 | 10.121452676919812319 | 7.500000000000000000 |
| Au | 12.968929636068059708 | 16.805151573320937786 | 7.500000000000000000 |
| Au | 10.328240993945895454 | 16.079910934252183097 | 7.500000000000000000 |
| Au | 12.213772319586556847 | 14.128220396457930264 | 7.500000000000000000 |
| Au | 9.579788737740823734 | 13.504975212455418543 | 7.500000000000000000 |
| Au | 11.399875193232402282 | 11.460977906103954282 | 7.500000000000000000 |
| Au | 13.325761851931128277 | 9.582001146202902220 | 7.500000000000000000 |
| Au | 15.602039398480231469 | 17.346439699598342088 | 7.500000000000000000 |

14

Au14(6)

| | | | |
|---|---|---|---|
| Au | 12.264629743783995508 | 18.282699564418983584 | 10.854669562597234034 |
| Au | 12.805766967422124836 | 17.185726645779400457 | 13.244437083575924419 |
| Au | 12.551770157437832864 | 15.843755747445264603 | 15.597829400567590596 |
| Au | 13.157746201786199691 | 15.706513084871934538 | 10.886551353940582487 |
| Au | 13.190255482700155909 | 14.374349040136490174 | 13.394639629649605084 |
| Au | 12.420237684237779519 | 13.097430294888011915 | 11.059804343173063046 |
| Au | 12.800330480988350246 | 11.710122208353894280 | 13.429622466599516528 |
| Au | 10.701324230563585616 | 17.392605923220525455 | 8.737955933109704532 |
| Au | 10.205131351817518848 | 16.414684564813494916 | 11.254480174073252741 |
| Au | 10.561102158767544879 | 15.407403459442237548 | 13.735685108279566791 |
| Au | 11.159725244658138976 | 14.745749883336571884 | 9.140850047187308647 |
| Au | 9.696945255794178209 | 13.795380420282276290 | 11.753037750811424544 |
| Au | 10.258797501305597422 | 12.205416068811851815 | 9.477464994847897728 |
| Au | 10.594779258737093031 | 11.132515669198541630 | 11.955817837586749874 |

14

Au14(7)

| | | | |
|---|---|---|---|
| Au | 15.812090123287923404 | 10.008547527043422321 | 7.500000000000000000 |
| Au | 18.399993392173257689 | 17.741208186439720862 | 7.500000000000000000 |
| Au | 20.121580049627439735 | 15.618949959751706302 | 7.500000000000000000 |
| Au | 17.501392245268547754 | 15.133741625290991095 | 7.500000000000000000 |
| Au | 19.376429462582340335 | 13.026852765437162418 | 7.500000000000000000 |
| Au | 16.643583896433252534 | 12.556467548438204673 | 7.500000000000000000 |
| Au | 18.505126772716923966 | 10.521363446196099645 | 7.500000000000000000 |
| Au | 15.744879075752972852 | 17.264884784823543384 | 7.500000000000000000 |
| Au | 13.158302511946001090 | 16.693448569052677044 | 7.500000000000000000 |
| Au | 10.568693092316291171 | 16.282065184678895520 | 7.500000000000000000 |
| Au | 14.873133836047133016 | 14.616903538740256607 | 7.500000000000000000 |
| Au | 12.067162062800207423 | 14.066604780844631506 | 7.500000000000000000 |
| Au | 13.960939382717981516 | 12.028317961247298129 | 7.500000000000000000 |
| Au | 11.381220120329709644 | 11.492618358015080915 | 7.500000000000000000 |

14

Au14(8)

| | | | |
|---|---|---|---|
| Au | 17.575969325158212087 | 15.298983695069690469 | 7.500000000000000000 |
| Au | 14.910200628871553619 | 14.774295827172831963 | 7.500000000000000000 |
| Au | 19.296571581480488788 | 13.199175393166818537 | 7.500000000000000000 |
| Au | 16.642152371652176868 | 12.710978203489016636 | 7.500000000000000000 |
| Au | 18.378519473515996907 | 10.621235074933277787 | 7.500000000000000000 |
| Au | 15.743700474086136154 | 10.057244729931476002 | 7.500000000000000000 |
| Au | 13.159579058826521347 | 16.873924657135948024 | 7.500000000000000000 |
| Au | 10.567999223552021704 | 16.295730247119028178 | 7.500000000000000000 |
| Au | 12.288742788258861793 | 14.207682006028401034 | 7.500000000000000000 |
| Au | 9.668981007359835900 | 13.538868027191433896 | 7.500000000000000000 |
| Au | 13.995892449322031581 | 12.103148802248492544 | 7.500000000000000000 |
| Au | 11.384853939149735425 | 11.506350953397021186 | 7.500000000000000000 |
| Au | 13.114678810260025799 | 9.421729356947304623 | 7.500000000000000000 |
| Au | 8.020674277506529393 | 15.627220198168918941 | 7.500000000000000000 |

14

Au14(9)

| | | | |
|---|---|---|---|
| Au | 12.260056811837328539 | 18.512233425484325267 | 11.341905370353265070 |
| Au | 12.267293573510423599 | 16.574045262598836814 | 13.228805836020756104 |
| Au | 13.619805679901888951 | 15.830033378261882504 | 15.393405170024035300 |
| Au | 12.961663312531326753 | 15.985286674975217380 | 10.491055941798551387 |
| Au | 12.669975142978184834 | 13.876080922124391037 | 13.861187617916483461 |
| Au | 13.354666829917556825 | 13.394303108547299885 | 11.097676644906179888 |
| Au | 13.329185151123168040 | 11.398931464458303964 | 12.997932289373142822 |
| Au | 10.289650618986881270 | 16.611330188718703482 | 11.229660614776150851 |
| Au | 9.126590448329448435 | 15.787219512842620261 | 8.907889542471519206 |
| Au | 10.036757818512839435 | 14.854449310564575626 | 13.512482233738554172 |
| Au | 11.323862662304255977 | 14.090378561482975073 | 9.422072278125488864 |
| Au | 9.052680802130559812 | 14.142127624573650024 | 11.079869241238249344 |
| Au | 9.715815761653107785 | 11.807774365562991647 | 9.838671544909903588 |
| Au | 10.918466920283256272 | 12.433346273803909554 | 12.201349875348018514 |

14

Au14(10)

| | | | |
|---|---|---|---|
| Au | 11.034404059358934092 | 16.045305928914601168 | 12.700933826453585596 |
| Au | 14.323706133372702709 | 15.332209671115473171 | 14.179226810462141728 |
| Au | 11.891251258407681490 | 17.044932551584619063 | 10.272150096044732948 |
| Au | 11.726669858450792461 | 14.567925551446725407 | 14.844434705873648994 |
| Au | 13.292781155721447917 | 14.726096217269438071 | 11.537390452345231751 |
| Au | 13.914933225510832671 | 13.334233803600509916 | 15.974432260639028769 |
| Au | 12.010582153225289659 | 12.318675950233455296 | 11.155762226187329489 |
| Au | 13.415155594339861267 | 12.683744677366433962 | 13.380806397032330324 |
| Au | 13.675325172557066367 | 17.257081919538510562 | 12.470452056298796606 |
| Au | 10.720724544408525958 | 14.651788747762855891 | 10.175893886114463882 |
| Au | 9.326043470205695129 | 14.215175851421054887 | 13.700290827628112567 |
| Au | 9.156221896154344009 | 12.662090776705710482 | 11.410095227586470656 |
| Au | 7.249324403409338480 | 12.610286361762598872 | 13.215548234511588888 |
| Au | 10.247087861877357540 | 12.216986374277658101 | 9.022460819822240907 |

14

Au14(11)

| | | | |
|---|---|---|---|
| Au | 13.928078798801555749 | 15.003328542471191298 | 15.615092837240045043 |
| Au | 14.043799115610347883 | 14.090459098859218656 | 11.061738355225042341 |
| Au | 12.698450035510225220 | 16.395423407115650605 | 10.669975370934636771 |
| Au | 11.490697720001843862 | 15.390166577841155160 | 14.440998798228818600 |
| Au | 10.347551638532156204 | 16.568617924483053372 | 12.301296249070098554 |
| Au | 10.323817941876543003 | 17.258412242750502230 | 9.659614356020693293 |
| Au | 10.501834133677562022 | 14.600719320602179252 | 10.258729519135346919 |
| Au | 13.221332389950912756 | 13.266494348180206231 | 13.684455218914440167 |
| Au | 14.198021502954967232 | 11.481948804041795853 | 11.920356090687022288 |
| Au | 13.912481677888962395 | 15.976409859928818236 | 13.041945319716951701 |
| Au | 11.706419662880660937 | 12.482720133750147440 | 11.414364363962404170 |
| Au | 8.945398107295968515 | 14.849929756902474409 | 13.832202672284388711 |
| Au | 9.094426743311037953 | 12.892862258738041703 | 11.797848266694636266 |
| Au | 6.953590062707543318 | 13.161844845334879750 | 13.346610039885179688 |

14
Au14(12)

| | | | |
|---|---|---|---|
| Au | 15.909040152717953021 | 13.142865302507773961 | 7.500000000000000000 |
| Au | 18.537813187425651762 | 20.807633972619171914 | 7.500000000000000000 |
| Au | 20.463216172075821220 | 18.851235473612405258 | 7.500000000000000000 |
| Au | 17.751002195918658799 | 18.266005000373905176 | 7.500000000000000000 |
| Au | 19.591753167746549735 | 16.276163544295403796 | 7.500000000000000000 |
| Au | 16.849816168546777106 | 15.674450821550772517 | 7.500000000000000000 |
| Au | 18.783411476399709983 | 13.780408248120179948 | 7.500000000000000000 |
| Au | 17.799093143572715547 | 11.309103589439445159 | 7.500000000000000000 |
| Au | 15.903149983101638298 | 20.405884720748545647 | 7.500000000000000000 |
| Au | 13.261129693605228752 | 19.728694600100080692 | 7.500000000000000000 |
| Au | 15.086185204045603925 | 17.753709959129533758 | 7.500000000000000000 |
| Au | 12.334614496960451646 | 17.217051461349448971 | 7.500000000000000000 |
| Au | 14.154958995014402134 | 15.162065145575985170 | 7.500000000000000000 |
| Au | 11.534276434868324657 | 14.686777509577211731 | 7.500000000000000000 |

14
Au14(13)

| | | | |
|---|---|---|---|
| Au | 18.460415753328970112 | 17.849345919031456731 | 14.798494000979102836 |
| Au | 16.062090291704770095 | 16.859824668077425969 | 15.519748610386203680 |
| Au | 20.247279408425686853 | 16.259732626561667956 | 13.797098790016317338 |
| Au | 17.876293061829255038 | 15.067174463029738618 | 14.458326785775348355 |
| Au | 15.302623873763156581 | 14.283658992996015158 | 15.307402295440695639 |
| Au | 16.707580757812394268 | 19.600723799536698522 | 15.811739182722325481 |
| Au | 17.043438683518289167 | 12.511428632277935336 | 14.258344231603782504 |
| Au | 14.311368823596367861 | 18.635590691833932908 | 16.578107361014200904 |
| Au | 11.829128519296782329 | 18.017723153312914519 | 17.425722533004876169 |
| Au | 13.551959076569216833 | 16.046039937858033397 | 16.349801491056137337 |
| Au | 11.105121830791826909 | 15.443958719118871414 | 17.301286707224782901 |
| Au | 12.772105912610408396 | 13.400810691664956309 | 16.187182313910575715 |
| Au | 10.270448667595120540 | 12.990793341657280280 | 17.213538737088530439 |
| Au | 14.547219586157872584 | 11.704887683043796898 | 15.075362306777048360 |

14
Au14(14)

| | | | |
|---|---|---|---|
| Au | 17.539146156427342760 | 15.327598242710941179 | 7.500000000000000000 |
| Au | 14.914709164579539546 | 14.683019506289712197 | 7.500000000000000000 |
| Au | 19.260030714127875484 | 13.233002401121492397 | 7.500000000000000000 |
| Au | 16.692942505083337323 | 12.640664991199075828 | 7.500000000000000000 |
| Au | 14.049215950333421077 | 12.018067621633779396 | 7.500000000000000000 |
| Au | 18.640104990047898781 | 10.608408076333986969 | 7.500000000000000000 |
| Au | 15.980788773547516257 | 10.019524259647315745 | 7.500000000000000000 |
| Au | 13.424006546008628860 | 9.372987703020243089 | 7.500000000000000000 |
| Au | 13.079626652915404605 | 16.669325404412440150 | 7.500000000000000000 |
| Au | 10.428118782498584594 | 16.074663298114529653 | 7.500000000000000000 |
| Au | 12.275154632638143326 | 14.050600607641730022 | 7.500000000000000000 |
| Au | 9.619850096821933150 | 13.518665387662455402 | 7.500000000000000000 |
| Au | 11.353326355002836934 | 11.436452886661411910 | 7.500000000000000000 |
| Au | 10.858834647967627873 | 8.822337566551016863 | 7.500000000000000000 |

14

Au14(15)

| | | | |
|---|---|---|---|
| Au | 13.251542701925892231 | 16.329825395546574640 | 13.600849179768754738 |
| Au | 13.945397953107237399 | 14.059414626388386793 | 12.003361723454062826 |
| Au | 12.909267375310461645 | 16.167846619414014242 | 10.768900273795940947 |
| Au | 11.627626157853994115 | 13.829493987883015649 | 13.759275282939276153 |
| Au | 10.515108942479040266 | 16.078566043439334265 | 12.438543241942895179 |
| Au | 10.569675308767278565 | 17.364854740490294205 | 9.989061014422034646 |
| Au | 11.002155773057912924 | 14.674804438239231530 | 9.507548468840861133 |
| Au | 14.079776085174366074 | 14.044170335899805124 | 14.891152879569313328 |
| Au | 13.366626804950122676 | 11.723483949297941464 | 13.318745271260150176 |
| Au | 12.158035003265052154 | 12.308839437814860673 | 10.903110799134084985 |
| Au | 9.914207434031505173 | 12.167876426041374316 | 9.339281661546742441 |
| Au | 9.009993566065309878 | 14.506174024839546988 | 14.072959790704542726 |
| Au | 9.730194307197265857 | 13.469456696936433104 | 11.694799671267142926 |
| Au | 11.930195918815764244 | 18.385589964769785354 | 12.334025588353073388 |

14

Au14(16)

| | | | |
|---|---|---|---|
| Au | 11.453106650267729449 | 17.932379739182771061 | 10.599875765730246258 |
| Au | 12.964002822114885305 | 17.417908954779246500 | 12.783456095810766584 |
| Au | 13.920859508276762995 | 16.195488929902293052 | 14.990221255128883371 |
| Au | 13.131078102258408435 | 15.715873521571081639 | 10.314659483338006751 |
| Au | 13.457380280652557758 | 14.625177741627153338 | 12.768950461561473730 |
| Au | 13.216674623616114559 | 13.585304961545505265 | 15.269429123418738925 |
| Au | 12.704975963305869868 | 13.084841388881468305 | 10.645682877026583668 |
| Au | 12.164752284294246820 | 12.121078307979063382 | 13.289779471861749016 |
| Au | 11.763303389044235558 | 16.811629339206206168 | 8.143916474968829533 |
| Au | 10.623218220202879536 | 15.979872453351243067 | 12.492832829199624456 |
| Au | 10.311663357368042071 | 15.256811362610090299 | 9.820209909270561255 |
| Au | 10.096497223392910669 | 13.429558537230471771 | 11.874514826385599520 |
| Au | 10.408503610285233520 | 12.611973582289078166 | 9.146767106593223673 |
| Au | 11.032847930921366952 | 10.847619149844256015 | 11.175112024705191160 |

14

Au14(17)

| | | | |
|---|---|---|---|
| Au | 18.196572919665310764 | 8.268679066254223997 | 7.500000000000000000 |
| Au | 14.929217370972068579 | 14.702059805997919284 | 7.500000000000000000 |
| Au | 19.391363661146911568 | 13.363198808646160387 | 7.500000000000000000 |
| Au | 16.818739691814155179 | 12.787722177865278539 | 7.500000000000000000 |
| Au | 18.789938922956118006 | 10.797506760991495867 | 7.500000000000000000 |

Au 16.120899457580957659 9.995889015394535448 7.500000000000000000
Au 13.039582925332208418 16.616660455513262207 7.500000000000000000
Au 10.411281725807736365 16.012969274465028491 7.500000000000000000
Au 12.340498860523776870 13.973767329828813644 7.500000000000000000
Au 9.733791824493222222 13.396200034332288453 7.500000000000000000
Au 14.247494211580610468 11.992635493116281964 7.500000000000000000
Au 11.481761609810078539 11.361225211341375640 7.500000000000000000
Au 13.536530331911539093 9.397364408488346754 7.500000000000000000
Au 10.993183320404948944 8.748345024764892841 7.500000000000000000
14
Au14(18)
Au 12.399077557225448132 18.517301056464511078 11.023509273712184253
Au 12.132331938319381948 17.253294763800067102 13.386603133342013905
Au 13.217477189669097015 15.895367763381965887 11.158074031901895395
Au 12.300013014436251169 14.891491529986794262 14.965700116931799357
Au 13.885881221630421933 13.912354449511690646 12.858919693464509137
Au 14.369209106823719324 13.271154481954411608 15.490142509886981159
Au 12.196800804961384301 13.413951919157529602 10.830200134029896120
Au 12.141280504073225899 12.179553718944170271 14.240185843123191134
Au 10.918817645342521416 17.031815452569233571 15.780480382094033587
Au 10.430995489443443347 16.607935242043708968 11.163725502933706579
Au 9.889412375086795493 15.508725406008901260 13.670423857763340791
Au 9.497799400656564472 14.118730189360297445 11.434573614167046429
Au 10.043781489339803414 12.218371077244658096 9.589163814333657498
Au 10.590493698991211957 11.608901830572474267 12.221346395315634936
14
Au14(19)
Au 14.032404584571246176 16.181203366023222401 14.857957722639845599
Au 11.750664000892363958 14.902728958310072827 14.109971589655652124
Au 13.993003633622477011 14.870355370289605190 12.443346538127235235
Au 10.169786844836526285 15.532706576520997288 11.873124817537032527
Au 12.482278439452191066 15.469462884234188138 10.305776960407875720
Au 10.665531186361711491 17.517709739323844786 10.081714717064301112
Au 9.942945204795666569 15.007198077110802004 9.148235225577643348
Au 13.972342717355040520 13.407691138588743840 14.864607752241765226
Au 9.076008256664572116 14.525843372951067423 14.129041863331973872
Au 11.398201833843806696 10.466518969680326023 11.353739844953919302
Au 12.885423568703366470 11.476713837539486818 13.440810001670278950
Au 10.556478837908214174 12.823163188448001293 12.602722084143330505
Au 12.894864800890532663 12.877095070582353031 10.974059323879826522
Au 10.333517060102382246 12.489409370397792642 9.815688788768826001
14
Au14(20)
Au 16.505933340799401776 19.789329075650790912 7.500000000000000000

| | | | |
|---|---|---|---|
| Au | 18.212817994207583183 | 17.838914080357273662 | 7.500000000000000000 |
| Au | 20.048614680527172993 | 15.888615168361029717 | 7.500000000000000000 |
| Au | 17.367176692702543050 | 15.253283099258764466 | 7.500000000000000000 |
| Au | 19.281958928103179574 | 13.321764197369224902 | 7.500000000000000000 |
| Au | 16.539402376004424866 | 12.522207675215870637 | 7.500000000000000000 |
| Au | 18.631256432248317623 | 10.812811312157563748 | 7.500000000000000000 |
| Au | 15.403456543034639736 | 17.322805414155983073 | 7.500000000000000000 |
| Au | 12.882151007924118602 | 16.544930238840343861 | 7.500000000000000000 |
| Au | 10.290986578694955256 | 16.111241331979822888 | 7.500000000000000000 |
| Au | 14.768287258372062354 | 14.637746177908889322 | 7.500000000000000000 |
| Au | 11.853192612547781337 | 13.946224548303222335 | 7.500000000000000000 |
| Au | 13.937376019343997058 | 12.087065439798976740 | 7.500000000000000000 |
| Au | 11.426564210736080085 | 11.310993558116800628 | 7.500000000000000000 |

14

Au14(21)

| | | | |
|---|---|---|---|
| Au | 14.648293265988332479 | 15.842843601620586469 | 14.653179162946594616 |
| Au | 13.924376928760441530 | 14.362604163934737045 | 12.478954315022766863 |
| Au | 13.829031513679145959 | 13.212325184188426164 | 15.004735048591159696 |
| Au | 11.902726907926297173 | 12.332275937747386507 | 13.369041415010039842 |
| Au | 11.983413500772249094 | 15.294607615041101667 | 14.393280626130248834 |
| Au | 10.543889376707015870 | 16.242545165392495932 | 12.340516651036569584 |
| Au | 9.665401601266324860 | 14.085924329247635711 | 13.784729201163960610 |
| Au | 8.016272610056834225 | 11.995497331136578012 | 13.449121755670679690 |
| Au | 12.796333571532167994 | 15.928496373828767929 | 10.647120237221868067 |
| Au | 12.386900923581915634 | 13.101249359164320651 | 10.686381320523434368 |
| Au | 10.402673759518505037 | 12.294144304817715252 | 8.902309467232566220 |
| Au | 10.576017162441758046 | 17.425396687291240028 | 9.857574151207497692 |
| Au | 10.245103292871686307 | 14.785587746500082673 | 9.899524875935417967 |
| Au | 9.718438080896778075 | 12.457905761088818863 | 11.494701857307491721 |

14

Au14(22)

| | | | |
|---|---|---|---|
| Au | 14.381971245805958759 | 11.633303652944089635 | 12.603184907597757558 |
| Au | 12.546420601421489849 | 16.355316031706969682 | 11.170105146976608168 |
| Au | 11.795600056542081546 | 12.256811752261899784 | 13.252142243395457655 |
| Au | 13.597506501328245676 | 14.301554592512612629 | 12.558210630041982014 |
| Au | 12.313836507630773198 | 10.019499080561066862 | 11.518929957672744635 |
| Au | 12.961389898868777237 | 12.507236533971568093 | 10.363671081768536908 |
| Au | 10.972595389625125151 | 14.961135046592666242 | 12.928692111236912865 |
| Au | 11.244003137067139164 | 16.965337221357373210 | 8.751715150862878545 |
| Au | 9.748470604620729674 | 16.483027799268672453 | 10.929243853381370499 |
| Au | 11.177120255185199582 | 14.350122165464300750 | 9.483239354591406212 |
| Au | 8.316498380597352025 | 15.515072999134297049 | 12.958141792033993411 |
| Au | 9.089741706678079680 | 12.899154554537718198 | 13.178959332194745357 |
| Au | 9.884416641869879783 | 10.295226281276859126 | 12.734660108577800841 |
| Au | 10.302334241759268707 | 12.087513132409274164 | 10.672033683668589177 |

14
Au14(23)

| | | | |
|---|---|---|---|
| Au | 11.949208991155920501 | 18.322327683205234194 | 11.437883264425224183 |
| Au | 11.595691087679115938 | 17.158365064257761645 | 8.961959800989758307 |
| Au | 13.563142790731037124 | 17.181763173369208886 | 13.297476904332011571 |
| Au | 12.797952271480326658 | 15.708321683583498896 | 11.081862821666181418 |
| Au | 13.389492132483965747 | 14.493462282598226665 | 13.439347365603591555 |
| Au | 12.245225844172988872 | 13.070580656529942942 | 11.276511216763346113 |
| Au | 13.700072710324818104 | 13.202022692502348278 | 15.726923814907081578 |
| Au | 12.258124227973414833 | 11.985917275319224018 | 13.837790339223326797 |
| Au | 9.961151940330065457 | 16.342597666019774039 | 10.981758698201474544 |
| Au | 7.966021818654441589 | 15.448679728536820477 | 12.448469314514651884 |
| Au | 11.231807098775096421 | 14.440778469382898663 | 9.088710385274486825 |
| Au | 9.351032123395299323 | 13.612847338228339922 | 11.150766519633158680 |
| Au | 10.229627025385806149 | 11.898379668297065948 | 9.265060481668989212 |
| Au | 10.522601284457438808 | 11.129148473169774292 | 11.959997228796517277 |

14
Au14(24)

| | | | |
|---|---|---|---|
| Au | 15.896182340102670949 | 14.591174823039390063 | 14.481932186835850374 |
| Au | 12.425048149762952576 | 15.372404662004036879 | 10.661466682714584309 |
| Au | 10.803654739085178704 | 15.195889284144938003 | 15.827208575033496984 |
| Au | 14.141520762152223867 | 14.270404665995526372 | 12.499188530303037226 |
| Au | 13.383839835022470766 | 14.903175912166556571 | 15.205530916701226474 |
| Au | 12.260098784454825704 | 12.714636500544552078 | 11.242950604944756066 |
| Au | 11.482997871152024416 | 14.600908863013353667 | 13.252207367549324601 |
| Au | 10.979624125017071634 | 17.517215589469287806 | 9.786667862056265932 |
| Au | 9.593369586375825264 | 15.694765252459040639 | 11.275043540832022870 |
| Au | 10.412197182956042241 | 15.017365088041556120 | 8.706954204719506762 |
| Au | 8.824270798978229635 | 14.872811564064408074 | 13.758810838993763781 |
| Au | 9.425801115379320194 | 13.021392911289241923 | 11.850553016070588086 |
| Au | 10.546781494037034577 | 10.591696163096218086 | 11.294520586976467769 |
| Au | 10.256518384524234477 | 12.466469564671362136 | 9.259894440269754057 |

14
Au14(25)

| | | | |
|---|---|---|---|
| Au | 14.116971097495619958 | 14.968989218652396289 | 7.500000000000000000 |
| Au | 18.575348837465345042 | 20.633397757718334731 | 7.500000000000000000 |
| Au | 20.395505345646697037 | 18.583544278869304378 | 7.500000000000000000 |
| Au | 17.715415267451909642 | 18.010417068077096303 | 7.500000000000000000 |
| Au | 19.570980577486718488 | 16.038164085983765972 | 7.500000000000000000 |
| Au | 21.284620115222587344 | 13.888543934995322360 | 7.500000000000000000 |
| Au | 18.696585582765724354 | 13.490134788077522643 | 7.500000000000000000 |
| Au | 16.750256465757871638 | 22.609933674328544839 | 7.500000000000000000 |
| Au | 14.054129908583862019 | 22.182841519301558009 | 7.500000000000000000 |
| Au | 15.886598011755637572 | 20.065341633025489898 | 7.500000000000000000 |
| Au | 13.199829902863328357 | 19.632580557585168179 | 7.500000000000000000 |
| Au | 15.052367022728523338 | 17.528048606166457546 | 7.500000000000000000 |
| Au | 12.374473226916698820 | 17.046795660853831578 | 7.500000000000000000 |
| Au | 16.777269169858794129 | 15.371510177365895089 | 7.500000000000000000 |

14

Au14(26)

| | | | |
|---|---|---|---|
| Au | 17.488602449910430892 | 15.368130161428570801 | 7.500000000000000000 |
| Au | 14.854647117724095295 | 14.860931345144615889 | 7.500000000000000000 |
| Au | 19.498264585538812810 | 13.303941594793437631 | 7.500000000000000000 |
| Au | 16.789464227785600769 | 12.797440489563571830 | 7.500000000000000000 |
| Au | 18.725516342287029659 | 10.779378260358939201 | 7.500000000000000000 |
| Au | 16.066941289039753826 | 10.171829405017728476 | 7.500000000000000000 |
| Au | 12.832992976357097348 | 16.671756830137795191 | 7.500000000000000000 |
| Au | 10.183998489554696221 | 15.942441324292179416 | 7.500000000000000000 |
| Au | 12.160571016074765538 | 14.090838497253372452 | 7.500000000000000000 |
| Au | 9.498394490185004813 | 13.335085731760583982 | 7.500000000000000000 |
| Au | 14.141210199975075668 | 12.205474925203182224 | 7.500000000000000000 |
| Au | 11.453076183920117614 | 11.469188385479130332 | 7.500000000000000000 |
| Au | 13.446034885215608412 | 9.615340126470453441 | 7.500000000000000000 |
| Au | 20.087875298431672633 | 15.885437739096278875 | 7.500000000000000000 |

13

Au13-(1)

| | | | |
|---|---|---|---|
| Au | 17.900422356624268616 | 18.634543758043783157 | 12.631853218771333047 |
| Au | 15.922181666758930874 | 17.019497086746259384 | 11.813801666178608940 |
| Au | 18.957434343414803379 | 18.715294950391225370 | 15.141108315261440964 |
| Au | 12.881225642806464649 | 13.594363591691417170 | 12.984129503346681034 |
| Au | 11.804226910509317605 | 11.877280385187017231 | 14.691759283729112440 |
| Au | 13.781996716304178108 | 13.475205359399081928 | 15.614204062316044741 |
| Au | 13.890096108880962689 | 15.278960250132790577 | 11.175610294963764346 |
| Au | 16.856024028529780878 | 15.301506890456369803 | 18.765090328788012641 |
| Au | 17.916725458757632339 | 16.997912377186882082 | 16.997099460166012364 |
| Au | 14.828829593365504991 | 13.558639878781297838 | 18.141502843185339344 |
| Au | 15.812428684540609680 | 15.224283875004934075 | 16.252717101458930671 |
| Au | 14.903832344473865845 | 15.300978370646831195 | 13.711207885034855991 |
| Au | 16.897586735033872429 | 16.958427682331286945 | 14.473253788799667063 |

13

Au13-(2)

| | | | |
|---|---|---|---|
| Au | 13.734354892054971842 | 14.061621231841799684 | 15.909148525847511380 |
| Au | 15.948437801287084881 | 15.499107381938083350 | 16.417825816194021371 |
| Au | 12.469100115953384744 | 14.147542979068260749 | 13.428968812196586669 |
| Au | 17.164455493283401921 | 15.355798613962264199 | 18.905740675239059101 |
| Au | 18.237124325490523091 | 16.979597537765656057 | 16.919099279442253447 |
| Au | 19.396307611847966257 | 16.802386082161191894 | 19.345229773094249737 |
| Au | 15.814327842310342831 | 17.193804047716803041 | 12.098607845773193148 |
| Au | 14.761396771181976462 | 15.619904611431573827 | 14.008702393666711217 |
| Au | 13.547422438417601853 | 15.733285666281984660 | 11.539513896001700743 |
| Au | 17.084532310675211164 | 17.141192467649062792 | 14.516223543260627693 |
| Au | 12.712091184636705066 | 12.493801874920121620 | 17.839755719783155996 |
| Au | 14.960818519492278256 | 13.919258570665192565 | 18.449326496572055589 |
| Au | 11.473179871368861527 | 12.583821998597811032 | 15.380652283928888835 |

13

Au13-(3)

| | | | |
|---|---|---|---|
| Au | 16.880756337139100509 | 15.472306699575428368 | 15.959995446659750229 |
| Au | 17.981814637688501080 | 14.166417346572101721 | 18.111280826286211720 |
| Au | 17.508093460758718862 | 12.837640868809650385 | 15.855915816641617866 |
| Au | 17.349359305130285236 | 16.909887424474501216 | 18.343508580389826790 |
| Au | 18.424299369854733044 | 15.524261974193674263 | 20.356722054207825323 |
| Au | 16.696417346297383943 | 19.486449624360925981 | 18.432072750523563087 |
| Au | 16.248217805799122004 | 18.083690634334018910 | 16.111754089597546624 |
| Au | 15.573500428352483382 | 20.750791991872731046 | 16.210532341869281936 |
| Au | 16.016805261543634487 | 22.075206092838417504 | 18.483011348068426827 |
| Au | 15.130842948850739660 | 19.402444706864333313 | 13.949874804754355040 |
| Au | 16.338160804437819706 | 14.187946661959694339 | 13.574700992847390069 |
| Au | 16.966374288575927665 | 11.612594281111508820 | 13.534065496676689122 |
| Au | 15.750557705571258893 | 16.761698693032840879 | 13.753036751477857536 |

13

Au13-(4)

| | | | |
|---|---|---|---|
| Au | 14.640465764631747092 | 15.791482739392026957 | 14.079007681359270165 |
| Au | 15.738648526082645418 | 17.289695161535224344 | 12.106332654153755257 |
| Au | 15.699645460512227757 | 16.088530207333828770 | 16.512897832306247636 |
| Au | 16.803394873804865739 | 17.593262886952246049 | 14.516234834050273150 |
| Au | 13.525293348199786791 | 15.463523158241986977 | 11.582157746589201608 |
| Au | 14.701972374243577946 | 17.001271245463978943 | 9.705143156083881451 |
| Au | 12.492981899619518771 | 14.025440384483067291 | 13.547786396979660140 |
| Au | 14.692740738881168383 | 14.655069677314475740 | 18.594895941457352251 |
| Au | 16.747267312055971189 | 16.379473374088021842 | 18.946685779707994612 |
| Au | 13.535149714114748321 | 13.149218924709744627 | 20.486488742272786823 |
| Au | 11.447690875789827913 | 12.549652535017719046 | 15.586927153102317689 |
| Au | 12.541272868470377588 | 12.891923884476895523 | 18.039079080197090832 |
| Au | 13.577856805594009870 | 14.326878910990629379 | 16.037679938739742624 |

13

Au13-(5)

| | | | |
|---|---|---|---|
| Au | 13.717572080312509186 | 13.101334056873239930 | 14.818690597088602701 |
| Au | 15.530873839669926184 | 13.496578401899196109 | 16.731706159388330946 |
| Au | 11.227773382947706438 | 13.699903972686900744 | 13.819381770819026301 |
| Au | 15.877001506451284385 | 14.752895238334755135 | 14.286903302651220216 |
| Au | 17.446775673705232634 | 18.426053236189243734 | 12.045158281099809017 |
| Au | 15.289220835827682876 | 16.886677725131484351 | 12.416614228998392733 |
| Au | 17.231516244564680562 | 17.309927170524083806 | 14.468120730716609756 |
| Au | 12.247965627693107038 | 15.382558558733666842 | 15.738830637432208448 |
| Au | 14.442040042195074534 | 16.980521335262274363 | 15.187503244158195415 |
| Au | 13.248563411804454404 | 15.331583612027035457 | 13.004950539230065942 |
| Au | 16.197706175381206606 | 14.781889718341336248 | 19.068240341128248616 |
| Au | 14.060616589153397626 | 15.777309837183617347 | 17.653314215997998105 |
| Au | 16.816059948294100224 | 16.108825551813502841 | 16.773161836291084370 |

13
Au13-(6)

| | | | |
|---|---|---|---|
| Au | 16.507813166472899979 | 18.842056607656388678 | 17.927828191199878916 |
| Au | 18.542203509003101658 | 17.152567160309104111 | 18.617414853925300378 |
| Au | 14.063804350910714902 | 19.880290840106759731 | 17.744351633374385813 |
| Au | 19.088504964475959014 | 19.015101031180492441 | 13.666373262619076812 |
| Au | 18.608879375268308820 | 18.478558297253297127 | 16.228786666893313395 |
| Au | 16.778455408799075599 | 17.775269574617848178 | 14.126650236383044401 |
| Au | 14.954266972361073940 | 15.772559153445733671 | 14.488945758049691648 |
| Au | 14.678136008102267596 | 18.137531775829522473 | 15.828424706201090544 |
| Au | 14.317132956128832078 | 13.212708810872618059 | 14.268910202714781832 |
| Au | 17.726392952824234328 | 14.352946430477889095 | 18.227176252692299130 |
| Au | 19.416494314901193263 | 15.176570412312248237 | 20.146573160670335767 |
| Au | 16.034345315330071458 | 13.702085863088745654 | 16.276378077817035717 |
| Au | 16.760469208421515219 | 16.308380901849428568 | 16.594743114460122513 |

13
Au13-(7)

| | | | |
|---|---|---|---|
| Au | 15.911050071150045326 | 17.123553401368063476 | 12.064993241389094791 |
| Au | 18.119512632153497833 | 18.661735959883809954 | 12.448840667484720512 |
| Au | 14.714076335836088205 | 15.403943088726364863 | 14.038043835296315720 |
| Au | 13.774244288432241135 | 15.695984357181625768 | 11.529002731030317364 |
| Au | 20.418383447167013145 | 18.508299647354593986 | 17.181651269487684175 |
| Au | 19.233283556426993499 | 18.542851044062484078 | 14.852152511539186364 |
| Au | 18.229362821167448772 | 16.922082994879062312 | 16.987674880345533523 |
| Au | 16.008552523637757048 | 15.433775370932714566 | 16.438015343270066637 |
| Au | 17.017242120179211895 | 17.007046340644023985 | 14.492455429565755054 |
| Au | 13.768757553568532259 | 13.861899544268670326 | 15.984562234996568364 |
| Au | 14.998574271234703659 | 13.798221141631422881 | 18.463994475667849571 |
| Au | 17.142567852448557630 | 15.303140230570761560 | 18.919911928512714638 |
| Au | 12.851715260597936563 | 12.289397403496490568 | 17.999257137414378604 |

13
Au13-(8)

| | | | |
|---|---|---|---|
| Au | 19.239598620346281876 | 18.483388680611369637 | 15.068177418920754462 |
| Au | 20.439671470548042009 | 18.194314882365883790 | 17.496621906302561200 |
| Au | 18.035113359960327273 | 18.750005002588661540 | 12.652836756142848529 |
| Au | 19.526479907921579837 | 16.443482343738040186 | 19.337641603241568333 |
| Au | 17.318871071211042079 | 14.900757219300805545 | 18.818567373325969072 |
| Au | 18.248504872746220684 | 16.661815980930335002 | 16.943620163059659234 |
| Au | 12.708362690303376397 | 13.908312543533364547 | 13.303393319685206109 |
| Au | 13.879187368842837103 | 13.612867403873858407 | 15.766113246224847444 |
| Au | 13.701012516232792038 | 15.671320525972848969 | 11.514986707195976834 |
| Au | 14.888505251153794973 | 15.400584479820647488 | 13.958660925953839893 |
| Au | 17.060633233506329987 | 16.947620876367928133 | 14.509799930982419269 |
| Au | 16.071833178840702061 | 15.125161687165116220 | 16.393103880840364184 |
| Au | 15.872995742386056151 | 17.217808389731320773 | 12.073895587123653073 |

13

Au13-(9)

| | | | |
|---|---|---|---|
| Au | 16.770552751719069562 | 16.141541545612945185 | 16.991926031703293631 |
| Au | 16.711564990802322939 | 17.651706449255364362 | 14.746729263027914314 |
| Au | 17.550575462320590958 | 14.202536722904463673 | 18.628229988847724741 |
| Au | 15.351217619768274147 | 15.124757355328453556 | 14.715543394714902803 |
| Au | 15.470063445330907825 | 16.582231164075164287 | 12.364625257847402295 |
| Au | 13.410628694781971504 | 14.962633662784499577 | 12.732838516230915005 |
| Au | 17.387024422496313747 | 18.454874799052952028 | 12.308480238005616414 |
| Au | 11.875735574637865000 | 17.912438837828041471 | 16.682156155213224480 |
| Au | 12.360223390974216073 | 15.767204646433221171 | 15.221145203979519422 |
| Au | 14.441341443979329284 | 17.371426520622144096 | 16.249080488984862569 |
| Au | 13.473388091443272430 | 13.185005740297025767 | 15.379607587993293194 |
| Au | 15.580193808532198929 | 13.505746007336423276 | 16.948705832858546927 |
| Au | 11.264220980213007550 | 13.670262664469166936 | 13.835876394592339977 |

13

Au13-(10)

| | | | |
|---|---|---|---|
| Au | 11.709646741217769161 | 16.434619171968201101 | 17.879831489523020593 |
| Au | 14.282138276564079860 | 15.776611528694937547 | 17.641064404501047846 |
| Au | 12.453064009882913510 | 15.367248422570188637 | 15.545160944886161403 |
| Au | 15.740613073430274937 | 13.605812293333508833 | 16.800545984446852543 |
| Au | 15.249201418514950035 | 11.166914411152466968 | 15.841500419479380923 |
| Au | 16.381000201308381747 | 14.836342689154799146 | 19.222299431911324064 |
| Au | 13.224924750899894832 | 15.164208740436730949 | 12.890941483403576484 |
| Au | 15.148408032849051352 | 15.703207753184502238 | 14.848519283885130093 |
| Au | 13.920733922879300337 | 13.185271115920844665 | 14.699577411809896077 |
| Au | 15.430826191912640155 | 16.834705669022078212 | 12.382055384913106977 |
| Au | 19.035707687529818344 | 17.744004593088927635 | 16.327760572226843294 |
| Au | 16.994613426657135591 | 16.143302484671128383 | 16.916587507531072987 |
| Au | 17.258590168353990180 | 17.305347057801718336 | 14.358811651482701777 |

13

Au13-(11)

| | | | |
|---|---|---|---|
| Au | 14.840959478878559707 | 11.042517858877559433 | 16.268532959384540248 |
| Au | 14.175969599948624378 | 12.991182078492686358 | 14.544775965171218957 |
| Au | 15.986995939802232769 | 13.455104758544484511 | 16.545439715469992592 |
| Au | 17.032444881440607531 | 16.032297484892389150 | 16.668210847897597660 |
| Au | 16.907159794925732399 | 17.893902050378887481 | 14.720928960677920116 |
| Au | 16.440101816111901911 | 14.643228117865904281 | 18.983596038473393719 |
| Au | 14.304797964038360902 | 17.146311264418823583 | 15.334124213052813701 |
| Au | 16.060080737655134442 | 15.124057794324006210 | 14.217535581484145979 |
| Au | 12.516007036061685298 | 15.115876807658892744 | 15.584868563294882904 |
| Au | 14.299964226384267008 | 15.574498971999151564 | 17.550010942119307344 |
| Au | 13.608954011860619815 | 15.155642338214708076 | 12.890584825888605991 |
| Au | 15.053219161245959157 | 17.424335743404888888 | 12.677517330391092187 |
| Au | 11.608867323646837377 | 13.403774511927286639 | 13.648685626694996742 |
| 13 | | | |
| Au13-(12) | | | |
| Au | 15.336530404494897795 | 17.493237165301458447 | 12.527647744384500683 |
| Au | 13.066907141894287037 | 18.215080671817872826 | 11.332006854262967721 |
| Au | 17.501707498221250603 | 16.719382892309862854 | 13.817134202853075919 |
| Au | 12.599579621483735892 | 14.634174829893327185 | 15.210052173153464139 |
| Au | 15.123103506175301192 | 15.352543033062438838 | 14.282955621812529756 |
| Au | 12.766048654810145280 | 16.401096096478916309 | 13.261129007766269794 |
| Au | 16.979071269055427962 | 16.177172943031141727 | 16.558315404395411718 |
| Au | 15.663784527501999122 | 17.020999957391229174 | 18.696113777288140056 |
| Au | 19.461041225852774517 | 15.739371396836332906 | 15.444572225117354947 |
| Au | 17.448257611671092349 | 13.904624509491650741 | 15.050443642173371828 |
| Au | 12.845344191136485179 | 12.920138231493719871 | 17.318854503404413236 |
| Au | 14.503545385901221110 | 15.121092070820282061 | 17.261451884889101649 |
| Au | 15.084686961801457628 | 12.897777020716324433 | 15.719697258499373049 |
| 13 | | | |
| Au13-(13) | | | |
| Au | 15.411224980802156637 | 17.511155350034343314 | 12.339892941236213275 |
| Au | 13.826376619885332175 | 17.602538472928245028 | 10.208375664676999861 |
| Au | 17.650220072181717512 | 18.938282540262502351 | 12.796797917984402915 |
| Au | 14.557928783465237998 | 15.613971257747211396 | 14.385815371107005589 |
| Au | 16.022024571878830557 | 15.363311525488445142 | 16.694974902964784036 |
| Au | 16.749725443893424881 | 17.116320923892960337 | 14.699864227731486821 |
| Au | 13.096064489959562849 | 15.869175254257600116 | 12.099836157133889714 |
| Au | 11.680239599833434738 | 12.424281123872185972 | 15.906993969867626149 |
| Au | 12.316386705994368356 | 14.143655742897621508 | 13.919234327373111171 |
| Au | 13.842277305170206603 | 13.849854262957299156 | 16.363478938978161636 |
| Au | 18.276331899282741489 | 16.836728581758691803 | 17.150999301832090538 |
| Au | 18.950401040217403903 | 18.511135981636101633 | 15.200570371697674332 |
| Au | 17.375822250435128069 | 15.066498620267198660 | 18.971261402416686792 |
| 13 | | | |
| Au13-(14) | | | |
| Au | 17.042895852539174939 | 16.190310643398881041 | 16.721855226213506285 |
| Au | 17.403988578223657413 | 17.249785249272495236 | 14.278167809157904955 |
| Au | 16.400188945680227448 | 14.998854587213759970 | 19.079541915646458961 |
| Au | 15.138225607601752287 | 11.153664355621328852 | 15.945433742786470077 |
| Au | 13.783645582521954154 | 13.162076387079022055 | 14.810775693769953065 |
| Au | 15.658093352116694064 | 13.644877750160121721 | 16.771163975461412576 |
| Au | 13.329107270145387432 | 15.232244745578025658 | 12.834509541070676875 |
| Au | 15.475117584676510774 | 16.741561856860769808 | 12.249856070470226399 |
| Au | 15.190885036533552821 | 15.708847384457891749 | 14.773867801662310839 |
| Au | 11.282226803404068249 | 13.677898616220343087 | 13.721987992411962765 |
| Au | 11.745897405771572153 | 16.718570715076012334 | 17.815573405528127182 |
| Au | 12.404331989761754329 | 15.421188137678722541 | 15.570918965762853503 |
| Au | 14.278368813023735839 | 15.904256143383081579 | 17.532059555058339129 |

13

Au13-(15)

| | | | |
|---|---|---|---|
| Au | 15.044514743749303420 | 15.422534181748964954 | 13.800069634457294399 |
| Au | 16.119933913119435687 | 15.249854623891060612 | 16.332148599016022672 |
| Au | 13.786701406324496588 | 15.390875950566138997 | 11.448823083625210728 |
| Au | 18.567215298660823208 | 16.714220265459363901 | 16.865947924230763277 |
| Au | 19.478970309534879846 | 18.119686621973212226 | 14.749947844355054016 |
| Au | 17.266809464930243934 | 16.726875337911746300 | 14.394614160149037829 |
| Au | 11.629648349073725910 | 12.542576548966140493 | 15.384090338967213896 |
| Au | 12.720542391484617184 | 12.361716397652136479 | 17.970353805618167087 |
| Au | 12.644998680062359497 | 13.947101503743894568 | 13.351029133322949605 |
| Au | 13.853411870322092980 | 13.877682153897612949 | 15.876903625022984912 |
| Au | 17.222008017395424417 | 15.105236534494107303 | 18.789760413139934059 |
| Au | 14.990771780704644556 | 13.750071562072697873 | 18.381470513373415088 |
| Au | 19.471141983638275974 | 16.404845189622943025 | 19.359140455722172902 |

13

Au13-(16)

| | | | |
|---|---|---|---|
| Au | 17.291785883459741768 | 13.760490167942496242 | 18.303846128566149787 |
| Au | 15.626756657144394325 | 12.538582453486103674 | 16.546738333524928066 |
| Au | 19.262139353101360939 | 15.372737348295270721 | 19.282409388064003508 |
| Au | 15.147149055668052142 | 15.165332725058945584 | 17.190433658896257185 |
| Au | 18.552265812992725813 | 17.716199974936252204 | 18.079661366871064843 |
| Au | 15.956937376021681274 | 17.693552815393978506 | 17.045053240951254736 |
| Au | 18.780745528611753059 | 15.373651595550914095 | 16.581193454316274227 |
| Au | 14.246798663127044904 | 13.855674817563720325 | 14.517852462250276702 |
| Au | 14.778456417061839545 | 16.543665539108452833 | 14.787822714062135887 |
| Au | 16.898339286465688502 | 14.445848289611294746 | 14.899022062506171338 |
| Au | 15.889074569984655838 | 18.998275323204342868 | 14.217078913911272764 |
| Au | 17.842018553582736473 | 19.475096424249461080 | 16.142247004867670057 |
| Au | 17.719804843777925640 | 17.025477125598520445 | 14.743823098213111322 |

13

Au13-(17)

| | | | |
|---|---|---|---|
| Au | 17.050755549571427139 | 16.927212294982176388 | 14.496733276580810568 |
| Au | 15.854288171970388532 | 17.198843263104265588 | 12.111930095402136942 |
| Au | 18.305335440013571713 | 16.689352068787368211 | 16.862401710054026438 |
| Au | 16.016492773917782699 | 15.301091564837259185 | 16.469796256012262603 |
| Au | 13.615059164392691926 | 13.919645080417245353 | 15.868118258971787071 |
| Au | 14.757136704254028459 | 15.565409612313395371 | 14.013247493993347348 |
| Au | 12.592138692751721152 | 12.351900225316160942 | 17.732001754882116273 |
| Au | 13.490506455089263937 | 15.797604782751767161 | 11.605714580008161363 |
| Au | 12.427566057910949482 | 14.199859407091286911 | 13.487591459440325892 |
| Au | 14.451914408987276417 | 17.336967606871898084 | 9.707067151173454533 |
| Au | 17.128533559691408783 | 14.925056261098989197 | 18.912798006635142656 |
| Au | 14.877065863135566559 | 13.646133180422028275 | 18.340706504461337545 |
| Au | 19.376766119313614922 | 16.280145158006448014 | 19.293784181385309040 |

13

Au13-(18)

| | | | |
|---|---|---|---|
| Au | 12.667189882726798800 | 12.357008018697566598 | 17.698837557999134162 |
| Au | 14.936700061708885201 | 13.696978032666745406 | 18.341103359750590585 |
| Au | 13.665449162300019381 | 13.937206913139927877 | 15.851287073193756072 |
| Au | 17.155419385940753330 | 15.072976106148830766 | 18.824876321276544644 |
| Au | 16.040609048588354568 | 15.388151861839803658 | 16.403910422023411542 |
| Au | 18.357243464118322862 | 16.790415238452901292 | 16.949391925780375345 |
| Au | 13.443117523280943004 | 15.794156329296287211 | 11.701756750577235522 |
| Au | 12.372455579224499900 | 14.145384191120664141 | 13.460036502126717650 |
| Au | 14.355480681319892966 | 17.400134849616815558 | 9.837183996201989800 |
| Au | 14.750836205181178329 | 15.582079917733329566 | 14.039923831062633752 |
| Au | 18.123194316625983902 | 18.609108222130764432 | 12.710034917743865179 |
| Au | 17.087079770997945616 | 17.012067322706492689 | 14.544771739749672079 |
| Au | 15.844957483986343405 | 17.323644441450401388 | 12.082092577514243814 |

13

Au13-(19)

| | | | |
|---|---|---|---|
| Au | 18.658463398484439466 | 16.853651191908635809 | 18.323053539328782335 |
| Au | 18.034931004313719427 | 18.275765780597613031 | 16.193549857459938579 |
| Au | 17.314111057822834994 | 15.581835223255543710 | 16.261094997052413902 |
| Au | 18.265012682317312454 | 13.998567530537622261 | 18.305269515367761812 |
| Au | 17.253194187625247480 | 12.810851848129129138 | 16.174441773726400129 |
| Au | 19.404554594479947127 | 15.304151824132672388 | 20.347464668513747910 |
| Au | 15.559679112866326989 | 18.991368027347736103 | 15.027807625806355318 |
| Au | 13.168635464932824775 | 18.262989520421527345 | 15.872896077954264271 |
| Au | 17.826366396885898524 | 19.553058085694733137 | 13.724615956139304629 |
| Au | 14.759639515411757671 | 16.301521405751518756 | 15.006221271529009087 |
| Au | 16.281021439377507676 | 14.325087627814275137 | 13.940112698320950457 |
| Au | 16.380302464839889609 | 11.683979749245303026 | 13.880116588545575951 |
| Au | 17.114509856642101226 | 16.917215745163808549 | 13.724884273255574030 |

13

Au13-(20)

| | | | |
|---|---|---|---|
| Au | 15.351315409623749630 | 11.348937334797319920 | 15.925662107968706849 |
| Au | 13.757102490495990210 | 13.124939695524652450 | 14.755944914616428321 |
| Au | 15.672580862718710648 | 13.857440766234896401 | 16.736183014240218370 |
| Au | 15.605612212662865801 | 16.714850680139317518 | 12.348512051773820275 |
| Au | 15.167044148874463616 | 15.764928512117871051 | 14.754166625179244932 |
| Au | 17.431123671044939982 | 17.398628926048882448 | 14.189253530789418534 |
| Au | 13.360529456061069453 | 15.142141132077762578 | 12.741101888307214907 |
| Au | 12.265823560420582083 | 15.401563187432419255 | 15.386573296596786520 |
| Au | 11.213560670862282365 | 13.412017509505712809 | 13.888961582186684751 |
| Au | 11.685330865773130071 | 16.930116841406245953 | 17.481915374329659585 |
| Au | 16.897090342519447859 | 16.319819550516225348 | 16.831819156792970915 |
| Au | 14.204164603195982508 | 16.098241213856365306 | 17.357404613442298569 |
| Au | 19.006422999746771296 | 17.821331294341735685 | 16.408957403776017969 |

13

Au13-(21)

| | | | |
|---|---|---|---|
| Au | 15.619755849897149602 | 18.348191353279339921 | 16.281358731183964039 |
| Au | 17.066066698031644222 | 20.054966589034354030 | 14.627619073807069938 |
| Au | 14.734164345934322782 | 16.364645440679606025 | 14.679567341082828236 |
| Au | 18.096675143868878877 | 15.798364628932539233 | 16.043396466905036846 |
| Au | 18.989877186248776297 | 16.886829123605014757 | 18.435457828859810547 |
| Au | 18.405382033862426994 | 18.536602135134245373 | 16.456394318681731193 |
| Au | 17.508215622306984471 | 14.425298983385713925 | 18.537471501476730396 |
| Au | 19.103008222281726347 | 15.226653150946763304 | 20.521143762892116058 |
| Au | 16.551918552741838653 | 13.397654685476739544 | 16.251691404085054415 |
| Au | 15.491779075088320639 | 15.801508305874166993 | 17.254322374891415848 |
| Au | 16.822703255199346728 | 14.792507419015107573 | 13.725213056796949473 |
| Au | 17.284611749765989686 | 17.437065833253917901 | 13.907812697583569417 |
| Au | 16.092207856772262176 | 12.221573847382925848 | 13.899491573754065854 |

13

Au13-(22)

| | | | |
|---|---|---|---|
| Au | 17.771721047740829391 | 12.991352741900739076 | 14.558015905997310213 |
| Au | 15.057454616291881422 | 12.950676358975277580 | 14.535937333232219260 |
| Au | 16.365467522874908468 | 14.518297756944948063 | 16.340275187573411131 |
| Au | 16.891197981817530405 | 16.827369546559410196 | 14.791523336441393610 |
| Au | 17.560270384808962518 | 16.712465695191536241 | 17.480379500395223147 |
| Au | 16.686499612406283433 | 17.385221757654775843 | 19.911100821170666819 |
| Au | 19.025126620945556510 | 15.299082258232521880 | 15.572209441042659250 |
| Au | 12.493347835604943441 | 15.137907731376339271 | 18.894808803449404166 |
| Au | 13.556658795529447303 | 14.579143183489614799 | 16.511490534993036761 |
| Au | 15.030578714784862626 | 15.911403881658319648 | 18.396996114117957433 |
| Au | 14.466821361167820470 | 15.663768763345878199 | 14.130181218460046821 |
| Au | 15.479775756097780004 | 17.658101298119103006 | 12.694857169215971737 |
| Au | 12.454620475929457868 | 13.753051683551369422 | 14.144329394910757358 |

13

Au13-(23)

| | | | |
|---|---|---|---|
| Au | 12.225062813371991410 | 15.160977440552418471 | 15.736334871155547077 |
| Au | 12.269278560722380433 | 17.698343417681858369 | 16.829045259785104349 |
| Au | 11.048573226977783435 | 12.936319401352086800 | 14.807013316349522114 |
| Au | 15.641482788798086290 | 17.272274924197546397 | 12.634668644008460348 |
| Au | 17.014976452374579452 | 17.194309063344409338 | 15.033546768044272923 |
| Au | 14.182225279552236330 | 16.959943324515872831 | 14.954344006727273708 |
| Au | 16.290873331969905280 | 12.192902225567729957 | 14.335395952082262028 |
| Au | 13.726878882230074908 | 13.113347204083593667 | 14.810748360528013023 |
| Au | 15.702875094097342057 | 14.886664487144363633 | 14.015670448102641288 |
| Au | 15.861402308403713235 | 14.454529940967953294 | 19.219413998377152808 |
| Au | 17.088904016275535014 | 16.164728932346928758 | 17.427207151625101034 |
| Au | 15.779163370598892513 | 13.800133749535646643 | 16.541583043126603769 |
| Au | 14.259400933626888985 | 15.873344191710362594 | 17.479665235088216235 |

13

Au13-(24)

| | | | |
|---|---|---|---|
| Au | 16.698129655597153231 | 17.286719982934585005 | 14.479482319162865878 |
| Au | 19.162628821247285060 | 16.886493078191584516 | 15.463853873334636901 |
| Au | 15.311706887362976914 | 14.873549137482410032 | 14.831124885542298131 |
| Au | 15.309560340285763402 | 19.260726975300091368 | 13.282851752619276198 |
| Au | 11.948219188037549898 | 17.433474846813020065 | 16.680529041120813361 |
| Au | 14.007517744913535296 | 17.493771241881475476 | 14.824797867390829609 |
| Au | 14.291608332636510426 | 16.113700590222968856 | 17.259113778769791026 |
| Au | 15.999780718735433283 | 15.335760311688144242 | 19.255141769677305064 |
| Au | 15.447470261916166123 | 13.443493882712044041 | 17.334058714583235883 |
| Au | 17.068224388294325422 | 15.898567399173199277 | 16.827054686017930862 |
| Au | 11.478025270273361613 | 13.011103182993862148 | 14.078656989967168300 |
| Au | 13.752740198342856104 | 12.647586203592693366 | 15.496397148777345976 |
| Au | 12.489502208357349033 | 15.171031017014104947 | 15.244231400037065782 |

13

Au13-(25)

| | | | |
|---|---|---|---|
| Au | 14.572847947511558431 | 17.294371020340598477 | 15.968701025790362991 |
| Au | 15.623242222806190682 | 19.808734369919619667 | 15.750814916173958125 |
| Au | 15.364584945371685620 | 15.114029218945169930 | 14.447810916965273975 |
| Au | 16.234414182056021048 | 18.304607238614892140 | 17.949837951009193660 |
| Au | 19.556530680797006738 | 15.439582788828897009 | 20.277286769629057517 |
| Au | 18.704790834503658203 | 17.140928713175370035 | 18.430854626913045990 |
| Au | 18.109454543370684831 | 14.283718826510513722 | 18.331876082984685183 |
| Au | 16.528159255188128895 | 13.320697245951077292 | 16.437223004952510763 |
| Au | 14.776155660659778590 | 12.531560639261087431 | 14.560440415327212449 |
| Au | 17.269388640674637969 | 15.962732183322454915 | 16.326628158810759572 |
| Au | 18.352742603499713425 | 18.569117962129528365 | 16.143975330328473206 |
| Au | 17.656817880631326290 | 20.012144461267428852 | 13.939114468063694474 |
| Au | 16.596627922929219068 | 17.504559870733139348 | 14.051378547051921686 |

13
Au13-(26)
Au 18.459765290075253574 16.970301574137323541 18.256504505406319083
Au 19.111177899321351248 15.508184878595796974 20.394352807405244477
Au 16.060470452433097677 17.972259481463119357 17.162218251658842405
Au 16.351350692160536937 14.456351109989405757 13.933075477132323527
Au 17.388252124835421597 16.956683821640215371 13.598406444557651795
Au 16.028618360670389364 11.806759589515358755 14.022643562181189125
Au 17.986939566192123863 14.126111302630215860 18.404635704588613976
Au 16.959374429536783424 12.889620553635873890 16.280132408237058428
Au 18.222181285970176390 15.469572609359540039 15.838373009212181941
Au 15.728786662220610282 15.235303618206026854 17.180730835427858949
Au 16.220444192009153994 19.277757181053996050 14.599323632666694195
Au 14.991070520028118906 16.700990381847788058 14.968277026899459159
Au 18.489379343547255985 18.268746425926039478 15.858061960626681497
13
Au13-(27)
Au 18.982764685983141817 17.729688104535700433 16.410262817188588258
Au 17.389241344784505117 17.406835824217804287 14.312811619832944032
Au 16.888991571657161472 16.181740878682973062 16.907890998442066888
Au 15.414464416366783439 16.774745195605486714 12.342353700130530569
Au 17.545380544384162391 18.372209939596249484 11.860287499832450564
Au 13.344345491983071739 15.158142014183296098 12.992798700770679332
Au 15.140783487034392252 15.729058524191076529 14.851419599778942882
Au 15.362085802878445051 11.234246174711472577 15.814973004760952691
Au 13.627454915838050908 13.049108061051500229 14.824812840438566397
Au 15.653983663496287448 13.713812081883194338 16.715343061242080580
Au 14.193335982850477706 15.903288715650990781 17.470209894235278369
Au 11.683120980211430862 16.734006298193833118 17.702734597826808027
Au 12.203790708532251230 15.178804612496600512 15.559745976519687716
13
Au13-(28)
Au 17.154698315041105872 17.412693044603621217 14.481893946303534548
Au 16.660705638077743629 16.044958394603678897 16.699968114095955940
Au 17.214441109975858524 18.556177846411145538 12.071302069124376644
Au 12.432196874708932199 17.524321594993963913 17.106203743904785597
Au 12.157387525475259693 15.086729405644604185 15.847686497953093365
Au 14.169285484902601979 15.518418527175485977 17.855586292844968455
Au 15.105785796960272904 16.962615020909225905 12.436180986724137298
Au 13.215285037796530077 15.298577882208991952 13.268423340372429209
Au 14.381212213334045202 17.004726500030120206 15.255215615181850097
Au 16.110886433425456232 12.106826821625997326 14.569756015006463556
Au 13.537652203370154069 13.044611422334897455 14.891600804836269134
Au 15.554456226341137892 13.483241109927806889 16.892988595896376580

| | | | |
|---|---|---|---|
| Au | 15.870235643590817176 | 14.812761413530491694 | 14.233108497755305066 |

13
Au13-(29)

| | | | |
|---|---|---|---|
| Au | 18.353888739345155301 | 16.827736942572872891 | 16.762350866867514299 |
| Au | 19.333659514645994193 | 18.457624725878886807 | 14.836345335364113041 |
| Au | 16.142711202167493667 | 15.357106668644581404 | 16.516222026636715015 |
| Au | 12.847405850469780120 | 12.352003789564317771 | 17.791430794807535420 |
| Au | 15.086379000121580063 | 13.694984656389472377 | 18.367199725415456868 |
| Au | 13.840945393306775202 | 13.982221676133599786 | 15.901617868051440396 |
| Au | 11.502092895939615147 | 12.504072500695688319 | 15.423890222307017339 |
| Au | 15.834363411415161238 | 17.277140920019380843 | 12.059718646299279143 |
| Au | 17.084954637997086024 | 17.052321805331910554 | 14.399630218175811791 |
| Au | 14.848934909387637049 | 15.629493374831742614 | 13.995727148042631072 |
| Au | 13.438431819980412030 | 15.738374070918260017 | 11.639232058238176748 |
| Au | 14.413377610254936911 | 17.350178681404766934 | 9.775372835817632833 |
| Au | 12.510820303967799205 | 14.159184697614335846 | 13.525139359976623865 |

13
Au13-(30)

| | | | |
|---|---|---|---|
| Au | 13.778784460318492577 | 13.217987774133556300 | 14.860846626092833134 |
| Au | 15.557211947086916126 | 11.314086741420890547 | 15.565397805359065586 |
| Au | 15.165446125331618177 | 15.755893515519117543 | 14.871429656122817420 |
| Au | 12.405946569159773674 | 15.373321662751010308 | 15.655550887403805405 |
| Au | 16.995135988835563978 | 16.207207250338225180 | 16.826929428560866597 |
| Au | 17.332307509115025823 | 17.319357887945688645 | 14.418936982309517703 |
| Au | 15.671835253589826209 | 13.613398093162873081 | 16.880508160500536263 |
| Au | 14.242887881967062569 | 15.859830855037259312 | 17.705768314194390456 |
| Au | 16.327553795906585066 | 14.913796832119219360 | 19.180282258311102339 |
| Au | 11.887118579042271094 | 17.076796672416989509 | 17.683334031558491262 |
| Au | 15.402277520510944342 | 16.877774171512193391 | 12.283055874312003297 |
| Au | 13.352408650998507156 | 15.311117657875602305 | 12.962701339701242276 |
| Au | 17.616934856137795862 | 18.401157190767708727 | 12.017903429573268070 |

13
Au13-(31)

| | | | |
|---|---|---|---|
| Au | 19.304861050740306894 | 18.349838619407648821 | 15.000905954266668729 |
| Au | 17.058600942565121272 | 17.006178508498045687 | 14.448747082438728739 |
| Au | 18.477838256272672623 | 16.745505075443247733 | 16.906108308183640077 |
| Au | 17.394256113090357019 | 14.995734998116191150 | 18.762201385178322255 |
| Au | 15.139835373015607445 | 13.650533340057300791 | 18.233967268972591569 |
| Au | 19.646828347417713445 | 16.338321742213892662 | 19.279804902312950077 |
| Au | 13.238626229619582020 | 15.723431492101958895 | 11.623588998437325515 |
| Au | 14.714330201348429839 | 15.505562683798702395 | 14.058323333110704922 |
| Au | 14.372583305344321047 | 17.480429763691393674 | 9.877552387471848050 |
| Au | 15.615306104981520718 | 17.162526710245224137 | 12.175248963126332313 |
| Au | 13.656410316715568598 | 13.751269708786580281 | 15.954134317278235500 |
| Au | 16.055860373189123180 | 15.283071742404866100 | 16.345680513397670808 |
| Au | 12.431856304699572036 | 14.114243012234814145 | 13.558538371825243019 |

13

Au13-(32)

| | | | |
|---|---|---|---|
| Au | 15.298500431525829057 | 16.886144849963859116 | 12.527857120877262886 |
| Au | 17.502427944515495994 | 18.320762553038715481 | 12.010101354726762324 |
| Au | 13.013071352190330288 | 15.467504780035179479 | 13.002328841522253100 |
| Au | 15.635834560937727389 | 11.601592014102353190 | 15.253897271255587142 |
| Au | 15.853203353613983140 | 13.663594260972741878 | 16.905318817306429935 |
| Au | 14.060981057744836420 | 13.674394540750656191 | 14.783585360165647060 |
| Au | 14.008132678711701757 | 15.592771726811120558 | 17.646472285318097306 |
| Au | 12.074049072116608272 | 15.314331685250690995 | 15.728160490428903628 |
| Au | 16.224807182160787988 | 14.947146218839385767 | 19.268455085349593503 |
| Au | 14.605390558765071418 | 16.452036199856351573 | 15.114625796182737716 |
| Au | 18.916762856983485364 | 17.809802960259098370 | 16.517789904603748852 |
| Au | 17.122544210971227585 | 17.532112406682838213 | 14.541785815204587351 |
| Au | 16.696132064763521186 | 16.385774271436496008 | 16.992957787058649899 |

13

Au13-(33)

| | | | |
|---|---|---|---|
| Au | 12.098719283538351021 | 14.090738903132258741 | 16.559638043316255107 |
| Au | 13.442257110719452839 | 14.788293101559903064 | 18.724809257445109267 |
| Au | 13.479958611734543794 | 15.082302573519253741 | 12.695087786287976073 |
| Au | 13.745009122664660239 | 13.057417794903242836 | 14.601712245166364568 |
| Au | 16.265092089607232850 | 12.248590343912942657 | 15.044332149115259512 |
| Au | 11.301255499969764529 | 13.938747452847097819 | 13.930956945312702899 |
| Au | 15.816410430855524893 | 14.952601849103652754 | 14.274501299233911311 |
| Au | 16.141478836906244965 | 15.334385781600683885 | 19.028313054694169892 |
| Au | 16.948638808876619777 | 16.100615314820554858 | 16.558088872919569212 |
| Au | 15.148495374387040968 | 13.939096575351642926 | 16.858531614968249812 |
| Au | 17.498714972890205388 | 18.434874560758355955 | 11.809435345929324868 |
| Au | 15.537936662791459241 | 16.638854898067634736 | 12.114866461865211988 |
| Au | 17.367233117058795244 | 17.220821915422519055 | 14.194416564745608866 |

13

Au13-(34)

| | | | |
|---|---|---|---|
| Au | 13.897125019618743025 | 17.447225330685029121 | 15.161412352290282612 |
| Au | 14.887290860239302148 | 17.866202639409188180 | 12.751307111230493874 |
| Au | 12.539619836113994111 | 14.996590136845551200 | 15.417080306502185749 |
| Au | 12.669466358293206554 | 16.269690196016703965 | 12.793884080204966480 |
| Au | 14.570315767995440481 | 15.672296580936816213 | 17.206741194233106995 |
| Au | 15.932684939136068536 | 13.366331798462580238 | 16.962481724263906102 |
| Au | 12.143195889336244164 | 17.015896340426269973 | 17.183772472228987738 |
| Au | 13.715707994375261336 | 12.514654053534762568 | 15.362299699687348209 |
| Au | 15.311447048661039005 | 10.809246176440824527 | 16.670322834500638010 |
| Au | 12.503680153396393493 | 13.572086161510089397 | 13.195857311810302193 |
| Au | 19.000353854223352101 | 17.506670507802908787 | 16.003019276382161706 |
| Au | 17.216835309056449432 | 15.689718732283626679 | 16.729915159768601995 |
| Au | 16.607933681555003602 | 17.742148714645935570 | 14.825617985896496265 |

13

Au13-(35)

| | | | |
|---|---|---|---|
| Au | 12.418256160921773201 | 17.217989313321908185 | 17.114890719198150748 |
| Au | 12.622507240878071855 | 14.999399435345276643 | 15.505307384521691105 |
| Au | 14.436316452190794735 | 15.379475716016843378 | 17.439358061117253840 |
| Au | 14.360508217158812982 | 17.373661996199402324 | 15.258107451898926143 |
| Au | 13.019599141907209017 | 16.291713868737975446 | 13.000323586075825943 |
| Au | 14.896155587085372574 | 18.177922873196365572 | 12.706094076566010642 |
| Au | 12.193640549237533932 | 13.726834999543680027 | 13.162788886387824405 |
| Au | 16.663116304913444310 | 17.067506442980565851 | 16.892906859388347129 |
| Au | 16.743010754797520434 | 18.570095164261928744 | 14.679375827483406525 |
| Au | 16.938364394271758329 | 14.741223244245150070 | 18.233214014188074970 |
| Au | 14.895521015974397372 | 10.403027059257498266 | 16.152456098727292044 |
| Au | 15.219708429176863618 | 12.823465616265773193 | 17.265234162454465405 |
| Au | 13.367071647486584496 | 12.434808298627533674 | 15.291227325993174802 |

13

Au13-(36)

| | | | |
|---|---|---|---|
| Au | 12.423361182568502414 | 15.303985278344955034 | 15.581585142570279601 |
| Au | 11.845310253834112402 | 16.916712238967786419 | 17.652093859438579670 |
| Au | 11.458888876085119080 | 13.556892484671495325 | 13.795743957920636902 |
| Au | 15.416935180805053207 | 11.140279504176954006 | 15.993122636860544361 |
| Au | 15.729570046468179356 | 13.661679138816708701 | 16.791344170523206714 |
| Au | 13.827899444183671562 | 13.026330286051509333 | 14.927122030341202219 |
| Au | 16.209422872115624159 | 14.931205169538500854 | 19.222696891735942870 |
| Au | 17.036106686457152648 | 16.113417886208832641 | 16.903928382925528240 |
| Au | 14.316504946489628125 | 15.946420169158031399 | 17.449290248451983842 |
| Au | 17.373171265109434103 | 17.061979418522216889 | 14.343571618132115120 |
| Au | 19.132313826145093572 | 17.625534787466843767 | 16.240327677306051868 |
| Au | 15.830447165391776210 | 16.682020889524672214 | 12.262806884837349486 |
| Au | 15.083325964347078951 | 15.548580073551462988 | 14.640100859956584500 |